\documentclass[aps,prd,10pt,superscriptaddress,twocolumn,nofootinbib,dvipsnames,longbibliography]{revtex4-2}
\usepackage[sort&compress]{natbib}
\usepackage[colorlinks=true,linkcolor=purple,filecolor=orange,urlcolor=blue,citecolor=purple,plainpages=false]{hyperref}
\usepackage{url}
\pdfoutput=1
\usepackage{ifpdf}
\usepackage[utf8]{inputenc}
\usepackage{color,hhline,psfrag,rotating}
\usepackage{dcolumn}
\usepackage{verbatim}
\usepackage{lipsum}
\usepackage{bm}
\usepackage{slashed}
\usepackage{braket}
\usepackage{rotating}
\usepackage{multirow,tabularx}
\usepackage{graphicx}
\usepackage[dvipsnames]{xcolor}
\usepackage{amsfonts,amssymb,amsmath}
\usepackage{booktabs}  

\mathchardef\mhyphen="2D

\begin{document}

\global\let\newpage\relax

\title{Standard Model predictions for $B\to K\ell^+\ell^-$, $B\to K\ell_1^- \ell_2^+$ and $B\to K\nu\bar{\nu}$ using form factors from $N_f=2+1+1$  lattice QCD}
\author{W.~G.~Parrott}
\email[]{w.parrott.1@research.gla.ac.uk}
\affiliation{SUPA, School of Physics and Astronomy, University of Glasgow, Glasgow, G12 8QQ, UK}
\author{C.~Bouchard}
\email[]{chris.bouchard@glasgow.ac.uk}
\affiliation{SUPA, School of Physics and Astronomy, University of Glasgow, Glasgow, G12 8QQ, UK}
\author{C.~T.~H.~Davies}
\email[]{christine.davies@glasgow.ac.uk}
\affiliation{SUPA, School of Physics and Astronomy, University of Glasgow, Glasgow, G12 8QQ, UK}

\collaboration{HPQCD collaboration}
\homepage{http://www.physics.gla.ac.uk/HPQCD}
\noaffiliation

\date{\today}

\begin{abstract}
We use HPQCD's recent lattice QCD determination of $B \to K$ scalar, vector and tensor form factors to determine Standard Model differential branching fractions for $B \to K \ell^+\ell^-$, $B\to K \ell_1^+\ell_2^-$ and $B \to K\nu \overline{\nu}$. These form factors are calculated across the full $q^2$ range of the decay and have smaller uncertainties than previous work, particularly at low $q^2$. For $B \to K \ell^+ \ell^-$ we find the Standard Model branching fraction in the $q^2$ region below the squared $J/\psi$ mass to exceed the LHCb results, with tensions as high as $4.2\sigma$ for $B^+\to K^+\mu^+\mu^-$. For the high $q^2$ region we see $2.7\sigma$ tensions. The tensions are much reduced by applying shifts to Wilson coefficients $C_9$ and $C_{10}$ in the effective weak Hamiltonian, moving them away from their Standard Model values consistent with those indicated by other $B$ phenomenology. We also update results for lepton-flavour ratios $R^{\mu}_e$ and $R^{\tau}_{\mu}$ and the `flat term', $F_H^{\ell}$ in the differential branching fraction for $\ell\in\{e,\mu,\tau\}$. Our results for the form-factor-dependent contributions needed for searches for lepton-flavour-violating decays $B\to K\ell^-_1\ell^+_2$ achieve uncertainties of 7\%. We also compute the branching fraction $\mathcal{B}(B\to K\nu\bar{\nu})$ with an uncertainty below 10\%, for comparison with future experimental results. 

\end{abstract}

\maketitle

\section{Introduction} \label{sec:intro}
The weak $B\to K$ transition involves a $b\to s$ flavour changing neutral current (FCNC) which proceeds at loop level in the Standard Model (SM), where contributions are suppressed by loop factors and associated elements of the Cabbibo-Kobayashi-Maskawa (CKM) matrix~\cite{Cabibbo:1963yz, Kobayashi:1973fv}.
This makes FCNC processes promising places to look for effects from new physics.

Tests for new physics require the accurate calculation in the SM of quantities that can be compared to experimental results. Because the weak transition occurs between hadronic bound states of the $b$ and $s$ quarks, key inputs to the SM calculation are the hadronic matrix elements of the weak currents, parameterised by form factors. The form factors are functions of the 4-momentum transfer, $q$, between $B$ and $K$ and depend on the strong interaction effects that bind the quarks inside the mesons. To calculate the form factors from first principles in quantum chromodynamics (QCD) requires the use of lattice QCD. Here we will update the phenomenology for $B\rightarrow K$ decays in the SM using improved form factors calculated using lattice QCD by the HPQCD collaboration~\cite{BtoK} and compare to existing experimental results. We also provide tables of results in a variety of $q^2$ bins that will be useful for comparison to future experimental analyses. 

The lattice QCD form factors that we use~\cite{BtoK} improve on previous calculations in a number of ways. They are the first to be calculated on gluon field configurations (generated by the MILC collaboration~\cite{MILC:2012znn}) that include $u$, $d$, $s$ and $c$ quarks in the sea ($N_f=2+1+1$). They use a fully relativistic discretisation of the QCD Dirac equation on the lattice~\cite{Follana:2006rc} which allows accurate normalisation of the lattice weak currents to their continuum counterparts~\cite{McLean:2019qcx,Hatton:2020vzp}, rather than the $\mathcal{O}(\alpha_s)$ perturbative normalisation used in earlier lattice QCD calculations~\cite{Bouchard:2013mia,Du:2015tda}. The calculation must be done at a range of masses for the heavy quark, from that of the $c$ quark upwards, and at a range of values of the lattice spacing. Very fine lattices must be used to get close to the $b$ quark mass for the heavy quark. Fits to the results then enable form factors for $B\to K$ to be determined and smoothly connected (since only the heavy quark mass changes) to those for $D\to K$. Our form factors for $D\to K$ agree well as a function of $q^2$ with those inferred from the experimental results for the tree-level weak decay $D\to K\ell \overline{\nu}$~\cite{Chakraborty:2021qav}, also giving a better-than-1\% determination of $V_{cs}$. We do not expect to see new physics in this decay, so this is a strong test of the agreement of (lattice) QCD with experiment for a SM weak process. Our $B\to K$ form factors build on this. They are also the first to cover the full  physical $q^2$ range of the $B\to K$ decay; this is possible because of the use of very fine lattices. This means in particular that our form factors have smaller uncertainties in the low $q^2$ region, allowing more accurate SM phenomenology here than has been possible before. 

The $B\to K$ process studied most to date is $B\to K\ell^+\ell^-$, for lepton $\ell=e$ or $\mu$. 
$B\to K\ell^+\ell^-$ experimental data has been collected by BaBar~\cite{Aubert:2008ps, Lees:2012tva, TheBaBar:2016xwe}, Belle~\cite{Wei:2009zv, BELLE:2019xld}, CDF~\cite{Aaltonen:2011qs} and LHCb~\cite{Aaij:2012cq, Aaij:2012vr, Aaij:2014pli, Aaij:2014tfa, Aaij:2014ora, LHCb:2016due, Aaij:2021vac}. We hope to see more experimental data for $\tau$ leptons in the final state in future.

The most direct comparison between the SM and experiment is for the differential branching fraction as a function of $q^2$. A complication is that of contamination by long-distance contributions from charmonium resonances in the intermediate $q^2$ region~\cite{Beneke:2009az}. These are not easily included in the SM analysis (and will not be included here) and so the key comparisons to be made are at low $q^2$, below the resonance region, or at high $q^2$, above the resonance region. Tensions between the SM and experimental results have been seen in earlier work~\cite{Bobeth:2007dw, Bobeth:2011nj, Bobeth:2012vn,Khodjamirian:2012rm,Bouchard:2013mia,Du:2015tda}. Here we will provide an improved analysis with smaller theory uncertainties from the form factors, particularly in the low $q^2$ region. We will also update SM results for the lepton-universality violating ratios, $R^{\mu}_e$ and $R^{\tau}_{\mu}$ (although form factor effects largely cancel in these ratios) and give results for the `flat term', $F^{\ell}_H$. 

The lepton-number-violating decay $B\to K\ell_1^-\ell_2^+$ is of interest for new physics constraints and upper limits on this decay have been obtained by
experiment~\cite{BaBar:2006tnv,BaBar:2012azg,BELLE:2019xld,LHCb:2019bix}. 
Here we provide the hadronic input needed for future analyses of this decay. 

The $B\to K\nu\bar{\nu}$ decay promises to be a useful mode for study in future, because it is free of charmonium resonance contamination. 
There is a small amount of experimental data on this from Belle~\cite{Grygier:2017tzo}, Belle-II~\cite{Belle-II:2021rof} and BaBar~\cite{Lees:2013kla}, with the promise of more to come~\cite{Halder:2021sgd}. 
Here we provide the SM predictions for the differential and total branching fractions for this decay mode with an improved level of accuracy compared to previous calculations. 

As well as comparing to experimental results, where available, we will also compare to earlier theoretical calculations~\cite{Bobeth:2007dw, Bobeth:2011nj, Bobeth:2012vn,Altmannshofer:2012az,Khodjamirian:2012rm,Wang:2012ab,Bouchard:2013mia,Du:2015tda}.
These use a variety of form factors from light-cone sum-rules~\cite{Ball:2004ye, Khodjamirian:2010vf}, lattice QCD (with $N_f=2+1$ sea quarks)~\cite{Bouchard:2013pna,Bailey:2015dka} and a combination of the two, also including dispersive bounds~\cite{Bharucha:2010im}.


The rest of the paper is organised as follows. 
Most of the paper is dedicated to the well-studied $B\to K\ell^-\ell^+$ decays. 
We begin with these decays in Section~\ref{sec:BtoKll}, which outlines the calculation of SM observables in Section~\ref{sec:obs} (including discussions of how we deal with resonances, and the effects from isospin, QED and lepton flavour), comparisons with experiment in Section~\ref{sec:BtoKll_expt} and comparisons with previous theoretical results in Section~\ref{sec:BtoKll_theory}. 
Section~\ref{sec:l1l2} looks at the lepton flavour number violating $B\to K\ell^-_1\ell^+_2$ decay, calculating several parameters relevant to this channel. 
In Section~\ref{sec:nunubar} we calculate the branching fraction for the $B\to K\nu\bar{\nu}$ decay, comparing with experimental bounds and other theoretical work. 
We conclude in Section~\ref{sec:conc}.

This version of our paper contains a correction of the scale used for $\alpha_{\rm EW}$, compared to the published version. Instead of $1/\alpha_{\rm EW}(M_Z)=127.952(9)$~\cite{Zyla:2020zbs}, we take $1/\alpha_{\rm EW}(4.2~\mathrm{GeV})=132.32(5)$, in line with the findings of~\cite{Bobeth:2003at}. This change affects our branching fractions for $B\to K \ell^+\ell^-$ and our results for $B\to K \ell_1^+\ell_2^-$, by an overall multiplicative factor of  $\alpha^2_{\rm EW}(4.2~\mathrm{GeV})/\alpha^2_{\rm EW}(M_Z)\approx0.94$. For most of our results, this leads to a change of approximately $1\sigma$, which does not significantly impact our conclusions. All affected figures and tables have been updated accordingly, and will appear as an erratum.

There is no effect on ratios of branching fractions such as $R^{\mu}_e$ and $F_H$, nor on $B\to K\nu\bar{\nu}$, as this uses $\alpha_{\rm EW}(M_Z)$~\cite{Buras:2014fpa}.

Additionally, we have removed a factor of $\eta_{\mathrm{EW}}=1.007(2)$, which was incorrectly applied to $G_{\mathrm{F}}$ in the original paper. This results in a modest reduction of 1.4\% to all branching fractions.

\section{$B\to K\ell^+ \ell^-$}\label{sec:BtoKll}
\begin{figure}
  \includegraphics[width=0.48\textwidth]{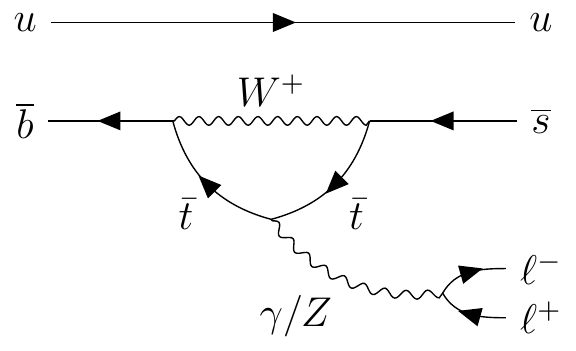}
  \caption{An example of a Feynman diagram for the process $B^+\to K^+\ell^+\ell^-$.}
  \label{fig:Feynman}
\end{figure}

The majority of theoretical and experimental work on $B\to K$ weak decays  has focused on $B\to K\ell^+\ell^-$. An example of a Feynman diagram for this process, sometimes called a penguin diagram, is shown in Figure~\ref{fig:Feynman}. The analysis of such decays in the SM starts from the effective weak Hamiltonian constructed from 4-fermion operators multiplied by Wilson coefficients ($C_1$--$C_{10}$), with the matrix elements of the 4-fermion operators expressed in terms of form factors~\cite{Buras:1998raa}. 
This section discusses the calculation of SM observables related to these decays in Sec.~\ref{sec:obs},  reviewing the inputs used in the evaluation of observables in Sec.~\ref{sec:inputs} and focusing on the Wilson coefficients in Sec.~\ref{sec:Wilson}. Section~\ref{sec:resveto} then discusses how we handle resonance contributions in a way that allows comparison to experiment. Sections~\ref{sec:isospin} and~\ref{sec:QED}  then discuss the uncertainties to account for the fact that our form factors are determined in a pure QCD calculation on the lattice in the strong-isospin limit ($m_u=m_d$). The discussion of uncertainties there is also relevant to our input form factors for Sections~\ref{sec:l1l2} and~\ref{sec:nunubar}. Section~\ref{sec:BtoKll_expt} gives our main results for SM observables and their comparison to experiment. Section~\ref{sec:BtoKll_theory} closes the section with a comparison to earlier theoretical results. 

\subsection{Calculating SM observables}\label{sec:obs}

The SM differential decay rate for $B\to K\ell^+\ell^-$ for lepton $\ell$ is constructed as follows, where we use the notation in~\cite{Bouchard:2013pna, Becirevic:2012fy}.
\begin{equation}
\label{eq:diffrate}
  \frac{d\Gamma_\ell}{dq^2} = 2a_\ell + \frac{2}{3}c_\ell ,
\end{equation}
where $a_\ell$ and $c_\ell$ are given by
\begin{equation}
\label{eq:ac}
  \begin{split}
    a_\ell &= \mathcal{C} \Big[ q^2 |F_P|^2 + \frac{\lambda(q, M_B, M_K)}{4}(|F_A|^2 + |F_V|^2)\\
    &+ 4m_\ell^2 M_B^2 |F_A|^2 + 2m_\ell(M_B^2 - M_K^2 + q^2){\rm Re}(F_PF_A^*) \Big], \\
    c_\ell &= -\frac{\mathcal{C} \lambda(q, M_B, M_K) \beta_\ell^2}{4}(|F_A|^2 + |F_V|^2),
  \end{split}
\end{equation}
with
\begin{align}
\label{eq:Cbeta}
    \mathcal{C} &= \frac{G_F^2 \alpha_{EW}^2 |V_{tb} V_{ts}^*|^2}{2^9\pi^5M_B^3}\beta_\ell \sqrt{\lambda(q, M_B, M_K)}, \nonumber \\
    \beta_\ell &= \sqrt{1-4m_\ell^2/q^2}, \\
  \lambda(a,b,c) &= a^4+b^4+c^4-2(a^2b^2+a^2c^2+b^2c^2).\nonumber 
\end{align}
$F_{P,V,A}$ are constructed from the scalar, vector and tensor form factors $f_0$, $f_+$ and $f_T$ respectively, by
\begin{equation}\label{eq:Fs}
  \begin{split}
    F_P &= -m_\ell C_{10} \Big[ f_+ - \frac{M_B^2-M_K^2}{q^2}(f_0-f_+) \Big], \\
    F_V &= C_9^{\rm eff,1} f_+ + \frac{2m_b^{\overline{\mathrm{MS}}}(\mu_b)}{M_B+M_K}C_7^{\rm eff,1}f_T(\mu_b), \\
    F_A &= C_{10}f_+ \, .
  \end{split}
\end{equation}
The Wilson coefficient $C_9^{\rm eff,1} = C_9^{\rm eff,0} + \Delta C_9^{\rm eff} + \delta C_9^{\rm eff}$ includes nonfactorisable corrections in $\Delta C_9^{\rm eff}$, as well as $\mathcal{O}(\alpha_s)$ and more heavily suppressed corrections in $\delta C_9^{\rm eff}$. 
Similarly, $C_7^{\rm eff,1} = C_7^{\rm eff,0} + \delta C_7^{\rm eff}$ contains $\mathcal{O}(\alpha_s)$ corrections in $\delta C_7^{\rm eff}$.
These corrections are discussed in detail in Appendix~\ref{app:corrs}.

$C_9^{\rm eff,0}$ ($\equiv C_9(\mu_b)+Y(q^2)$) is a function of $q^2$ through
\begin{equation}\label{eq:Y}
  \begin{split}
    Y(q^2) &= \frac{4}{3}C_3 + \frac{64}{9}C_5 + \frac{64}{27}C_6 \\
    &- \frac{1}{2}h(q^2,0)\left( C_3 + \frac{4}{3}C_4 + 16C_5 + \frac{64}{3}C_6 \right)   \\
    &+h(q^2,m_c)\left( \frac{4}{3}C_1 + C_2 + 6C_3 + 60C_5 \right) \\
    &- \frac{1}{2}h(q^2,m_b)\left( 7C_3 + \frac{4}{3}C_4 + 76C_5 + \frac{64}{3}C_6 \right), 
  \end{split}
\end{equation}
where
\begin{equation}\label{eq:h}
  \begin{split}
    h(q^2,m) &= -\frac{4}{9}\left( \ln \frac{m^2}{\mu^2} - \frac{2}{3} - x \right) -  \frac{4}{9}(2+x)  \\
    &\times\hspace{-0.04in}
      \begin{cases} 
	\sqrt{x-1} \arctan \frac{1}{\sqrt{x-1}} \hspace{-0.1in}&\!\!, x>1 \vspace{0.02in}\\
	\sqrt{1-x} \left(\ln \frac{1+\sqrt{1-x}}{\sqrt{x}} - \frac{i\pi}{2}\right)  \hspace{-0.1in}&\!\!, x\leq1,
      \end{cases}
  \end{split} 
\end{equation}
with $x=4m^2/q^2$.

\subsubsection{Inputs}\label{sec:inputs}
\begin{figure}
\includegraphics[width=0.48\textwidth]{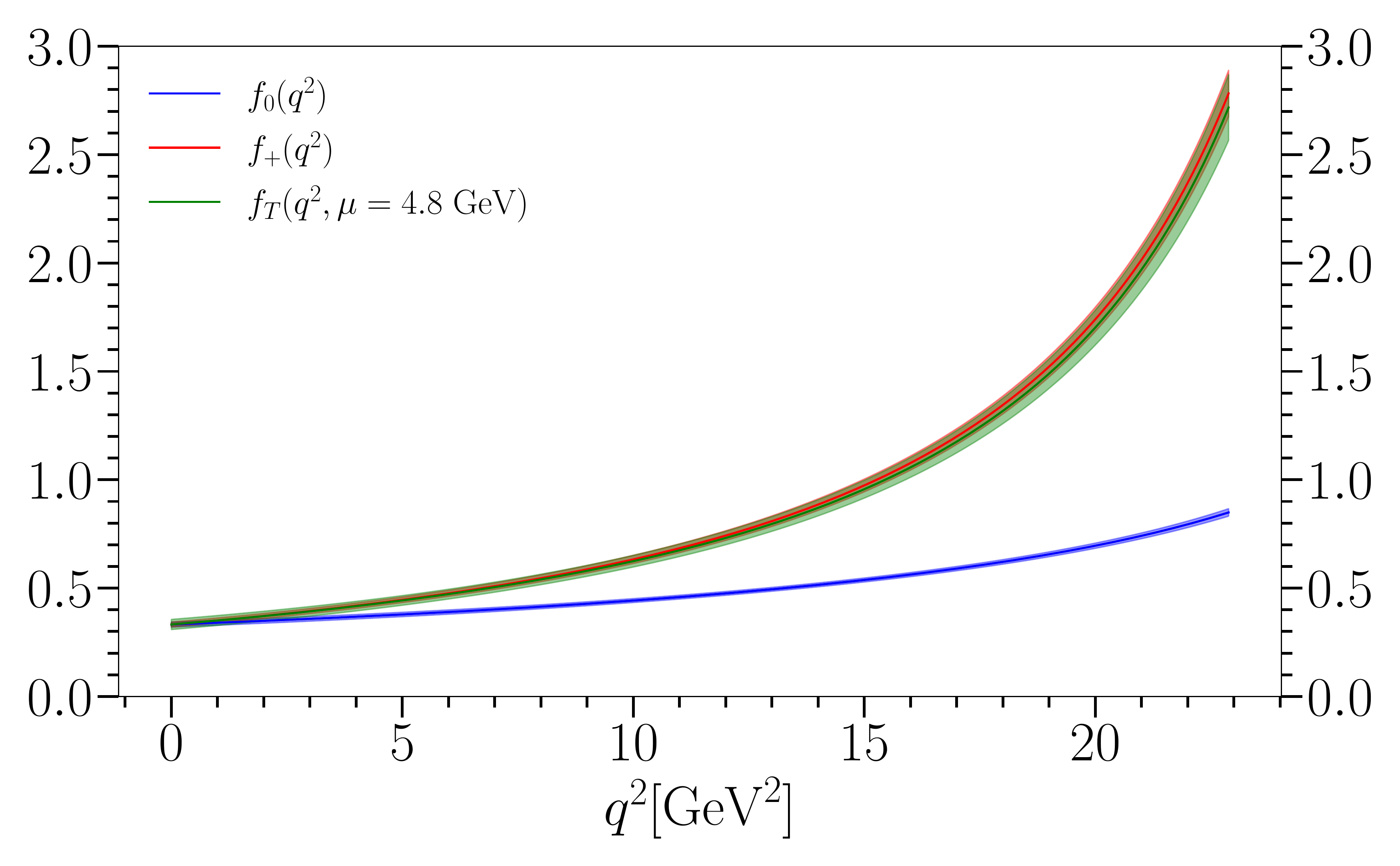}
\caption{Form factors used in this analysis, from~\cite{BtoK}.}
\label{fig:FFs}
\end{figure}

%
\begin{table}
  \begin{tabular}{ccc}
    \hline\hline	
    Parameter  		& Value							& Reference    \\ 
    \hline
    $G_F$		& $1.1663787(6)\times10^{-5}$\,$\text{GeV}^{-2}$	& \cite{Zyla:2020zbs}\\
    $m_c^{\overline{\mathrm{MS}}}(m_c^{\overline{\mathrm{MS}}})$& 1.2719(78)\,GeV& See caption	\\
    $m_b^{\overline{\mathrm{MS}}}(\mu_b)$& 4.209(21)\,GeV& \cite{Hatton:2021syc}	\\
    $m_c$		& 1.68(20)\,GeV				&    -  \\
    $m_b$		& 4.87(20)\,GeV				&    - \\
    $f_{K^+}$          & 0.1557(3)\,GeV& \cite{FLAG2021,FermilabLattice:2014tsy,Dowdall:2013rya,Carrasco:2014poa}\\
    $f_{B^+}$             & 0.1894(14)\,GeV                        &  \cite{Bazavov:2017lyh}\\
    $\tau_{B^0}$	& 1.519(4)\,ps				& \cite{HFLAV:2019otj}	\\
    $\tau_{B^\pm}$  	& 1.638(4)\,ps				& \cite{HFLAV:2019otj}	\\
    $1/\alpha_{\rm EW}(\mu_b)$ 	& 132.32(5)				&- \\
    $|V_{tb}V_{ts}^*|$	& 0.04185(93)				& \cite{Dowdall:2019bea}	\\
    $C_1(\mu_b)$		& -0.294(9)  			        & \cite{Blake:2016olu} 	\\
    $C_2(\mu_b)$		& 1.017(1)  				& \cite{Blake:2016olu} 	\\ 
    $C_3(\mu_b)$		& -0.0059(2)  			        & \cite{Blake:2016olu} 	\\
    $C_4(\mu_b)$		& -0.087(1)  			        & \cite{Blake:2016olu} 	\\
    $C_5(\mu_b)$		& 0.0004  		         	& \cite{Blake:2016olu}	\\
    $C_6(\mu_b)$		& 0.0011(1)  		                & \cite{Blake:2016olu} 	\\
    $C_7^{\rm eff,0}(\mu_b)$	& -0.2957(5)  			        & \cite{Blake:2016olu} 	\\
    $C_8^{\mathrm{eff}}(\mu_b)$      & -0.1630(6)  		        & \cite{Blake:2016olu} 	\\
    $C_9(\mu_b)$	& $4.114(14) $  	        & \cite{Blake:2016olu} 	\\
    $C_9^{\rm eff,0}(\mu_b)$	& $C_9(\mu_b) + Y(q^2)$  	        & - 	\\
    $C_{10}(\mu_b)$  & -4.193(33)  			        & \cite{Blake:2016olu} 	\\    
\hline\hline
  \end{tabular}\caption{Input parameters used to calculate SM observables. Details of the error on the Wilson coefficients are given in~\cite{Blake:2016olu}, and these are quoted at $\mu_b=4.2\,\mathrm{GeV}$. For more details on Wilson coefficients and their uncertainties, see Section~\ref{sec:Wilson}. $m_c^{\overline{\mathrm{MS}}}(m_c^{\overline{\mathrm{MS}}})$ is obtained using $m_c^{\overline{\mathrm{MS}}}(3~\mathrm{GeV})=0.9841(51)$~\cite{Hatton:2020qhk} and running the scale to its own mass. $m_c$ and $m_b$ are the $c$ and $b$ quark pole masses obtained from the masses in the $\overline{\mathrm{MS}}$ scheme at three loops (see Appendix~\ref{app:polemass} for details).}
\label{tab-SMparams}
\end{table}
Key inputs to our determination of the differential rate are our recently calculated scalar, vector and tensor form factors~\cite{BtoK}. These are plotted as a function of $q^2$ in Fig.~\ref{fig:FFs}. The plot shows the $\pm 1\sigma$ error bands, smaller than those for previous results, especially in the low $q^2$ region. See~\cite{BtoK} for details on how these form factors were calculated. 
Note that we run $f_T$ (calculated in~\cite{BtoK} at scale 4.8 GeV) to a scale $\mu_b=4.2$\,GeV, to match the scale used in the Wilson coefficients of Table~\ref{tab-SMparams}. This is achieved by following~\cite{Hatton:2020vzp} and multiplying our form factor result $f_T(\mu_b=4.8\,\mathrm{GeV})$ by a factor of $1.00933(16)$. The value for $\alpha_{\rm EW}(\mu_b=4.2\,\mathrm{GeV})$ is run from $1/\alpha_{\rm EW}(4.8\,\mathrm{GeV}) =132.14(4)$~\cite{Pivovarov:2000cr}.

Further external input parameters needed for Eq.~\eqref{eq:diffrate} are provided in Table~\ref{tab-SMparams}. 
%
%
Except in the case of Equation~(\ref{eq:Fs}), all expressions above (and in Appendix~\ref{app:corrs}) use the pole masses $m_c$ and $m_b$ given in Table~\ref{tab-SMparams}. 
These pole masses are derived from the masses given in the same table in the $\overline{\mathrm{MS}}$ scheme, determined to high accuracy using lattice QCD calculations~\cite{Hatton:2020qhk,Hatton:2021syc}.  We convert to the pole mass using the standard perturbative matching factor at three-loop order~\cite{Chetyrkin:1999ys,Melnikov:2000qh}. For more details, see Appendix~\ref{app:polemass}.
The perturbation series in this expression suffers from the presence of a renormalon in the pole mass. 
To account for renormalon-induced uncertainties~\cite{Beneke:1994sw} in the pole mass, we allow an error of 200\,MeV which dwarfs any other uncertainty in the mass. 
We assess the impact of this uncertainty by shifting the central values of $m_c$ and $m_b$ down by 200\,MeV. In the case of $m_b$, we find almost no effect on the branching fractions in the important $1.1\leq q^2/\mathrm{GeV}^2 \leq 6$ and $15\leq q^2/\mathrm{GeV}^2 \leq 22$ regions. In the case of $m_c$, a maximal effect of $0.3\,\sigma$ is seen.

To obtain the branching fraction $\mathcal{B}$, we note that $\mathcal{B}_{\ell}=\Gamma_{\ell}\tau_B$, where $\tau_B$ is the lifetime of the $B$ meson. The values for $\tau$ used are given in Table~\ref{tab-SMparams}. 

The value of $|V_{tb}V^*_{ts}|$ that we use comes from the recent HPQCD lattice QCD calculation~\cite{Dowdall:2019bea} of the mixing matrix elements for the $B_s$ meson, the first to be done with $N_f=2+1+1$ quarks in the sea. $|V_{tb}V_{ts}^*|$ is obtained by combining the lattice QCD result with the experimentally-determined $B_s$ oscillation rate~\cite{Zyla:2020zbs}. The oscillation rate can be determined to high precision, so the uncertainty in $|V_{tb}V_{ts}^*|$ is dominated by that from the lattice QCD result, where the uncertainty is smaller than that for previous lattice QCD calculations. The value of $|V_{ts}|$ quoted in~\cite{Dowdall:2019bea} used $|V_{tb}|=0.999093$ from~\cite{CKMfitter}. The value of $|V_{ts}|$ that we use is about 1$\sigma$ higher than values based on CKM unitarity that do not use the information from the $B_s$ oscillation rate, for example that in~\cite{CKMfitter21}. 

\subsubsection{Wilson coefficients}\label{sec:Wilson}
The Wilson coefficients we use are from Table I of~\cite{Blake:2016olu}. They are quoted at $\mu_b=4.2~\mathrm{GeV}$ with uncertainties given for higher order corrections, as well as for a variation (doubling and halving) of the matching scale at the top quark mass, $m_t$. These uncertainties do not allow for possible scale dependence in $\mu_b$, however. Whilst our differential decay rates should be scale independent, we test this over a reasonable variation of $\mu_b$ and include an uncertainty to account for any variation. We look at the effect of changing $\mu_b$ to $4.8~\mathrm{GeV}$ on branching fractions (for both meson charges and $\ell\in\{e,\mu\}$) in the key ranges $1.1~\mathrm{GeV}^2\leq q^2\leq 6~\mathrm{GeV}^2$ and $15~\mathrm{GeV}^2\leq q^2\leq 22~\mathrm{GeV}^2$. This involves the following changes:

1) Removing the running of $f_T(\mu_b)$ described above, so we have $f_T(\mu_b=4.8~\mathrm{GeV})$. 

2) Running the $m_b^{\overline{\mathrm{MS}}}(\mu_b)$ given in Table~\ref{tab-SMparams} to $m_b^{\overline{\mathrm{MS}}}(\mu_b=4.8~\mathrm{GeV})=4.108(24)~\mathrm{GeV}$.

3) Shifting the central values of the Wilson coefficients to the $4.8~\mathrm{GeV}$ values given Table I of~\cite{Blake:2016olu}.

4) Setting $1/\alpha_{\rm EW}(4.8\,\mathrm{GeV}) =132.14(4)$~\cite{Pivovarov:2000cr}.

After making these changes, we observe a change of the same absolute size and sign in the branching fractions for both $q^2$ regions. This corresponds to a change of between $1.4\%$ and $2.6\%$ on the branching fraction, a maximum of $0.3\sigma$. Though a modest effect, we account for this by including a $2.5\%$ uncertainty on $a_{\ell}$ and $c_{\ell}$ across the full $q^2$ range to allow for reasonable variations in $\mu_b$.  

\subsubsection{Resonances and vetoed regions}\label{sec:resveto}
From Fig.~\ref{fig:Feynman} it is easy to see how a resonance that couples to the photon can affect the differential rate for $B\to K\ell^+\ell^-$, causing a spike at $q^2=M^2$ for a resonance of mass $M$. These `long-distance' resonance effects are not included in the `short-distance' SM calculation of Eqs.~\eqref{eq:diffrate}--\eqref{eq:h}. The resonances of concern are $u\overline{u}$ and $d\overline{d}$ resonances (the $\rho$ and $\omega$), which appear at very low $q^2$, below $1~\mathrm{GeV}^2$, the $s\overline{s}$ $\phi$ close to $1~\mathrm{GeV}^2$ and $c\overline{c}$ resonances, which affect the intermediate $q^2$ region. The $c\bar{c}$ resonances that dominate are the $J/\psi$ and $\psi(2S)$ and experimentalists typically discard their data  in the vicinity of these resonances for a $B\to K\ell^+\ell^-$ analysis. Whilst the regions blocked out vary somewhat from experiment to experiment, we will use the choices $8.68\leq q^2/\text{GeV}^2 \leq 10.11$ and $12.86\leq q^2/\text{GeV}^2 \leq 14.18$, which are adopted by several experiments, and mark these as `vetoed regions' (greyed out in our plots). In these two regions and between them, the resonance contributions are large~\cite{LHCb:2016due} and there are strong violations of quark hadron duality~\cite{Beneke:2009az}. The decay rate is dominated by $B\to K\Psi\to K\ell^+\ell^-$~\cite{Beylich:2011aq} and is approximately two orders of magnitude larger than the short-distance rate of Eqs.~\eqref{eq:diffrate}--\eqref{eq:h}. 

To avoid these resonances in the comparison between the SM short-distance calculations and experiment it is common practice to restrict the $q^2$ range to two `well-behaved' regions,  giving separate results for the integrals over $1.1\leq q^2/\text{GeV}^2 \leq 6$ and $15\leq q^2/\text{GeV}^2 \leq 22$ (or similar). The lower $q^2$ region here is above the light-quark resonances but below the $c\overline{c}$ resonances. The higher $q^2$ region is above the dominant $c\overline{c}$ resonances, but small contributions are seen in the experimental data from higher resonances in this region~\cite{LHCb:2016due}. Following~\cite{Du:2015tda}, we allow for an uncertainty of 2\% on the differential rate in this high $q^2$ region to allow for these additional resonance contributions. We do this by applying a factor of $1.00(2)$ to $a_{\ell}$ and $c_{\ell}$ (Eq.~\eqref{eq:diffrate}) above $q^2=14.18~\mathrm{GeV}^2$.
We will adopt these two $q^2$ regions here and our key results will be a comparison with experiment for these two regions. This allows us to test for the presence of new physics in as clean a way as possible. 

It is also informative, however, to quote the total branching fraction for the decay, integrating over the full $q^2$ range. This means that some approach must be adopted for the $q^2$ regions where the resonances sit so that a comparison can be made between the short-distance SM branching fraction and experiment. In the very low $q^2$ region this is not an issue; the light-quark resonances contribute only at the 1\% level~\cite{LHCb:2016due}. The narrow $\phi$ resonance can be cut out of the experimental data, where the data is sufficiently accurate to warrant it~\cite{Aaij:2014pli}. 
The $c\overline{c}$ resonances in the intermediate $q^2$ region have a bigger impact over a larger $q^2$ range. The integration over this region must then be handled carefully in a way that is consistent with what is done in experimental analyses. 
To cover  the $c\overline{c}$ resonance region, we linearly interpolate $a_\ell$ and $c_\ell$ across the whole of the two vetoed regions and the gap between them, i.e. for $8.68 \leq q^2/\text{GeV}^2 \leq 14.18$. 
This is only necessary for the case $\ell \in \{e, \mu\}$, as $\ell = \tau$ sits mostly above this region. 
This approach is used in~\cite{Du:2015tda} and similar interpolations are performed in experimental papers (see~\cite{Aaij:2014pli}, the interpolation of which will be compared with ours below). This means that the total branching fraction is then also comparable between theory and experiment. 
Of course, comparisons with other theoretical calculations that are based on Eqs.~\eqref{eq:diffrate}--\eqref{eq:h} can be done at any $q^2$ value. 

Finally, it should be noted that the $m_c$ value we use (Table~\ref{tab-SMparams}) is such that $4m_c^2 = 11.34\,\mathrm{GeV}^2$. This sits between the two vetoed regions and therefore inside the region in which we perform the linear interpolation described. This means that the cusp arising at this point from the function $h$ of Eq.~\eqref{eq:h} is not visible in our plots. 

\subsubsection{Isospin and lepton flavour}\label{sec:isospin}
The $B\to K$ decays we wish to study are for $B^0\to K^0\ell^+\ell^-$ and $B^+\to K^+\ell^+\ell^-$.
The charged and neutral $B$ meson cases differ only in the flavour of the light spectator quark, which can either be an up or a down. 
The masses of the $B^{0/+}$ and $K^{0/+}$ differ slightly as a result of this (along with QED effects). We take these masses from the Particle Data Tables~\cite{Zyla:2020zbs} and use these in the equations of Sec.~\ref{sec:obs} for each case; it has negligible effect on the differential rates. The difference in branching fractions between the charged and neutral cases is much larger, however, because of the difference in $B^+$ and $B^0$ lifetimes, which we take from~\cite{HFLAV:2019otj} and give in Table~\ref{tab-SMparams}.

We must also include an uncertainty for the fact that our form factors~\cite{BtoK} were calculated in pure lattice QCD, with no QED effects included and with degenerate $u$ and $d$ quark masses (as were previous lattice QCD calculations of these form factors~\cite{Bouchard:2013pna, Bailey:2015dka}). Instead of having separate values for $m_u$ and $m_d$ we used $m_u=m_d=m_l$, where $m_l$ is denoted `the light quark mass'. For this reason, our form factors represent the case with $M_B = (M_{B^0}+M_{B^+})/2$ and $M_K = (M_{K^0}+M_{K^+})/2$ with the light quark mass tuned to the physical point $m_s/m_l = 27.18(10)$ corresponding to $m_l=(m_u+m_d)/2$. In order to work out how much our form factors would change if the $u$ and $d$ quark masses were different we follow the procedure adopted in HPQCD's $D \to K$ analysis~\cite{Chakraborty:2021qav}. 

We can repeat our analysis to fit the $B\to K$ form factors from~\cite{BtoK} but changing the physical value of $m_l$ so that it corresponds to either $m_u$ or $m_d$ 
(this means taking $m_s/m_l=27.18\times3/2$ or $27.18\times3/4$ respectively). Note that this will give an overestimate of the effect because this change refers to all the light quarks in the simulation, whereas in fact we just want to change the valence quark in the $B$ and $K$. At the same time as making this change to $m_l$ we also change the meson masses that appear in the fit function to their appropriate values corresponding to a $u$ or $d$ spectator quark. For details on the fit function for the form factors see~\cite{BtoK}. 

Looking at the effects of these tests on the $f_0$, $f_+$ and $f_T$ form factors across the $q^2$ range, we find a maximal effect of $\approx0.5\%$~\cite{BtoK}. To allow for the full effect as an uncertainty on our form factors for the analysis here, we apply an uncorrelated multiplicative factor of $1.000(5)$ to each of our form factors.
We include this uncertainty in our form factors for all of the calculations discussed here, both in this Section on $B \to K \ell^+\ell^-$ and Sections~\ref{sec:l1l2} and~\ref{sec:nunubar}.

The lepton masses, $m_e$, $m_{\mu}$ and $m_{\tau}$ as appropriate, also appear in the equations to determine the differential rate (see Eqs.~\eqref{eq:ac} and~\eqref{eq:Cbeta}). One of the notable changes between leptons is in the minimum $q^2$ value, since $q_{\text{min}}^2 = 4m_{\ell}^2$. In practice, in the case of the differential decay rate, our final result is almost identical for $\ell=e$ and $\ell=\mu$ because their masses are both so small. We sometimes treat these cases together, simply using the $\ell=\mu$ case where a lepton is not explicitly stated. Larger differences are seen for the $\ell=\tau$ case and we show results for that case in Sec.~\ref{sec:BtoKll_theory} where we compare to other theory calculations.

\subsubsection{QED effects}\label{sec:QED}
The largest QED effects in the decay rate for $B\to K \ell^+\ell^-$ are expected to be from final-state interactions of the charged leptons. The photon radiation produced can cause several issues within the experimental analysis, particularly in the hadronic environment of the LHC. These are handled using the PHOTOS software~\cite{Davidson:2010ew} to correct the differential rate to a fully photon-inclusive rate that can be compared with theory. It is also used in the LHC context to obtain a value for $q^2$. Theoretical tests~\cite{Isidori:2020acz,Isidori:2022bzw} indicate that this is a robust procedure. 

We should allow an uncertainty for QED effects in our comparison with experiment, however, and we will base it on the size of QED final-state radiative corrections~\cite{Isidori:2020acz,Isidori:2022bzw}. This will then comfortably include the uncertainty from other QED effects missing from our calculation, such as that from the fact that the quarks inside the $B$ and $K$ mesons have electric charge and that there are additional structure-dependent radiative effects from the $K$ mesons. We will quote the uncertainty for QED effects as a separate uncertainty in Tables where we quote values so that its impact is explicit. We will not include it in plots of the differential rate, nor in comparison with other theory results that do not include a QED uncertainty. 

From~\cite{Isidori:2020acz,Isidori:2022bzw}, we take an overall multiplicative 5\% (2\%) uncertainty to allow for missing QED effects in the case of final state electrons (muons). This is a conservative move, but we shall see that this uncertainty has minimal effect on our results. We take the QED uncertainty for final state $\tau$ to be negligible because the $\tau$ is so heavy.

For the case of the lepton-flavour-universality-violating ratio $R^{\mu}_e$, there has been a lot of analysis of QED effects~\cite{Bordone:2016gaq,Mishra:2020orb,Isidori:2022bzw}. There is agreement that QED effects are well accounted for using PHOTOS~\cite{Davidson:2010ew}. An additional 1\% uncertainty for QED effects in the comparison between theory and experiment is suggested in~\cite{Bordone:2016gaq} and we adopt that here when discussing $R^{\mu}_e$ in Section~\ref{sec:Rmue}.

\subsection{Comparison to experiment}\label{sec:BtoKll_expt}
\begin{table}
  \begin{tabular}{cc}
    \hline\hline
    Label & Reference \\
    \hline
    Experiment &\\
    \hline
    BaBar '08 & \cite{Aubert:2008ps}\\
    BaBar '12 & \cite{Lees:2012tva}\\
    BaBar '16 & \cite{TheBaBar:2016xwe}\\
    Belle '09 & \cite{Wei:2009zv}\\
    Belle '19 & \cite{BELLE:2019xld}\\
    CDF '11 & \cite{Aaltonen:2011qs}\\
    LHCb '12A & \cite{Aaij:2012cq}\\
    LHCb '12B & \cite{Aaij:2012vr}\\
    LHCb '14A & \cite{Aaij:2014pli}\\
    LHCb '14B & \cite{Aaij:2014tfa}\\
    LHCb '14C & \cite{Aaij:2014ora}\\
    LHCb '16 & \cite{LHCb:2016due}\\
    LHCb '21  & \cite{Aaij:2021vac}\\
    \hline
    Theory &\\
    \hline
    AS '12 & \cite{Altmannshofer:2012az}\\
    BHP '07 & \cite{Bobeth:2007dw}\\
    BHDW '11 & \cite{Bobeth:2011nj}\\
    BHD '12 & \cite{Bobeth:2012vn}\\
    FNAL/MILC '15 & \cite{Du:2015tda}\\
    GRDV '22 & \cite{Gubernari:2022hxn} \\
    HPQCD '13 & \cite{Bouchard:2013mia}\\
    HPQCD '22 & This work\\
    KMW '12 & \cite{Khodjamirian:2012rm}\\
    WX '12 & \cite{Wang:2012ab}\\
    \hline\hline
\end{tabular}\caption{Labels used in figures and the corresponding references.}
\label{tab:lookup}
\end{table}

In this section we will give our results for the SM short-distance differential branching fractions (and integrals thereof) for $B\to K \ell^+\ell^-$ using Eqs.~\eqref{eq:diffrate}--\eqref{eq:h} and the improved $B\to K$ form factors of~\cite{BtoK}. We will compare to experimental results and show where tensions exist, and how significant they are. 

There are numerous experimental results in the literature from BaBar~\cite{Aubert:2008ps,Lees:2012tva,TheBaBar:2016xwe}, Belle~\cite{Wei:2009zv,BELLE:2019xld}, CDF~\cite{Aaltonen:2011qs} and LHCb~\cite{Aaij:2012cq,Aaij:2012vr,Aaij:2014pli,Aaij:2014tfa,Aaij:2014ora,LHCb:2016due,Aaij:2021vac}, for decays $B^0\to K^0\ell^+\ell^-$ and $B^+\to K^+\ell^+\ell^-$ where $\ell\in \{e,\mu\}$. There are few for $\ell=\tau$, which we will treat separately. 
Experiments are labelled in figures and Table~\ref{tab:lookup} provides a correspondence between references and labels. 
Naturally, different analyses from one particular experiment are likely to be correlated, and should be viewed as such. 
One could take the most recent data from each of the experiments. However, owing in part to the large number of different choices made with regard to $q^2$ limits, meson charge and lepton flavour, we find it most informative to display all available results in general, even in the case where older measurements have been superseded by more recent analyses. 
This allows the reader to observe the progress of experimental results, in comparison with our own numbers.  
In the discussion of branching fraction $\mathcal{B}$ (for $\ell\in \{e,\mu\}$) that follows, we will group experimental data by $B/K$ charge, as our results are sensitive to this. 
We will not group results by $e$ or $\mu$ lepton flavour because our results (in the SM) are not sensitive to this (see Appendix~\ref{sec:numres} for tables including our results for different $\ell$). 
In the case of lepton flavour, experiments sometimes give $e$ and $\mu$ values separately and sometimes both together. In this latter scenario we will call the lepton $\ell$ and take an average of $e$ and $\mu$. 
All these will be treated together for a given $B/K$ charge. 
Similarly, results are given for a fixed meson charge, as well as for the average of the charged and uncharged case. 
We will use the following labelling conventions, which are adopted in many experimental papers. 
Where a charge is specified, it will be labelled, with the unlabelled case indicating the average of the charged and neutral cases. 
Similarly, $e$ and $\mu$ leptons will be labelled explicitly, with $\ell$ indicating an average of the two. 
Where possible in plots comparing values with experiment, one set of charges or lepton flavours will be given in red, the other in blue, and the average in purple. 
\begin{figure}

\includegraphics[width=0.48\textwidth]{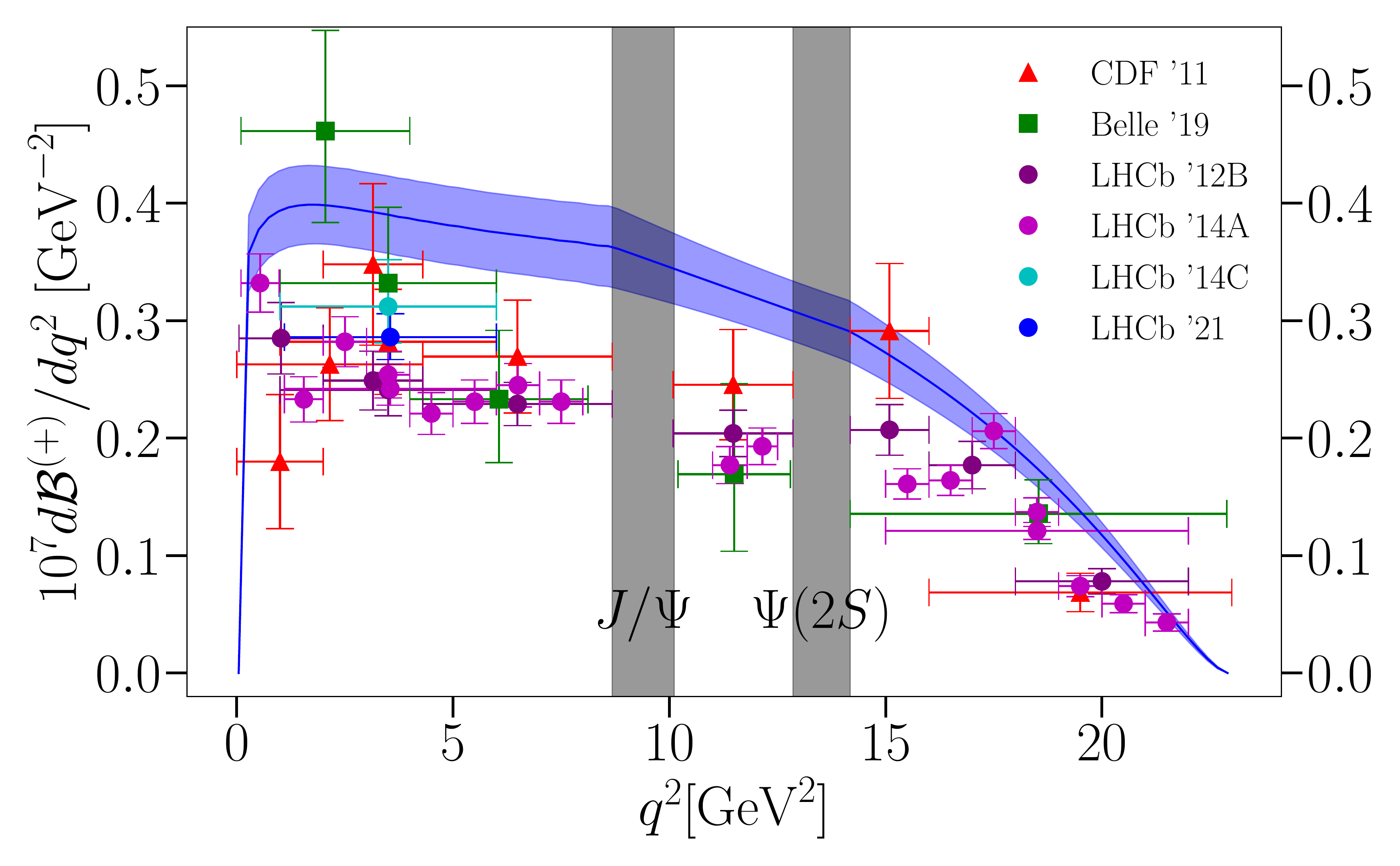}
\caption{Differential branching fraction for $B^+\to K^+\ell^+\ell^-$, with our result in blue, compared with experimental results~\cite{Aaltonen:2011qs,Aaij:2012vr,Aaij:2014pli,Aaij:2014ora,BELLE:2019xld,Aaij:2021vac}. Note that Belle '19, and LHCb '14C and '21 have $\ell=e$, whilst otherwise $\ell=\mu$. Horizontal error bars indicate bin widths.}
\label{fig:dBdqemup}
\end{figure}
\begin{figure}

\includegraphics[width=0.48\textwidth]{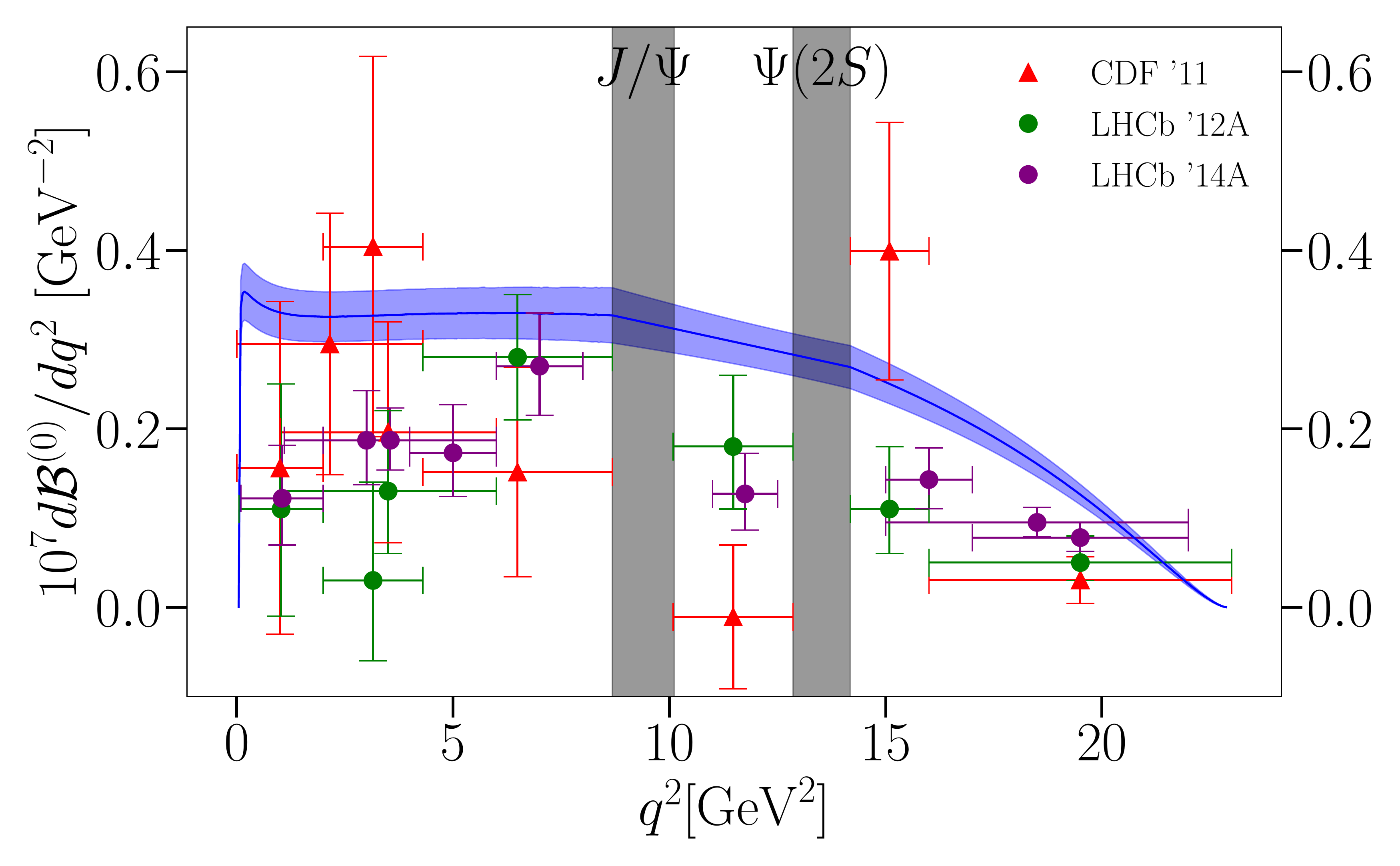}
\caption{Differential branching fraction for $B^0\to K^0\ell^+\ell^-$, with our result in blue, compared with experimental results~\cite{Aaltonen:2011qs,Aaij:2012cq,Aaij:2014pli}. All experimental results take $\ell=\mu$. Horizontal error bars indicate bin widths.}
\label{fig:dBdqemu0}
\end{figure}
\begin{figure}

\includegraphics[width=0.48\textwidth]{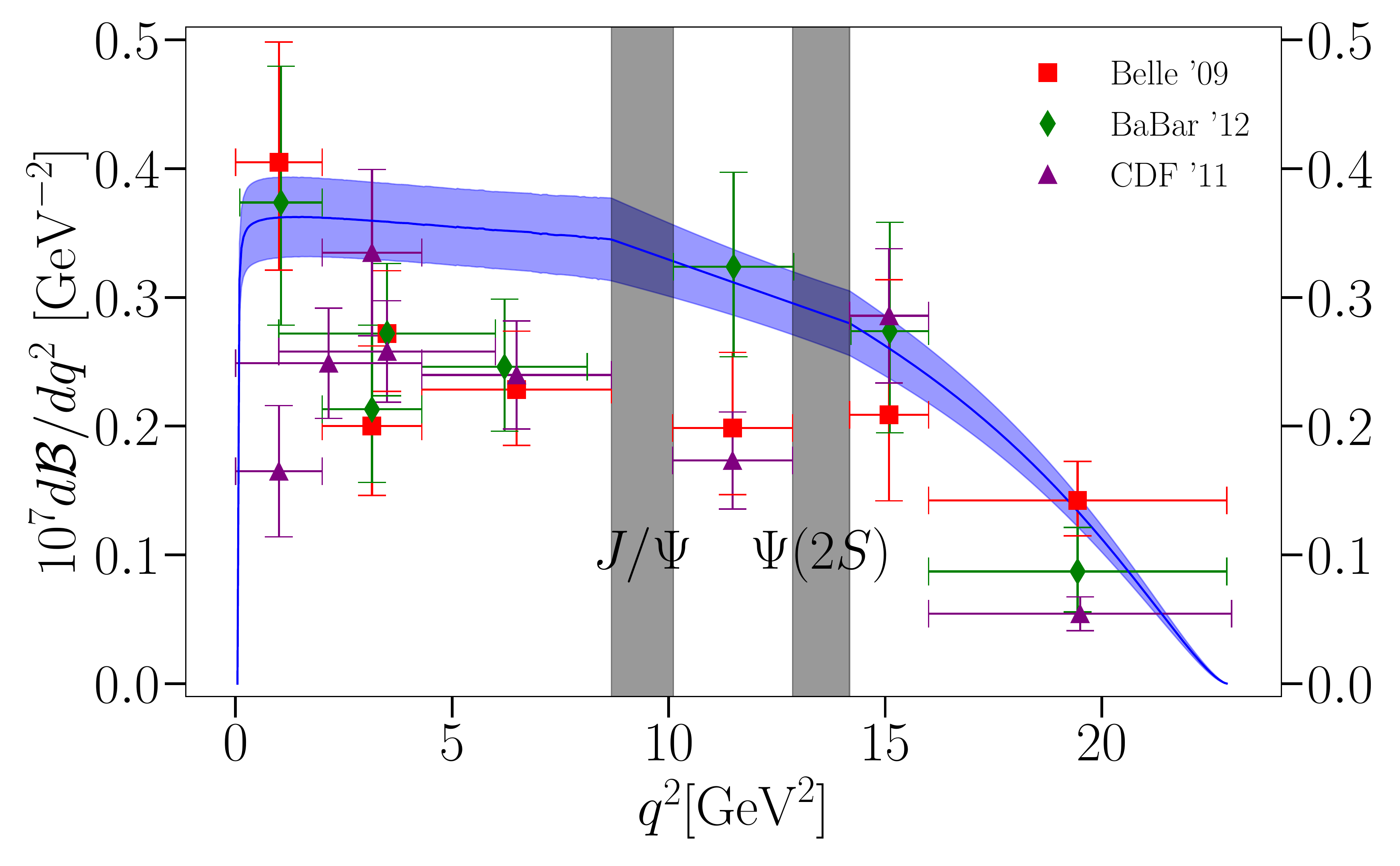}
\caption{Differential branching fraction for $B\to K\ell^+\ell^-$, with our result in blue, compared with experimental results~\cite{Wei:2009zv,Lees:2012tva,Aaltonen:2011qs}. CDF '11 takes $\ell=\mu$, whilst Belle '09 and Babar '12 do not differentiate $e$ from $\mu$. Horizontal error bars indicate bin widths.}
\label{fig:dBdqemu}
\end{figure}
\subsubsection{Differential branching fraction and integrals over low and high $q^2$ regions avoiding $c\overline{c}$ resonances}
\label{sec:diffbrvsexp}

Figures~\ref{fig:dBdqemup},~\ref{fig:dBdqemu0} and~\ref{fig:dBdqemu} show our calculation (as a blue band) of the differential branching fractions for the cases of positively charged mesons, neutral mesons, and the average of the two, respectively, across the full physical $q^2$ range. Note that the different charges give shapes which vary not just because of the overall factor of the lifetime $\tau_{B^{0/+}}$, but also because of the charge dependence of the corrections to $C_9^{\mathrm{eff},0}$ detailed in Appendix~\ref{app:corrs}. The black vertical bands indicate regions vetoed because of the $J/\psi$ and $\psi(2S)$ resonances, not included in our calculation.  As discussed in Section~\ref{sec:resveto}, our result is interpolated across these regions and the gap between them.

Experimental results for decays to electrons, muons or both (averaged) are displayed in each case as coloured points, with the results shown for each experimental $q^2$ bin. The horizontal error bars on the experimental results reflect the width of the bin. Some of the experimental results are for $\ell=e$ and some for $\ell=\mu$; our results are insensitive to the difference. The experiments ignore data taken in the black vetoed regions, but there are results in between these regions. However, we cannot make a reliable comparison between our short-distance SM results and the experimental results between the vetoed regions.

Figures~\ref{fig:dBdqemup},~\ref{fig:dBdqemu0} and~\ref{fig:dBdqemu} show that our results are somewhat higher than experiment in most cases, particularly in the region $4\leq q^2/\mathrm{GeV^2} \leq 8.68$. 
This is most clearly visible in Figure~\ref{fig:dBdqemup}, where the tension between our result and the most precise data from LHCb is obvious. 

\begin{table*}
  \begin{center}
    \begin{tabular}{ccccc}
      \hline
      Channel & Result & $q^2/\mathrm{GeV}^2$ range& $\mathcal{B}\times 10^{7} $ & Tension with HPQCD '22\\
      \hline
      $B^+\to K^+e^+e^-$ & LHCb '21 & $(1.1,6)$& $1.401^{+0.074}_{-0.069}\pm 0.064$& $-2.7\sigma~(-2.4\sigma)$\\
      $B^+\to K^+e^+e^-$ & HPQCD '22 & $(1.1,6)$& $1.91\pm0.16 (\pm0.095)_{\mathrm{QED}}$& -\\
      \hline
      $B^+\to K^+e^+e^-$ & Belle '19 & $(1,6)$& $1.66^{+0.32}_{-0.29}\pm 0.04$& $-0.8\sigma~(-0.8\sigma)$\\
      $B^+\to K^+e^+e^-$ & HPQCD '22 & $(1,6)$& $1.94\pm0.16 (\pm0.097)_{\mathrm{QED}}$& -\\
      \hline
      $B^0\to K^0\mu^+\mu^-$ & LHCb '14A & $(1.1,6)$& $0.92^{+0.17}_{-0.15}\pm 0.044$& $-3.1\sigma~(-3.1\sigma)$\\
      $B^0\to K^0\mu^+\mu^-$ & HPQCD '22 & $(1.1,6)$& $1.60\pm0.14(\pm0.032)_{\mathrm{QED}}$& -\\
      \hline
      $B^0\to K^0\mu^+\mu^-$ & LHCb '14A & $(15,22)$& $0.67^{+0.11}_{-0.11}\pm 0.035$& $-2.7\sigma~(-2.7\sigma)$\\
      $B^0\to K^0\mu^+\mu^-$ & HPQCD '22 & $(15,22)$& $1.070\pm0.095(\pm0.021)_{\mathrm{QED}}$& -\\
      \hline
      $B^+\to K^+\mu^+\mu^-$ & Belle '19 & $(1,6)$& $2.30^{+0.41}_{-0.38}\pm0.05$& $+0.8\sigma~(+0.8\sigma)$\\
      $B^+\to K^+\mu^+\mu^-$ & HPQCD '22 & $(1,6)$& $1.95\pm0.16(\pm0.039)_{\mathrm{QED}}$& -\\
      \hline
      $B^+\to K^+\mu^+\mu^-$ & LHCb '14A & $(1.1,6)$& $1.186\pm0.034\pm 0.059$& $-4.2\sigma~(-4.1\sigma)$\\
      $B^+\to K^+\mu^+\mu^-$ & HPQCD '22 & $(1.1,6)$& $1.91\pm0.16(\pm0.038)_{\mathrm{QED}}$& -\\
      \hline
      $B^+\to K^+\mu^+\mu^-$ & LHCb '14A & $(15,22)$& $0.847\pm0.028\pm 0.042$& $-2.8\sigma~(-2.7\sigma)$\\
      $B^+\to K^+\mu^+\mu^-$ & HPQCD '22 & $(15,22)$& $1.16\pm0.10(\pm0.023)_{\mathrm{QED}}$& -\\
      \hline
          \end{tabular}\caption{Comparison of branching fractions with recent experimental results~\cite{Aaij:2014pli,BELLE:2019xld,Aaij:2021vac} in well behaved regions of $q^2$. Note that the $B^+\to K^+e^+e^-$ result quoted here from LHCb '21 is obtained using the $B^+\to K^+\mu^+\mu^-$ result from LHCb '14A, combined with the ratio determined in LHCb '21. In the fifth column, the tension is given by mean(Experiment - HPQCD)/$\sigma$(Experiment - HPQCD). An additional uncertainty for QED of 5\% for $e$ and 2\% for $\mu$ is included in brackets on each branching fraction (see Section~\ref{sec:QED}). Additionally, the final column gives the tensions with and without QED uncertainty - those including this uncertainty are in brackets.}
    \label{tab:comparisons}
  \end{center}
\end{table*}
\begin{figure}
  \includegraphics[width=0.48\textwidth]{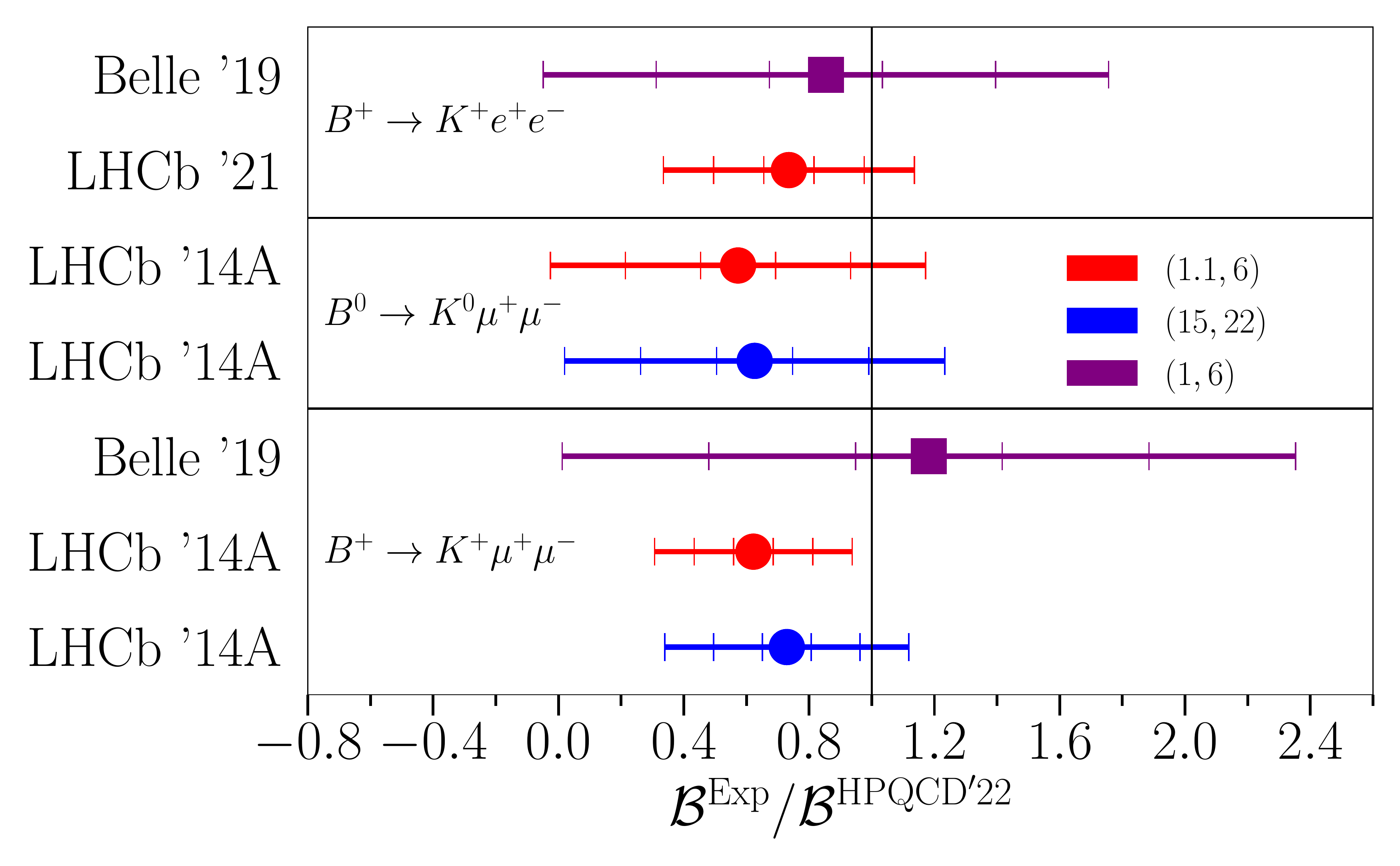}
  \caption{Comparison of branching fractions with recent experimental results~\cite{Aaij:2014pli,BELLE:2019xld,Aaij:2021vac} in low and high regions of $q^2$ away from the charmonium resonance region. Here we show the ratio of the experimental branching fraction to our results, compared to the black vertical line at the value 1. The error bars are $5\sigma$ long, with markers at 1, 3 and 5$\sigma$. Note that the $\sigma$ here are for the ratio, so not the same as those calculated for the difference in Table~\ref{tab:comparisons}. On the right, labels indicate the colours of the $q^2$ bins in units of $\mathrm{GeV}^2$. No uncertainty from QED is included in this plot.}
  \label{fig:headline}
\end{figure}

To examine this tension in more detail, we integrate over two well-behaved $q^2$ regions, one above and one below the $c\overline{c}$ resonances, as discussed in Section~\ref{sec:resveto}. For these regions we can make a reliable comparison with experiment. We show the results in Table~\ref{tab:comparisons}; these constitute our main numerical results.   
In Table~\ref{tab:comparisons}, we compare our branching fractions with the most recent experimental results available for $B\to Ke^+e^-$ and $B\to K\mu^+\mu^-$. Note that our relative uncertainties are comparable to those from the experiments for most of the values. We have larger uncertainties than those for LHCb '14A for $B^+\to K^+\mu^+\mu^-$ but smaller uncertainties than those from Belle '19. Our uncertainties are dominated by those from the form factors, followed by those from the CKM elements $|V_{tb}V_{ts}^*|$.

We find our partial branching fractions to be significantly higher than the experimental values, typically with tensions exceeding $2.7\sigma$ and with LHCb'14A~\cite{Aaij:2014pli} in the low $q^2$ region with a tension reaching $4.2\sigma$ for $B^+\to K^+\mu^+\mu^-$. We do not see any tension with the Belle'19 results~\cite{BELLE:2019xld}, which are themselves in tension with LHCb '14A for $B^+\to K^+\mu^+\mu^-$. Belle '19 have considerably larger uncertainties than those from LHCb and those from our results. The uncertainties we allow for QED make little difference to the tensions that we see. 

The same results, plotted in the form of a ratio of the experimental branching fraction to our result, are shown in Figure~\ref{fig:headline}, with error bars indicating 1, 3 and 5$\sigma$ differences from a value for the ratio of 1. This makes the tensions graphically clearer and also shows the difference in the size of uncertainties in the comparison with different experiments. Note that the tensions in the comparison of the ratio to the value 1 are not exactly the same as in the differences between our result and experiment shown in Table~\ref{tab:comparisons}. 

It is worth noting that the corrections to $C_7^{\mathrm{eff},0}$ and $C_9^{\mathrm{eff},0}$ (discussed in Appendix~\ref{app:corrs}) shift our results by amounts of between 0.3 and 1$\sigma$, depending on the $q^2$ region. In the high $q^2$ region, this is mostly driven by corrections to $C_7^{\mathrm{eff},0}$. In the low $q^2$ region the size and direction of the shift depends on the meson charge.  
The impact of the corrections on the tension between our SM theory and experiment given in Table~\ref{tab:comparisons} then amounts at most to 0.6$\sigma$. 

Our numerical results for integration of our differential branching fraction for a variety of $q^2$ ranges are given in Table~\ref{tab:diffrateBKll} in Appendix~\ref{sec:numres}. 

In Section~\ref{sec:conc} we discuss the effect on our results, and the tensions with experimental values, of changing the values of the Wilson coefficients $C_9$ and $C_{10}$ that we use in Eq.~\eqref{eq:Fs}. This tests the impact of possible new physics at a high scale that would change the effective weak Hamiltonian. 

\subsubsection{Total branching fractions} 
\label{sec:totalbr}

In this section we discuss the total branching fraction, i.e. $d\mathcal{B}/dq^2$ integrated across the physical $q^2$ range from $q^2_{\text{min}} = 4m_{\ell}^2$ to $q^2_{\text{max}} = (M_B-M_K)^2$. 
By convention, experimentally measured branching fractions are extrapolated to the full physical $q^2$ range, ignoring resonances. 
This is achieved using correction factors based on an assumed $q^2$ distribution (see e.g.~\cite{Ali:1999mm}) to the branching fraction integrated over the experimentally measured bins (see discussion in~\cite{Aaij:2014pli}). This achieves the same result as our interpolation across the $c\overline{c}$ resonance region (discussed in Section~\ref{sec:resveto}) and this should make our short-distance theory results comparable to experiment here.

\begin{figure}
\includegraphics[width=0.48\textwidth]{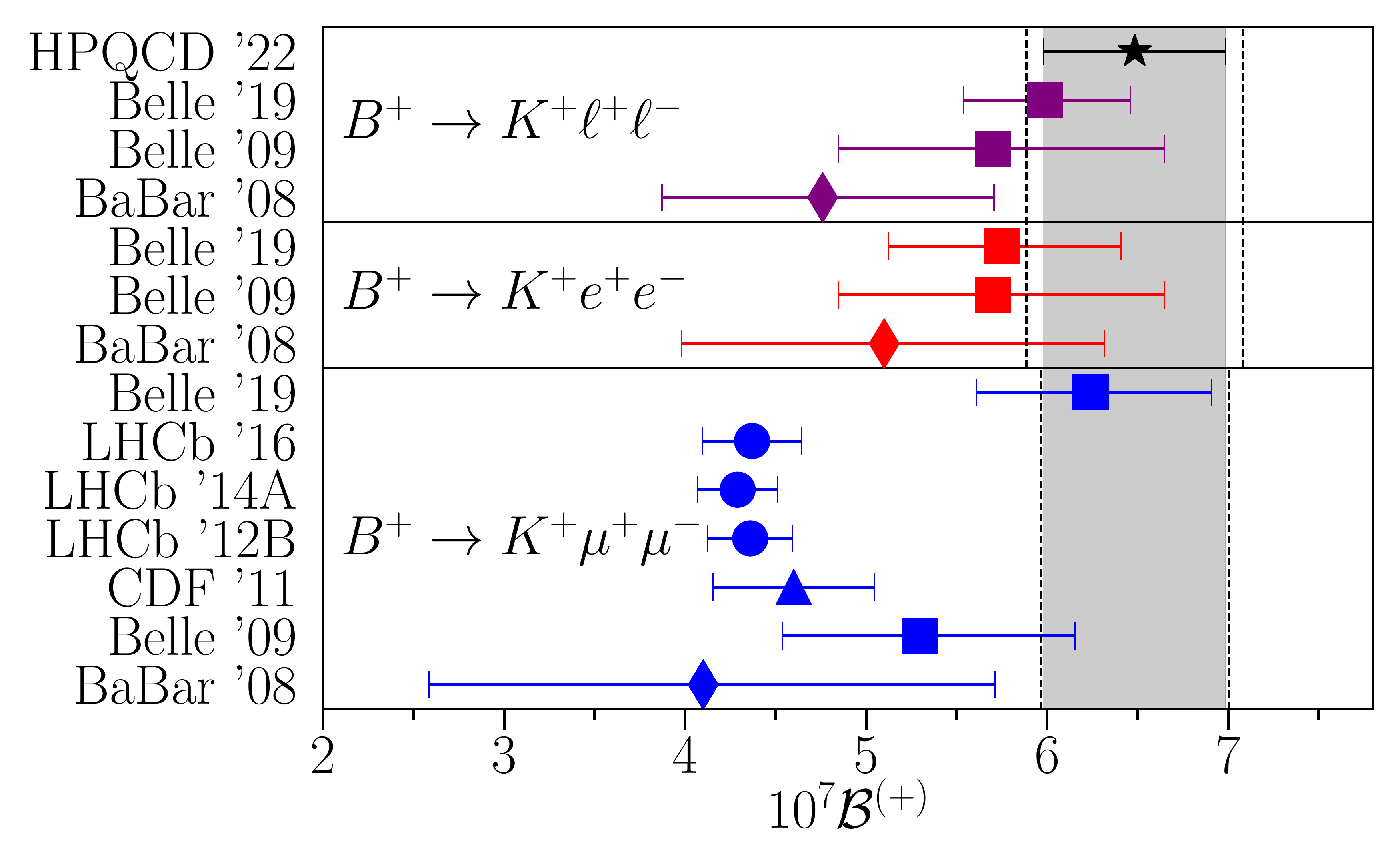}
\caption{The total branching fraction for  $B^+\to K^+\ell^+\ell^-$. 
Our result (HPQCD '22) is given by the black star and grey band, as compared with experimental results~\cite{Aubert:2008ps,Wei:2009zv,Aaij:2014pli,Aaij:2012vr,Aaltonen:2011qs,LHCb:2016due,BELLE:2019xld}. Dashed lines indicate the effect of adding QED uncertainty (see Section~\ref{sec:QED}) to our result.}
\label{fig:Bp}
\end{figure}

\begin{figure}
\includegraphics[width=0.48\textwidth]{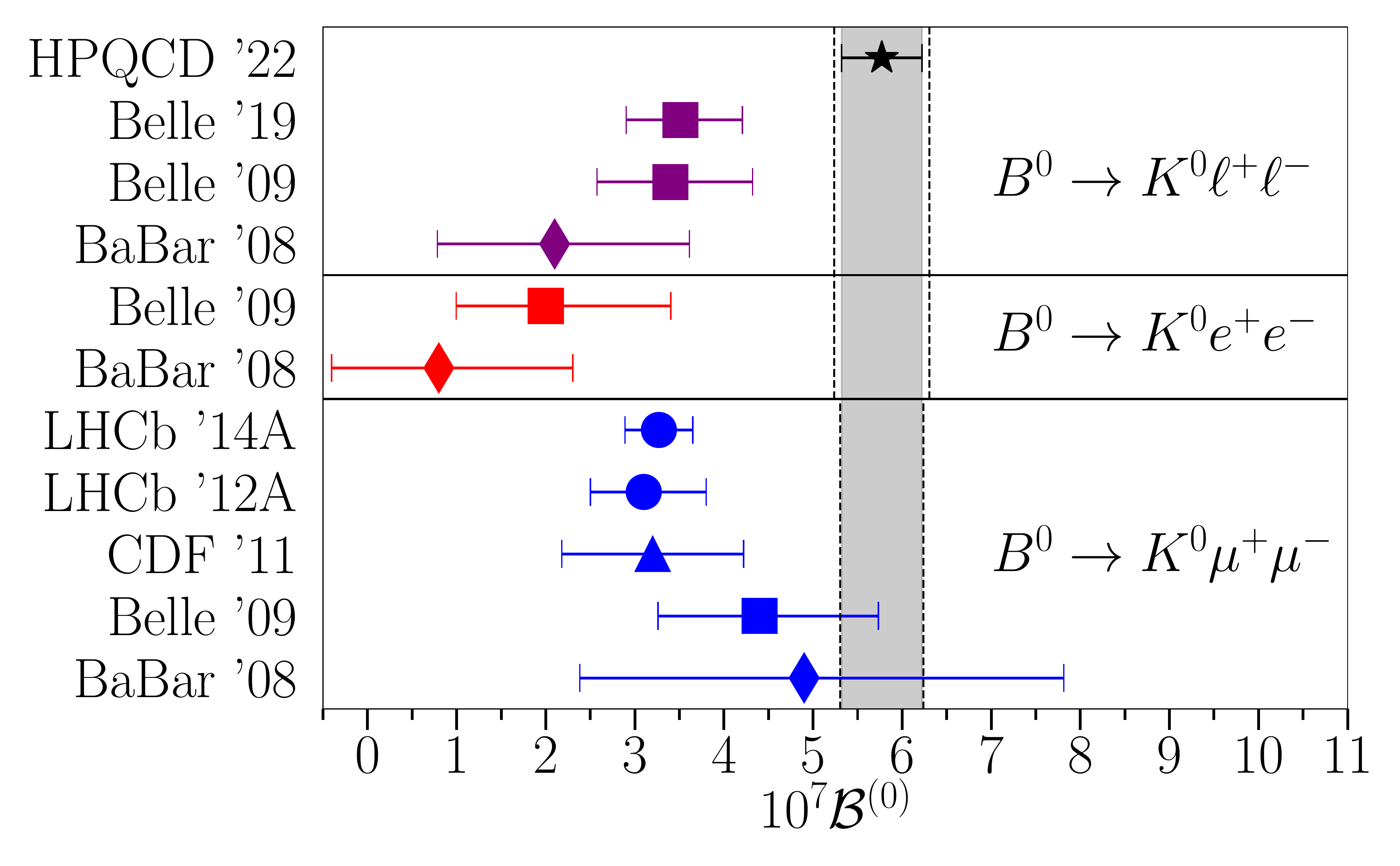}
\caption{The total branching fraction for  $B^0\to K^0\ell^+\ell^-$. 
Our result (HPQCD '22) is given by the black star and grey band, as compared with experimental results~\cite{Aubert:2008ps,Wei:2009zv,Aaij:2014pli,Aaij:2012cq,Aaltonen:2011qs,BELLE:2019xld}. Dashed lines indicate the effect of adding QED uncertainty (see Section~\ref{sec:QED}) to our result.}
\label{fig:B0}
\end{figure}

\begin{figure}
\includegraphics[width=0.48\textwidth]{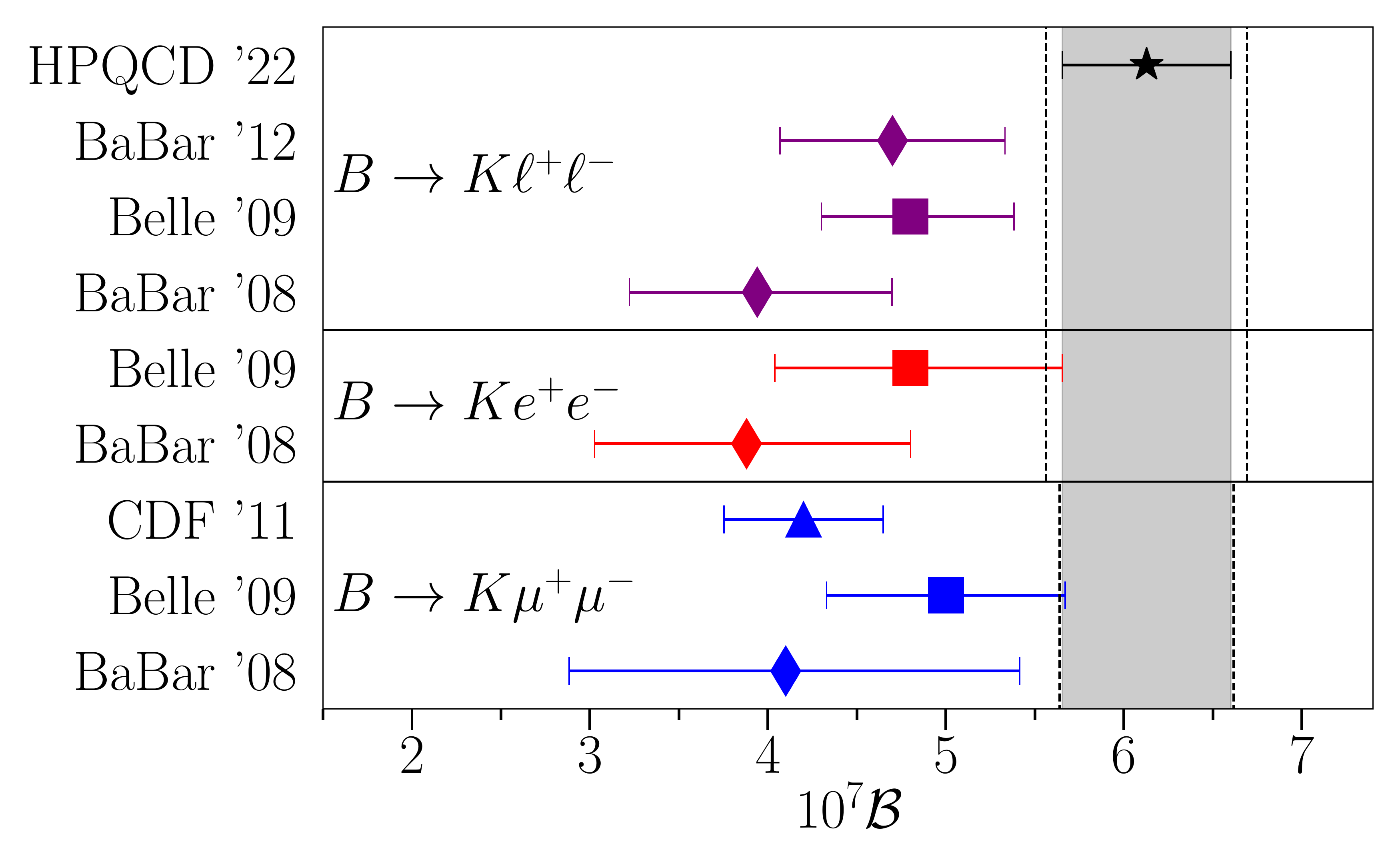}
\caption{The total branching fraction for  $B\to K\ell^+\ell^-$. 
Our result (HPQCD '22) is given by the black star and grey band, as compared with experimental results~\cite{Lees:2012tva,Aubert:2008ps,Wei:2009zv,Aaltonen:2011qs}. Dashed lines indicate the effect of adding QED uncertainty (see Section~\ref{sec:QED}) to our result.}
\label{fig:B}
\end{figure}

Figures~\ref{fig:Bp},~\ref{fig:B0} and~\ref{fig:B} show this comparison for the $\ell\in \{e,\mu\}$ case.
Again, results are split up by meson charge and our result (HPQCD '22) is given at the top, denoted by a black star. 
The grey band carries our result down the plot.
The experimental results are split up according to lepton flavour, but as discussed earlier, our results are insensitive to this distinction. 
As we do not have information about correlations between the various results for each experimental collaboration we do not give an experimental average here for comparison. 

Our result is in general larger than the experimental values, reflecting the pattern in Figures~\ref{fig:dBdqemup},~\ref{fig:dBdqemu0} and~\ref{fig:dBdqemu}, which do not contain exactly the same experiments. 
This is clearest in the case of the neutral meson decay, but both the charged and neutral LHCb '14A results with $\mu$ in the final state in Figures~\ref{fig:Bp} and~\ref{fig:B0} show clear tension: $3.9(4.0)\sigma$ for $B^+$ and $4.2(4.2)\sigma$ for $B^0$, with(without) QED uncertainty.

To test the effect of the LHCb '14A $q^2$ interpolation across the vetoed regions, we compare our total branching fraction (without QED uncertainty) with that when we exclude the regions vetoed by LHCb '14A ($8\leq q^2/\text{GeV}^2\leq 11$ and $12.5\leq q^2/\text{GeV}^2 \leq 15$), we obtain ratios $1.3814(84)$ and $1.3951(84)$ for the $B^+$ and $B^0$ cases, respectively. This compares very well with the factor of $1.39$ used by LHCb, and indicates that the interpolation used by them is consistent with our linear interpolation approach. This adds confidence that the comparison between theory and experiment here is a reliable one. 

We conclude that, in agreement with what was  seen in the low and high $q^2$ regions in Sec.~\ref{sec:diffbrvsexp}, significant tension (exceeding 4$\sigma$) is also seen between the total short-distance SM branching fraction and the most accurate (LHCb) results for $B\to K \ell^+\ell^-$.

For the decay with $\tau$ in the final state, experimental results are much less mature. 
The BaBar collaboration reported the first measurement of the total $B^+\to K^+\tau^+\tau^-$ branching fraction in~\cite{TheBaBar:2016xwe}. 
Their result of $(1.31^{+0.66+0.35}_{-0.61-0.25})\times 10^{-3}$, (where errors are statistical and systematic respectively) is consistent with no signal. It is also consistent with our 7\%-accurate SM result of $\mathcal{B}^{(+)}_{\tau}=1.68(12)\times 10^{-7}$, where we do not include an uncertainty for QED effects since the $\tau$ is so heavy. Note that the $q^2$ range available for the case with the $\tau$ is much smaller than for the lighter leptons. We will discuss the $\tau$ case further in Sec.~\ref{sec:BtoKll_theory}.

Our numerical results for the total branching fraction for $\ell\in \{e,\mu\}$ are given in Table~\ref{tab:diffrateBKll} in Appendix~\ref{sec:numres}. Results for $\tau$ in the final state are given in Table~\ref{tab:BKtau} in the same Appendix.

\begin{figure}

\includegraphics[width=0.48\textwidth]{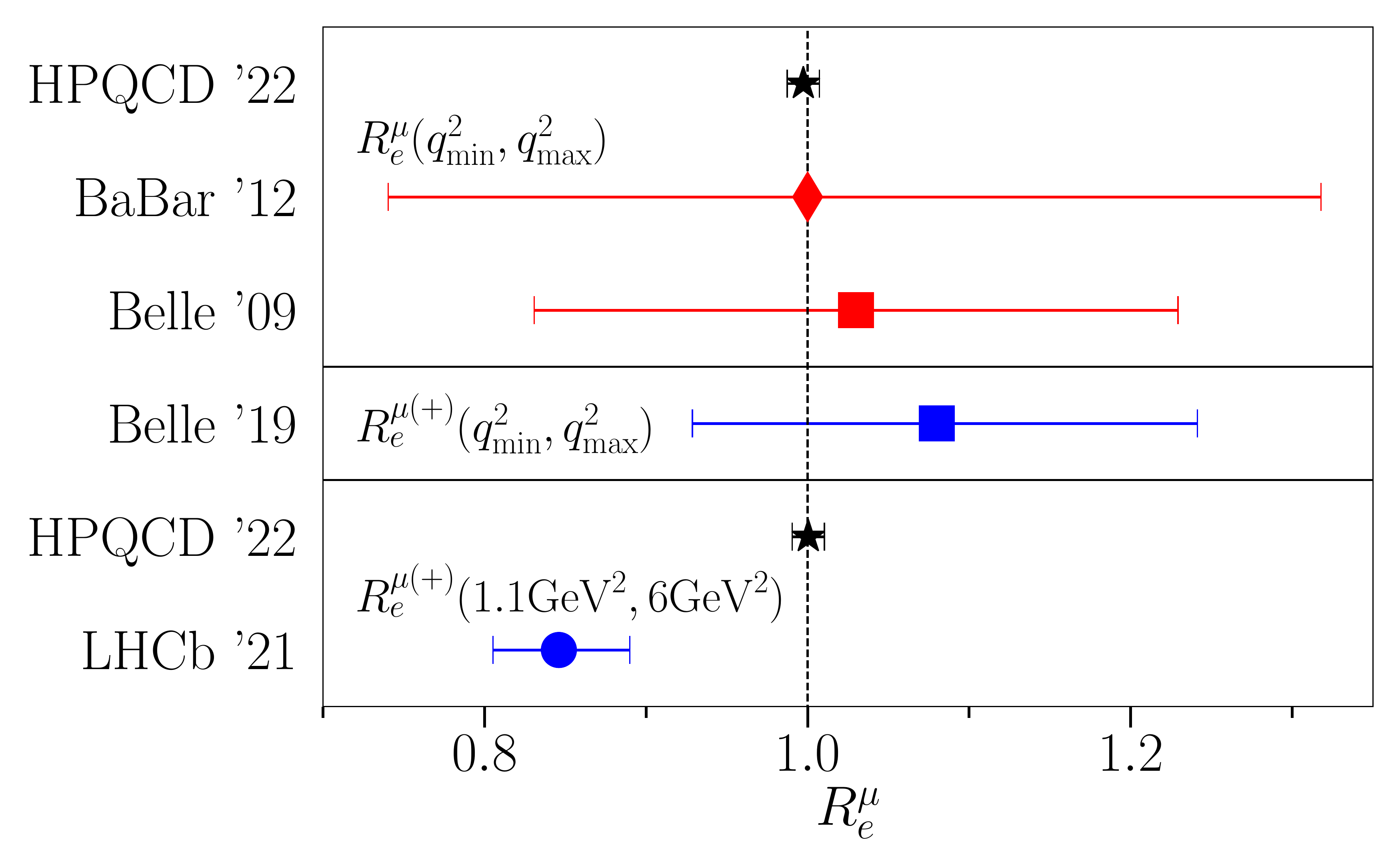}
\caption{The ratio of branching fractions to muons and electrons, as defined in Equation~(\ref{eq:R}), with various $q^2$ limits. Our result (HPQCD '22) has negligible uncertainty compared to experiment~\cite{Wei:2009zv, Lees:2012tva, BELLE:2019xld, Aaij:2021vac}, and does not differ visibly when integrated over different $q^2$ ranges. Note that for the ratios of total branching fractions (top two panes), we have used a lower limit on the $q^2$ integral, $q^2_{\text{min}}$, that differs between $e$ and $\mu$ cases in the ratio. This was done to match what was done in the experiment, but has no visible effect. We include a 1\% uncertainty to allow for possible QED corrections~\cite{Bordone:2016gaq} as discussed in the text. This completely dominates our theoretical error here. The vertical dotted line is at 1.}
\label{fig:Rmueexp}
\end{figure}
\begin{figure}

\includegraphics[width=0.48\textwidth]{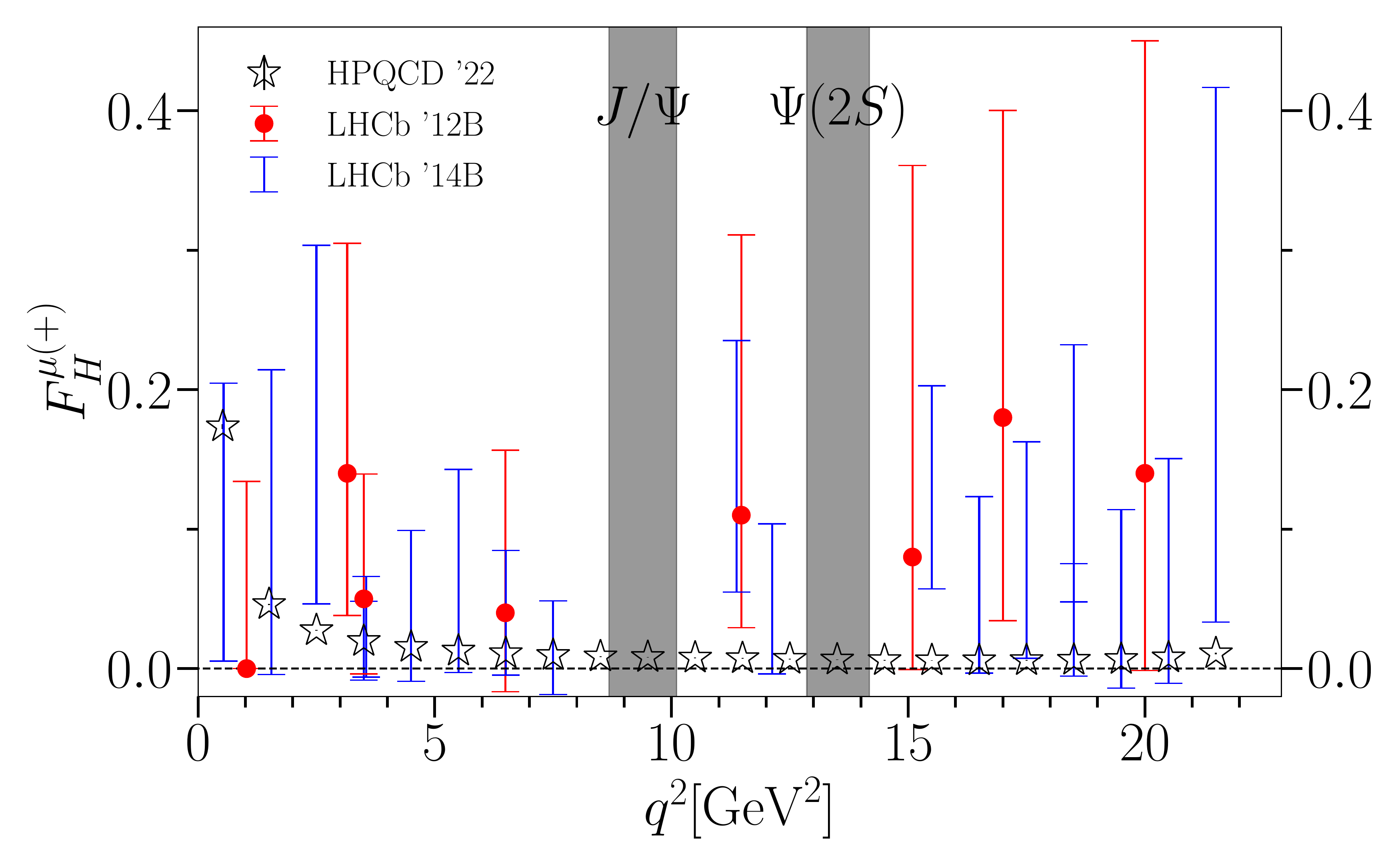}
\caption{The flat term $F_H^{\ell}$ for the $B^+\to K^+\mu^+\mu^-$ decay. 
Our result (HPQCD '22) is given for bins between integer values of $q^2$, with the uncertainty included but too small to be visible. 
Experimental results~\cite{Aaij:2012vr, Aaij:2014tfa} are included for comparison.}
\label{fig:F_H}
\end{figure}

\subsubsection{$R^{\mu}_e$}
\label{sec:Rmue}

We define the ratio of branching fractions for decays to different leptons,
\begin{equation}\label{eq:R}
    R^{\ell_1}_{\ell_2}(q^2_{\text{low}},q^2_{\text{upp}})=\frac{\int_{q^2_{\text{low}}}^{q^2_{\text{upp}}}\frac{d\mathcal{B}_{\ell_1}}{dq^2}dq^2}{\int_{q^2_{\text{low}}}^{q^2_{\text{upp}}}\frac{d\mathcal{B}_{\ell_2}}{dq^2}dq^2}.
\end{equation}
We calculate this separately for the charged and neutral meson cases, $R^{\mu (0/+)}_e$. We also give the charge-averaged result, $R^{\mu}_e$, which is obtained from performing a charge average in each of the numerator and denominator of Eq.~\eqref{eq:R}. 

As discussed above, the effect of changing lepton flavour between electrons and muons on our branching fraction is very small. The factors of $\beta$
in Eqs.~\eqref{eq:ac} and~\eqref{eq:Cbeta} mean that $R^{\mu}_e <1$ at very small values of $q^2$ and the factors of $m_\ell$ mean that $R^{\mu}_e >1$ at large values of $q^2$. However, for almost the entire $q^2$ range $R^{\mu}_e$ is very close to 1. This also means that effects from the form factors and from the resonances cancel between numerator and denominator and the theory uncertainties in $R^\mu_e$ are small. Our results are tabulated in Table~\ref{tab:R-mue-vals} in Appendix~\ref{sec:numres} for a variety of $q^2$ ranges; we find $R^{\mu}_e$ to differ from 1 by a few parts in 1000. 

Radiative corrections from QED (see Sec.~\ref{sec:QED}) affect the $\mu$ and $e$ cases differently but are well understood~\cite{Bordone:2016gaq,Isidori:2020acz}. Tests show that the corrections applied on the experimental side using PHOTOS~\cite{Davidson:2010ew} should be reliable up to $\pm$1\%~\cite{Bordone:2016gaq}. We will take a 1\% uncertainty on our results to allow for missing QED effects in the comparison of our results with experiment. However, where this uncertainty is already included in the experimental results this will amount to a double-counting of this source of error. This QED uncertainty dominates all other sources of uncertainty in our determination of $R^\mu_e$. 

Figure~\ref{fig:Rmueexp} shows a comparison of our result for $R^\mu_e$ (including the 1\% uncertainty) with those from experiment. The most notable experimental results are the recent values from LHCb '21~\cite{Aaij:2021vac} with relatively small uncertainties for the charged meson decay, integrated over the low $q^2$ region. The LHCb '21 result differs from 1 (taken as the SM result when QED corrections are removed~\cite{Bordone:2016gaq}) by 3.1$\sigma$. The relative uncertainty from theory is negligible so having improved form factors, as we do here, makes no difference to this tension. When we come to compare with other theoretical work in Section~\ref{sec:BtoKll_theory}, potential deviations of $R^{\mu}_e$ from 1 will be studied in more detail.

\subsubsection{$F^{\ell}_H$}
\label{sec:flat}

Two additional experimentally measured quantities with which we can compare are based on studying the angular distribution of decay products. The differential distribution in angle is given by~\cite{Bobeth:2007dw} 
\begin{equation}
\label{eq:flatdef}
  \frac{1}{\Gamma_{\ell}}\frac{d\Gamma_{\ell}}{d\cos\theta}=\frac{3}{4}(1-F_H^{\ell})(1-\cos^2\theta)+\frac{1}{2}F_H^{\ell}+A^{\ell}_{\text{FB}}\cos\theta \, ,
\end{equation}
where $\theta$ is the angle between $B$ and $\ell$ as measured in the dilepton rest frame. Both the flat term, $F_H^\ell/2$, and the forward-backward asymmetry, $A^{\ell}_{\text{FB}}$, are small, and sensitive to new physics. $A^{\ell}_{\text{FB}}=0$ in the SM, up to QED corrections, so we will not consider it here.
We can construct $F^{\ell}_H$ from~\cite{Bobeth:2007dw}
\begin{equation}\label{eq:F_H}
  F^{\ell}_{H}(q^2_{\text{low}},q^2_{\text{upp}})=\frac{\int_{q^2_{\text{low}}}^{q^2_{\text{upp}}}(a_{\ell}+c_{\ell})dq^2}{\int_{q^2_{\text{low}}}^{q^2_{\text{upp}}}(a_{\ell}+c_{\ell}/3)dq^2} .
\end{equation}

Figure~\ref{fig:F_H} shows $F^\ell_H$ for the $B^+\to K^+\mu^+\mu^-$ decay. 
The black stars show our result, calculated in bins between integer values of $q^2/\text{GeV}^2$, with the exception of the first bin, which starts at $q^2_{\text{min}}$. 
Experimental results from~\cite{Aaij:2012vr} are included for comparison, as are those from~\cite{Aaij:2014tfa}. 
In the latter case, central values are not given for the data, and so in order to add the systematic error to the statistical confidence interval, this interval is assumed Gaussian, and the central value taken as the middle of the interval. 
The effect of this approximation is negligible, owing to the small systematic errors, and does not affect the comparison with our results, as the overall confidence interval is broad.

As with $R^\mu_e$, the cancellation of correlated uncertainties in Eq.~\eqref{eq:F_H} means that our result here is very precise compared with experiment (our error bars are included in Figure~\ref{fig:F_H} but not visible). We can see that a large reduction in experimental uncertainty will be required in order to reveal any meaningful inconsistency (if one exists) between SM theory and experiment. 

Our results for $F_H^\ell$ for $\ell\in \{e,\mu\}$ are tabulated in Table~\ref{tab:FHl} in Appendix~\ref{sec:numres} for a variety of $q^2$ ranges; we will discuss the flat term for the $\tau$ case in Section~\ref{sec:BtoKll_theory}. 

\subsection{Comparison to theory}\label{sec:BtoKll_theory}

In this section we compare to previous theoretical results, both from previous lattice QCD calculations~\cite{Bouchard:2013mia, Du:2015tda} and others~\cite{Bobeth:2007dw,Bobeth:2011nj,Bobeth:2012vn,Khodjamirian:2012rm,Altmannshofer:2012az,Wang:2012ab,Gubernari:2022hxn}. Again, Table~\ref{tab:lookup} provides a correspondence between references and labels in figures. 
Different theory calculations differ both in the form factors they use and in their input parameters used to determine the differential rates. In~\cite{BtoK} we compare the improved lattice QCD form factors that we use here with earlier lattice QCD form factors. 
Here we will show comparisons of differential and total branching fractions, flat terms etc. as for the comparison with experiment in Section~\ref{sec:BtoKll_expt}. There are many more theory results for the case of $\tau$ leptons in the final state, so we can make more detailed comparison of that case. 

\subsubsection{Differential and total branching fractions}
\label{sec:diffratetheory}

\begin{figure}

\includegraphics[width=0.48\textwidth]{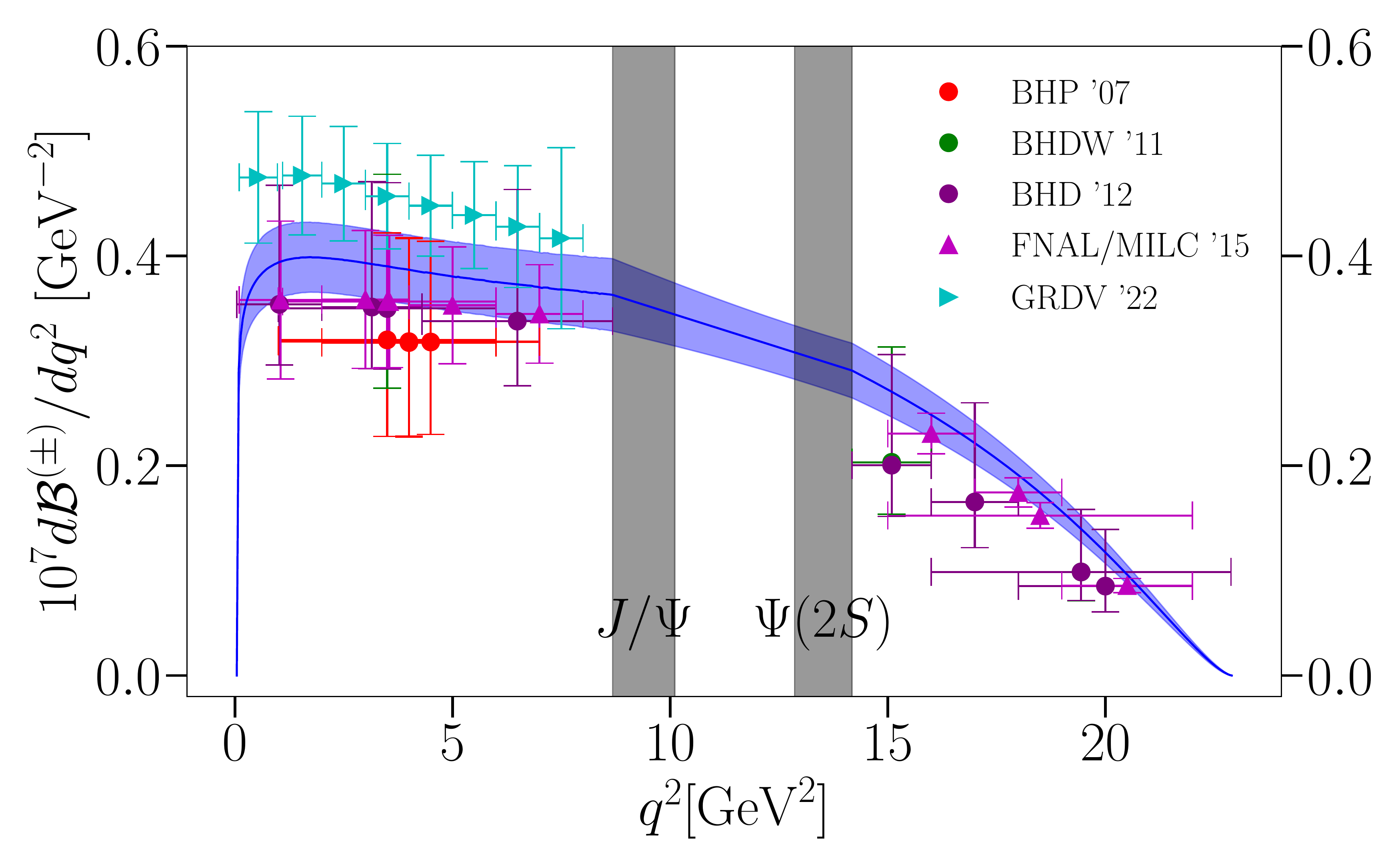}
\caption{Differential branching fraction for $B^{\pm}\to K^{\pm}\ell^+\ell^-$, with our result in blue, compared with earlier work~\cite{Bobeth:2007dw,Bobeth:2011nj,Bobeth:2012vn,Du:2015tda}. Error bars in $q^2$ indicate bin widths.}
\label{fig:dB_the_p}
\end{figure}

\begin{figure}

\includegraphics[width=0.48\textwidth]{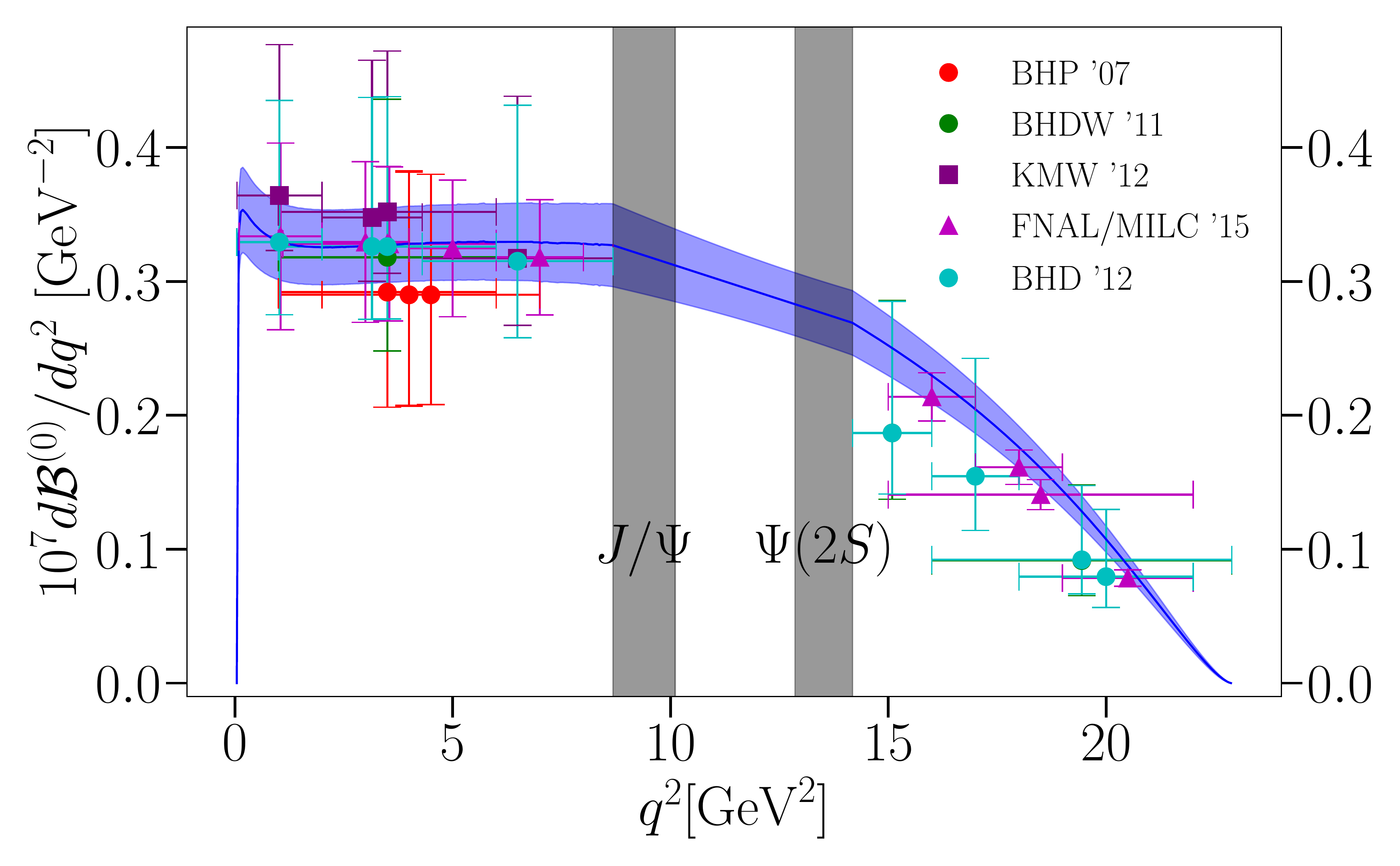}
\caption{Differential branching fraction for $B^0\to K^0\ell^+\ell^-$, with our result in blue, compared with earlier work~\cite{Bobeth:2007dw,Bobeth:2011nj,Bobeth:2012vn,Du:2015tda,Khodjamirian:2012rm}. Error bars in $q^2$ indicate bin widths.}
\label{fig:dB_the_0}
\end{figure}

\begin{figure}

\includegraphics[width=0.48\textwidth]{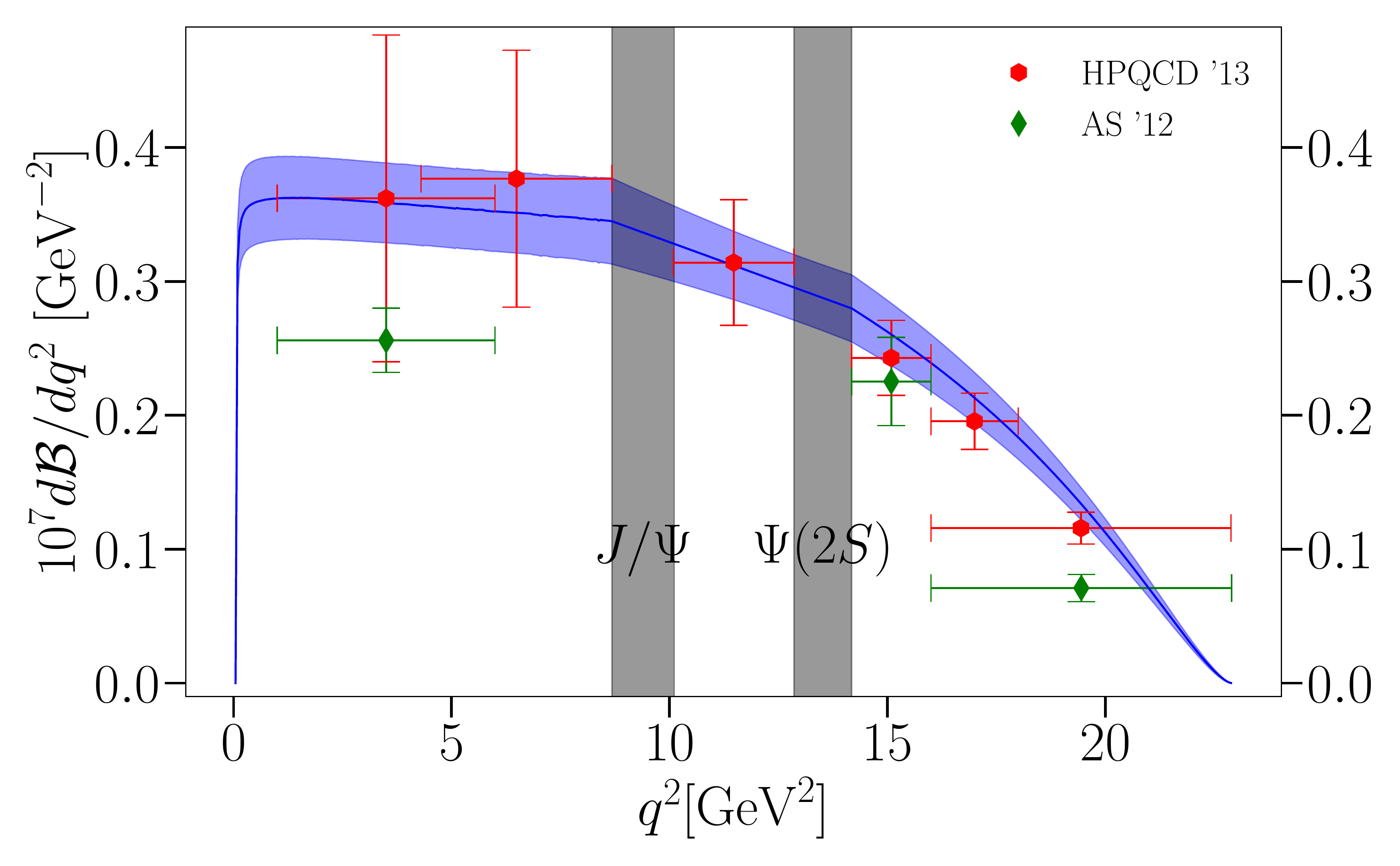}
\caption{Differential branching fraction for $B\to K\ell^+\ell^-$ (the average of that for charged and neutral $B$), with our result in blue, compared with earlier work~\cite{Bouchard:2013mia,Altmannshofer:2012az}. Error bars in $q^2$ indicate bin widths.}
\label{fig:dB_the}
\end{figure}

We compare differential branching fractions for $\ell\in\{e,\mu\}$ in the same manner that we did for experiment, allowing us to factor in the different $q^2$ bin choices made by different authors. 
As before, lepton flavour makes an insignificant difference for $\ell\in\{e,\mu\}$, so we treat all cases together. 
However, we split the cases up by meson charge.
The three charge cases (charged $B$, neutral $B$ and the average of the two) are shown in Figures~\ref{fig:dB_the_p},~\ref{fig:dB_the_0} and~\ref{fig:dB_the}. Again, note the effect of the charge dependence of the corrections to $C_9^{\mathrm{eff},0}$ detailed in Appendix~\ref{app:corrs}.

We see good agreement with previous work, although our uncertainties are generally a lot smaller. We do not agree with the results of AS'12~\cite{Altmannshofer:2012az} at low $q^2$ shown in Figure~\ref{fig:dB_the}. Our central values are slightly higher than most previous results. In the comparison with FNAL/MILC '15~\cite{Du:2015tda}, this $\approx0.5\sigma$ effect can be traced to the larger value of $|V_{ts}|$ that we use, based on more recent $B$-mixing results (see Section~\ref{sec:inputs}); we agree well on the form factor central values (see~\cite{BtoK}). GRDV '22~\cite{Gubernari:2022hxn} fix their form factors from a combination of earlier lattice QCD results (at high $q^2$) and values derived from light-cone sum rules (at low $q^2$). Our form factors have lower central values than theirs but this is partly offset by the difference of values of $|V_{ts}|$ used. Note that GRDV '22 also include non-local form factors in their calculation that account for charmonium resonance contributions. The uncertainty in their differential branching fraction is dominated by that from their local form factors.

\begin{figure}

\includegraphics[width=0.48\textwidth]{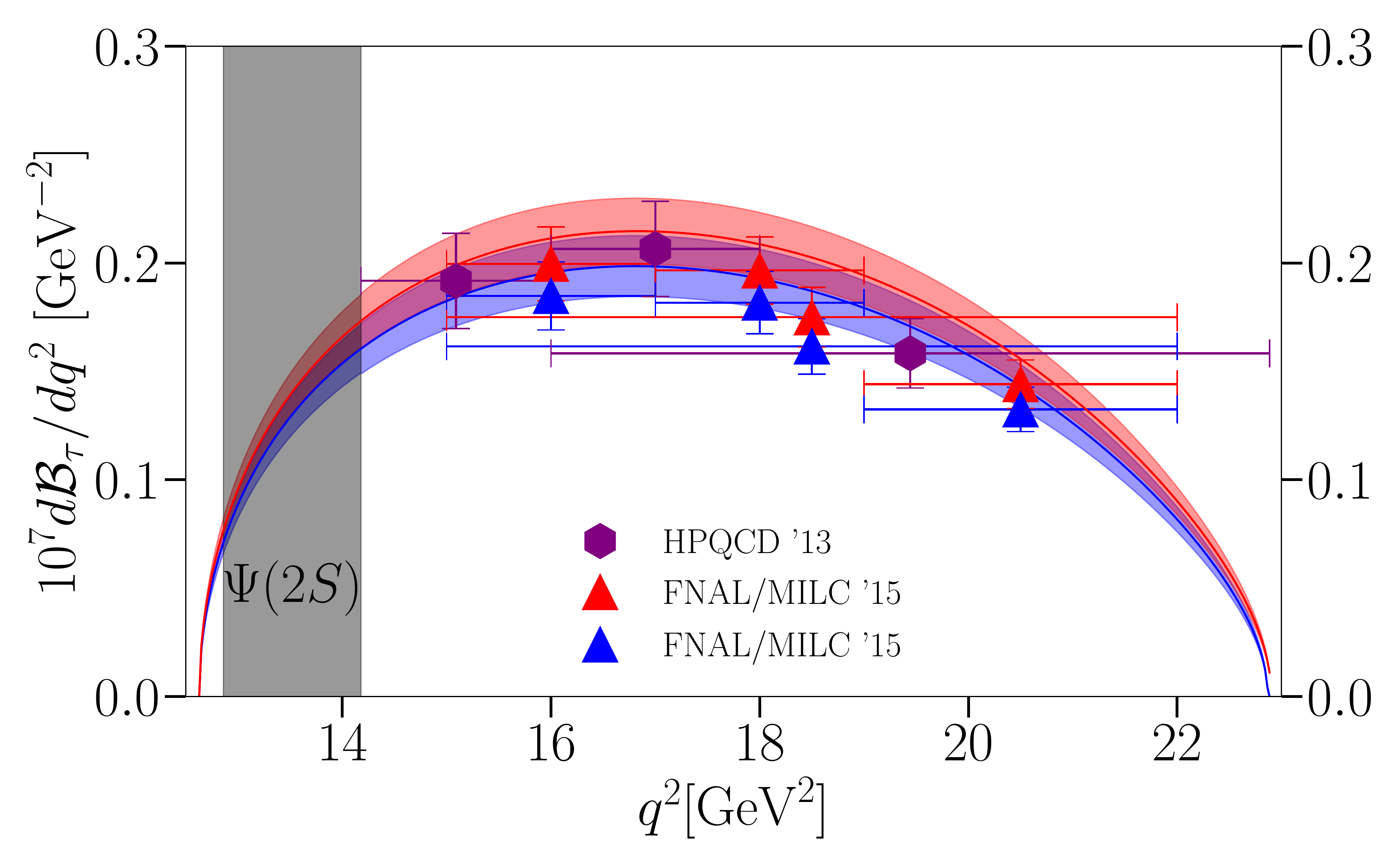}
\caption{Differential branching fraction for $B\to K\tau^+\tau^-$, with our result as a blue band ($B^0\to K^0$) and a red band ($B^+\to K^+$), compared to earlier results from FNAL/MILC '15~\cite{Du:2015tda} (blue and red filled triangles). Results from HPQCD '13~\cite{Bouchard:2013mia} are for the average of the charged and neutral $B$ meson cases. Error bars in $q^2$ indicate bin widths.}
\label{fig:dB_the_tau}
\end{figure}

We compare our results for the differential branching fraction for the case $\ell=\tau$ to those of FNAL/MILC '15~\cite{Du:2015tda} and HPQCD '13~\cite{Bouchard:2013mia} in Figure~\ref{fig:dB_the_tau}. 
The $q^2$ range is limited by the much larger value of $q^2_{\text{min}}$ in this case (set by $m_{\tau}$), and so no interpolation across the vetoed $c\overline{c}$ resonance region is required. 
We include both the charged and neutral meson cases on the same plot, in red and blue respectively. 
We see again that our results are somewhat, but not significantly, higher than the earlier results largely because of the change in $|V_{ts}|$. 

\begin{figure}
\includegraphics[width=0.48\textwidth]{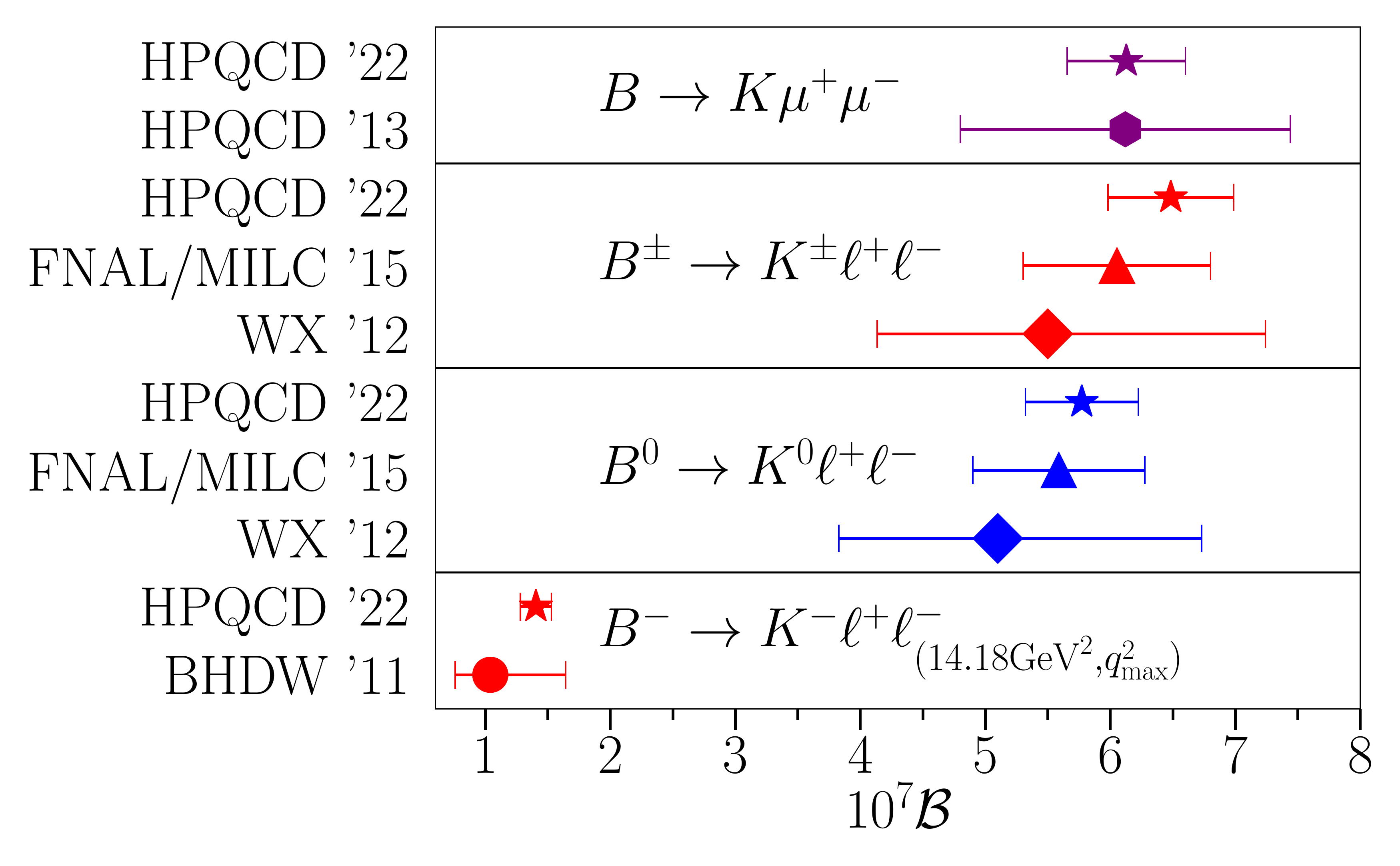}
\caption{The total branching fraction for  $B\to K\ell^+\ell^-$ ($\ell\in\{e,\mu\}$), integrated from $q^2_{\text{min}}$ to $q^2_{\text{max}}$ compared with~\cite{Bobeth:2011nj,Bouchard:2013mia,Du:2015tda,Wang:2012ab}. 
Different meson charges are treated separately. 
The lowest panel shows a comparison to BHDW '11 in which the integration over the differential branching fraction starts at $q^2_{\text{min}}=14.18\,\text{GeV}^2$.}
\label{fig:Bemu_the}
\end{figure}
\begin{figure}

\includegraphics[width=0.48\textwidth]{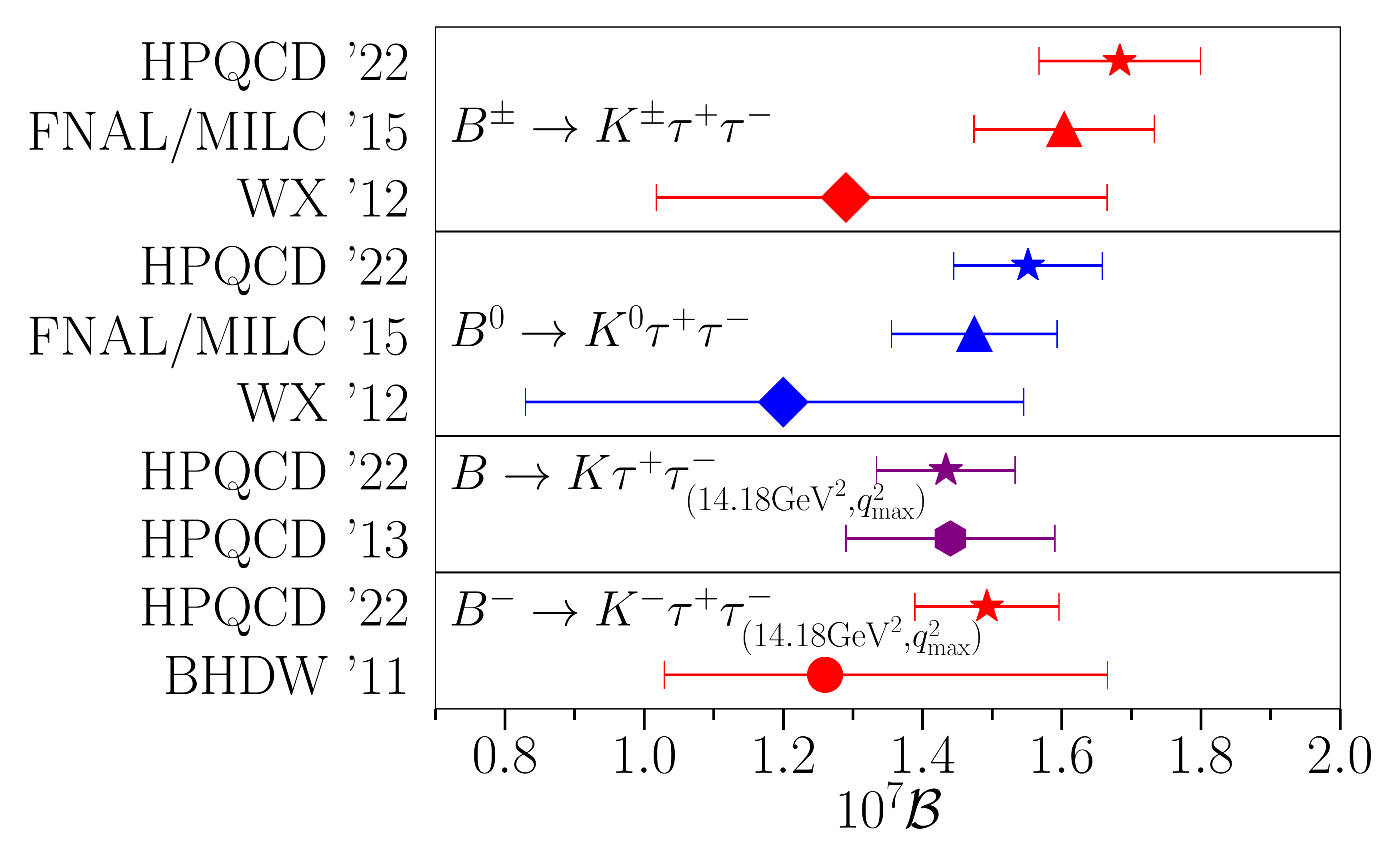}
\caption{The total branching fraction for  $B\to K\tau^+\tau^-$, integrated from $q^2_{\text{min}}$ to $q^2_{\text{max}}$ compared with~\cite{Bobeth:2011nj,Bouchard:2013mia,Du:2015tda,Wang:2012ab}. 
Different meson charges are treated separately. 
The lowest two panels show a comparison to BHDW '11 and HPQCD '13 in which the integration over the differential branching fraction starts at $q^2_{\text{min}}=14.18\,\text{GeV}^2$.}
\label{fig:Btau_the}
\end{figure}

Figures~\ref{fig:Bemu_the} and~\ref{fig:Btau_the} show the total branching fraction for the $\ell\in\{e,\mu\}$ and $\ell=\tau$ cases respectively. 
The figures are broken down according to meson charge, with the integral extending from $q^2_{\text{min}}$ to $q^2_{\text{max}}$, except as indicated otherwise in the plot. 
In HPQCD '13~\cite{Bouchard:2013mia} the branching fraction to electrons is also given, but (as is the case for us) is insignificantly different from the muon case shown, so we omit it. 
Again, our central values are somewhat higher than previous work, but not by a significant margin. Our results provide an improvement in uncertainty over earlier values. 
\begin{figure}

\includegraphics[width=0.48\textwidth]{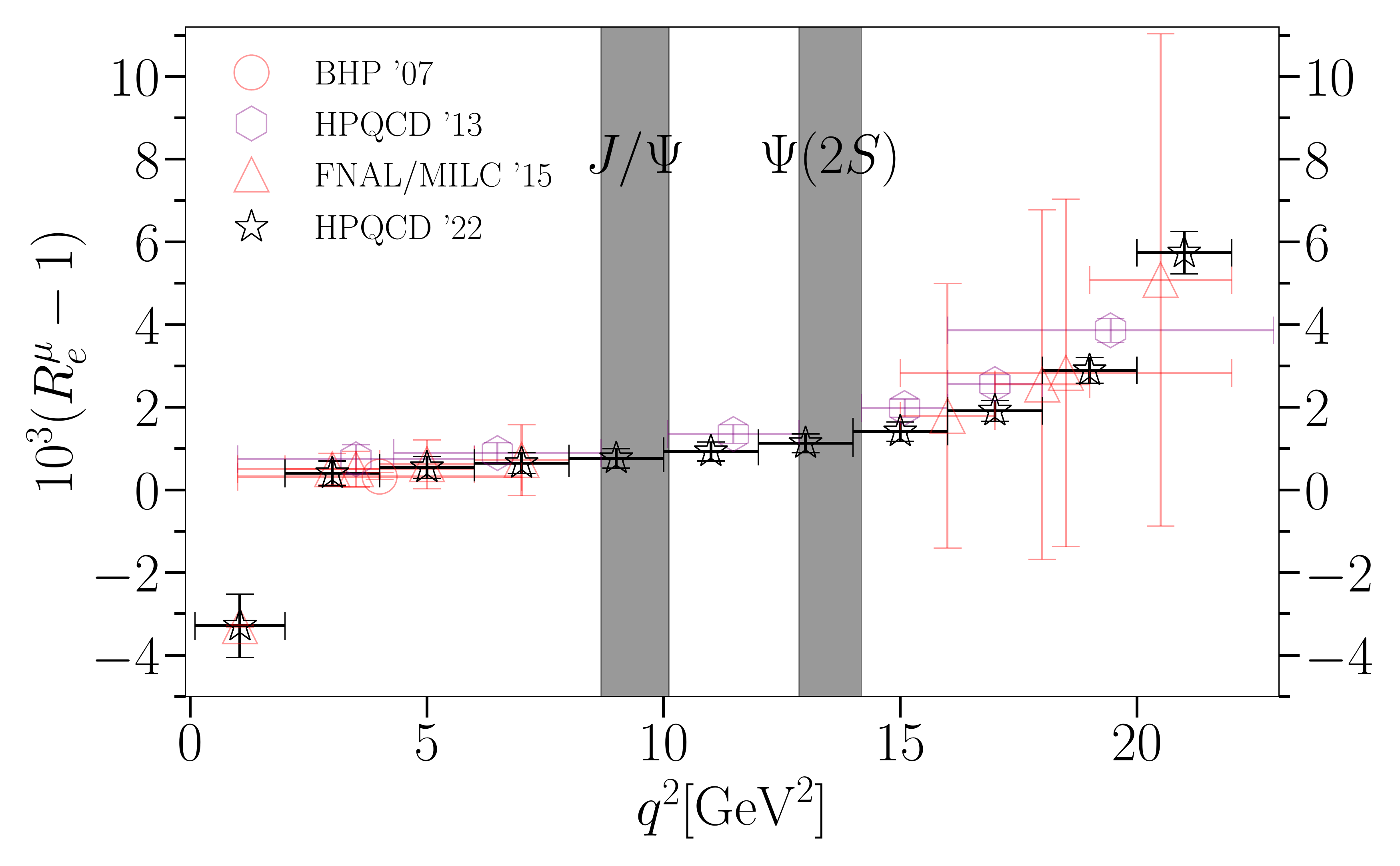}
\caption{The deviation of the ratio $R^{\mu}_{e}$ (Eq.~(\ref{eq:R})) from 1. 
Our results (HPQCD '22) are shown as open black stars for the charged $B$ case, evaluated in $q^2$ bins defined by $0.1, 2, 4, 6, \dots , 22\,\text{GeV}^2$.  Error bars in $q^2$ indicate bin widths. Earlier theory results are from~\cite{Bobeth:2007dw,Bouchard:2013mia,Du:2015tda}. 
We do not include any uncertainty from possible QED corrections that would affect experimental results. 
}
\label{fig:Rbybin_the_emu}
\end{figure}

\subsubsection{$R^\mu_e$ and $R^\tau_\mu$} 
\label{sec:Rtheory}

We now look in more detail at the ratio of branching fractions to different flavours of leptons, defined in Eq.~\eqref{eq:R}. 
Figure~\ref{fig:Rbybin_the_emu} shows our results (as black open stars) for the deviation of the ratio $R^{\mu}_{e}$ from 1.0, as a function of $q^2$. We use $q^2$ bins of width 2 $\text{GeV}^2$, following the pattern 2--4 $\text{GeV}^2$, 4--6 $\text{GeV}^2$ etc, except for the first bin which covers 0.1--2 $\text{GeV}^2$. We plot our results for the charged $B$ case; values for the neutral $B$ case are indistinguishable, except in the lowest $q^2$ bin (numerical values for each bin are given in Table~\ref{tab:R-mue-vals} in Appendix~\ref{sec:numres}). As discussed earlier, the ratio is very well determined theoretically in a pure QCD calculation; uncertainties from possible QED corrections that would affect experiment are not included in Fig.~\ref{fig:Rbybin_the_emu}. We see that $R^\mu_e$ differs from 1 at the level of $10^{-3}$, with a larger deviation at large $q^2$. There is also a sizeable effect in the smallest $q^2$ bin where the kinematic cut-off from the $\mu$ mass starts to have an effect. 

Figure~\ref{fig:Rbybin_the_emu} also shows a comparison to earlier theory results. Good agreement is seen between the different calculations.

\begin{figure}

\includegraphics[width=0.48\textwidth]{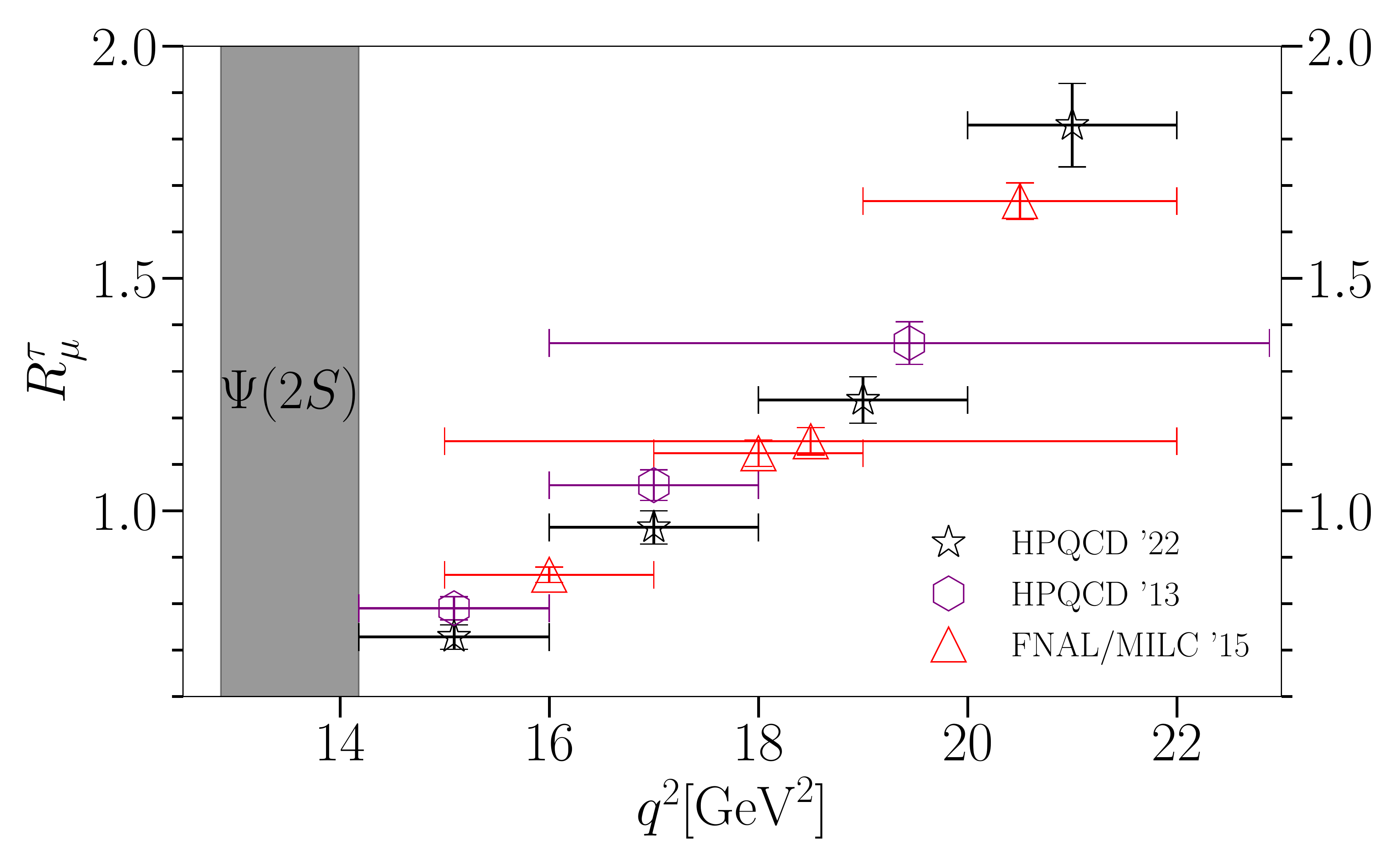}
\caption{The ratio $R^{\tau}_{\mu}$ (Equation~(\ref{eq:R})). 
Our results (HPQCD '22, open black stars) are evaluated in $q^2$ bins defined by $14.18, 16, 18, 20, 22\,\text{GeV}^2$, for the charged $B$ meson case. 
We compare with HPQCD '13~\cite{Bouchard:2013mia} and FNAL/MILC '15~\cite{Du:2015tda}. No uncertainty is included for QED corrections that would affect experiment.}
\label{fig:Rbybin_the_tau}
\end{figure}
\begin{figure}

\includegraphics[width=0.48\textwidth]{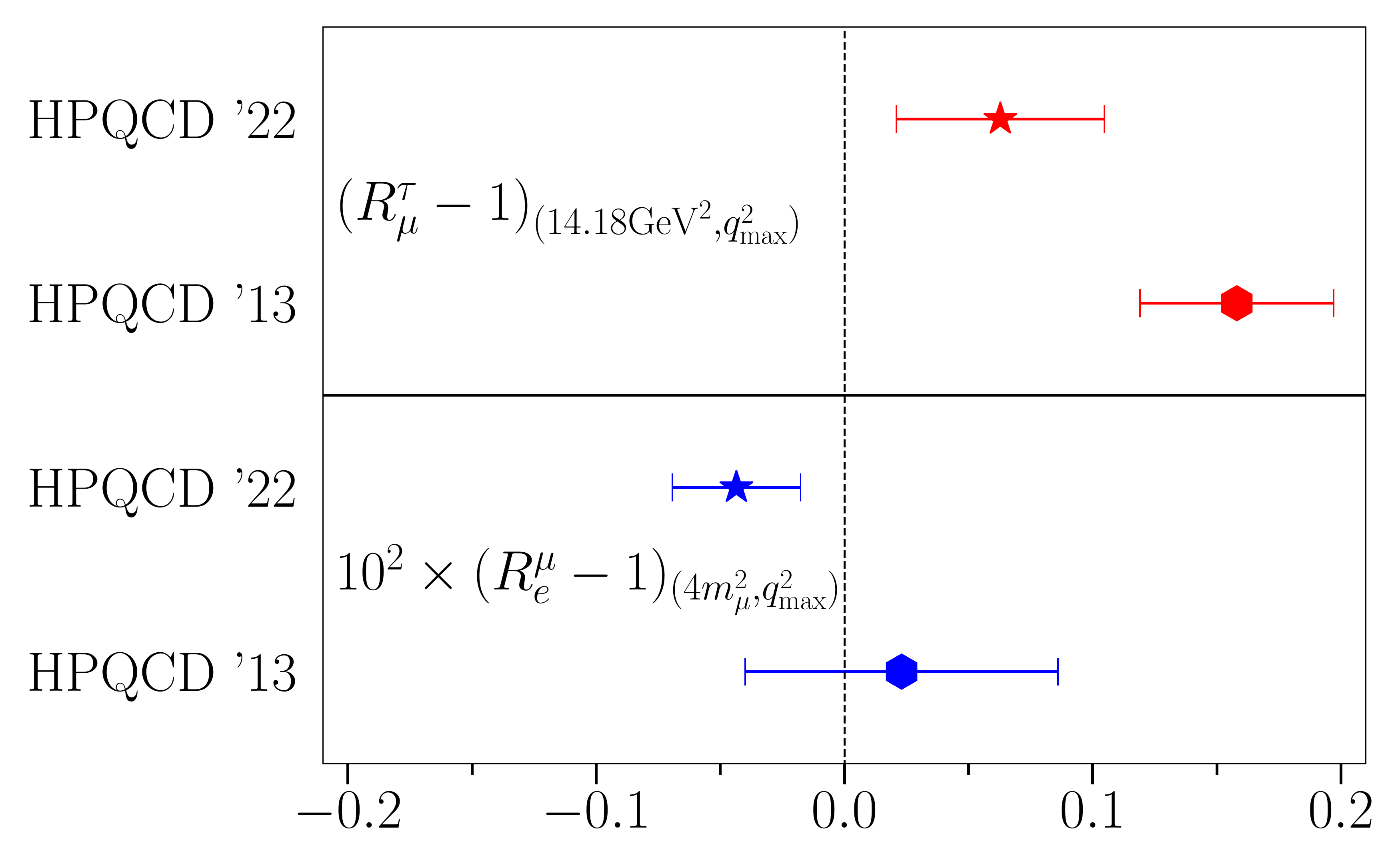}
\caption{Our result for $R^{\mu}_{e}$ (filled blue star) determined from Eq.~\eqref{eq:R} using the full $q^2$ range and shown as the deviation from 1 in units of $10^{-2}$. The filled red star gives our result for $R^{\tau}_{\mu}$ using the $q^2$ range from 14.18 $\text{GeV}^2$ to $q^2_{\text{max}}$. We compare to results from~\cite{Bouchard:2013mia} (filled hexagons). No uncertainty for QED corrections is included here.}
\label{fig:R_the}
\end{figure}

Figure~\ref{fig:Rbybin_the_tau} shows our results for the ratio $R^{\tau}_{\mu}$ (Eq.~(\ref{eq:R})) as open black stars, calculated in $q^2$ bins of width 2 ${\text{GeV}}^2$. The lowest bin is from the top of the $c\overline{c}$ resonance region at 14.18 ${\text{GeV}}^2$ (slightly above the kinematic end-point of $4m_\tau^2$) to 16 ${\text{GeV}}^2$, the next 14--16 ${\text{GeV}}^2$ and so on. Results are shown for the charged meson case, but the neutral meson results are indistinguishable, see the numerical values given in Table~\ref{tab:Rtau} in Appendix~\ref{sec:numres}. 
 $R^\tau_\mu$ deviates from 1.0 much more strongly than $R^\mu_e$; given the lepton masses involved this is not surprising.
The results show a similar pattern to $R^\mu_e$ with lepton mass effects pushing $R$ values upwards as $q^2$ increases. $R^\tau_e$ results would not differ visibly and so are not shown. Again, the comparison to earlier results shows good agreement. 

In Fig.~\ref{fig:R_the} we give our result for $R^\mu_e$ using the full $q^2$ range from $4m_\mu^2$ to $q^2_{\text{max}}$. This is shown as the deviation from 1 in units of $10^{-2}$. Note that in forming this ratio we integrate both the numerator and denominator of Eq.~\eqref{eq:R} through the vetoed region as discussed in Sec.~\ref{sec:resveto}. We also give our result for $R^\tau_\mu$ using the range from 14.18 $\text{GeV}^2$ (the upper edge of the vetoed region) to $q^2_{\text{max}}$. Both of our results are consistent with zero deviation from 1 at the 2$\sigma$ level. We compare to HPQCD '13~\cite{Bouchard:2013mia} and see reasonable agreement; our result for $R^\mu_e$ has half the theory uncertainty of their earlier value.

\begin{figure}
\includegraphics[width=0.48\textwidth]{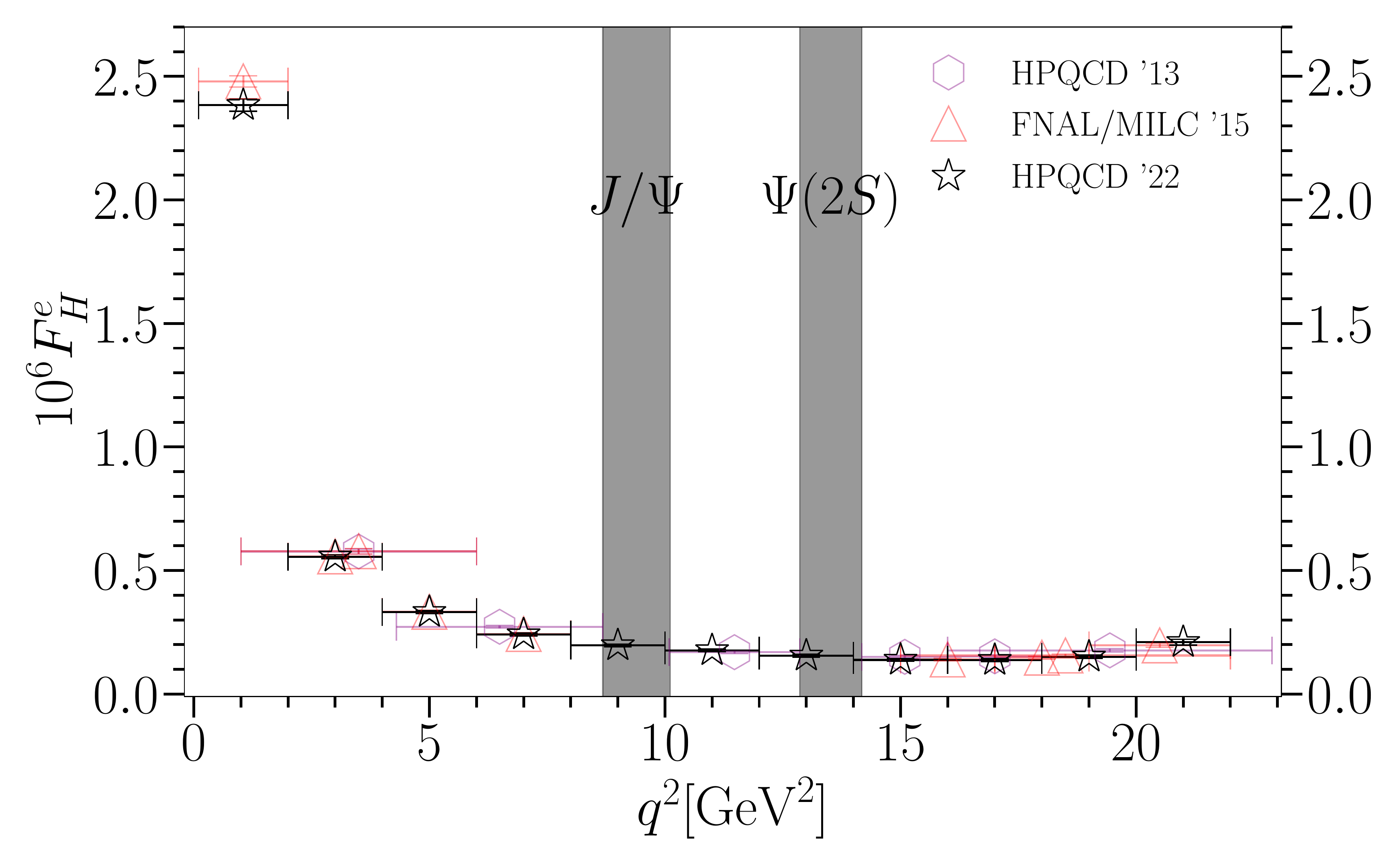}
\caption{Open black stars show our results for $F^{e}_H$ for the charged $B$ case in bins of $q^2$ defined by $0.1, 2, 4, 6, \dots , 22\,\text{GeV}^2$. We compare to HPQCD '13~\cite{Bouchard:2013mia} and FNAL/MILC '15~\cite{Du:2015tda}. 
}
\label{fig:FHbybin_the_e}
\end{figure}
\begin{figure}

\includegraphics[width=0.48\textwidth]{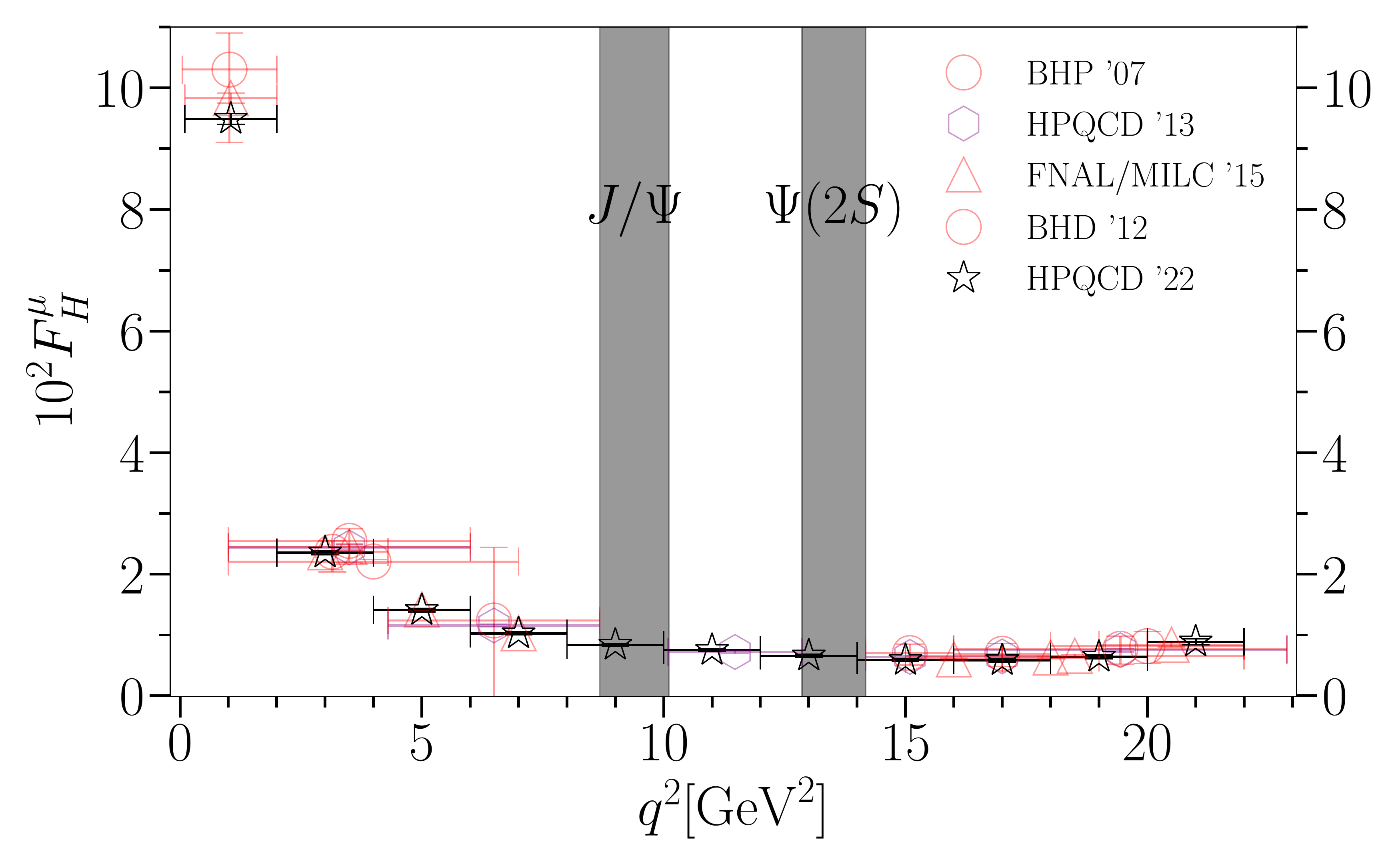}
\caption{Open black stars show our results for $F^{\mu}_H$ for the charged $B$ case in bins of $q^2$ defined by $0.1, 2, 4, 6, \dots , 22\,\text{GeV}^2$. We compare to HPQCD '13~\cite{Bouchard:2013mia}, FNAL/MILC '15~\cite{Du:2015tda} and BHD '12~\cite{Bobeth:2012vn}. 
}
\label{fig:FHbybin_the_mu}
\end{figure}
\begin{figure}

\includegraphics[width=0.48\textwidth]{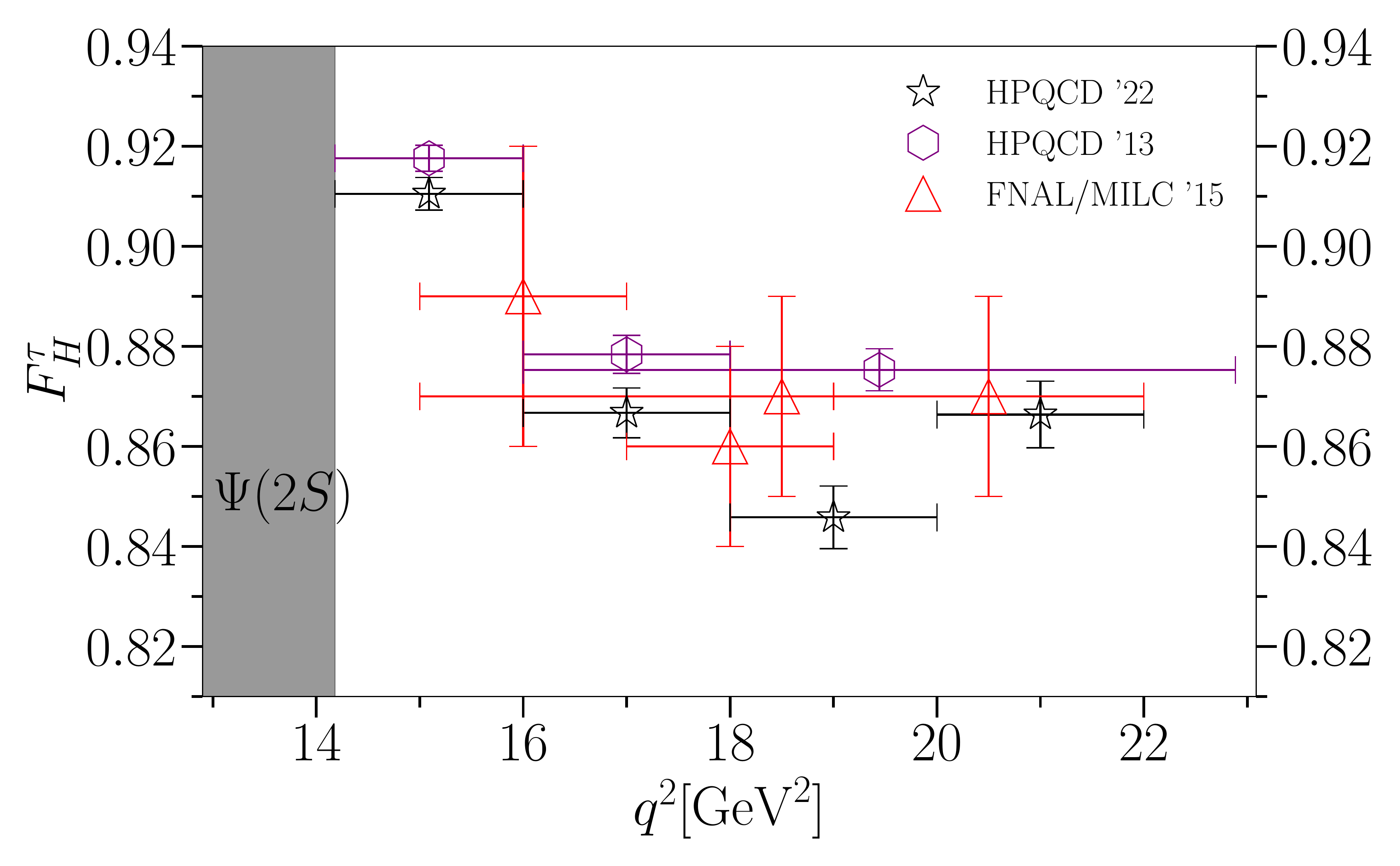}
\caption{Open black stars show our results for $F^{\tau}_H$ for the charged $B$ case in bins of $q^2$ defined by $14.18, 16, 18, 20, 22\,\text{GeV}^2$. We compare to HPQCD '13~\cite{Bouchard:2013mia} and FNAL/MILC '15~\cite{Du:2015tda}.
}
\label{fig:FHbybin_the_tau}
\end{figure}

\subsubsection{$F_H^e$, $F_H^\mu$ and $F_H^\tau$}
\label{sec:FHtheory}

Here we examine in more detail the quantity $F_H^\ell$, twice the flat term of Eq.~\eqref{eq:flatdef}, for different lepton flavours. $F_H^\ell$ is determined from the ratio of Eq.~\eqref{eq:F_H} in bins of $q^2$. 
 Figures~\ref{fig:FHbybin_the_e},~\ref{fig:FHbybin_the_mu} and~\ref{fig:FHbybin_the_tau} show our results for $e$, $\mu$ and $\tau$ cases respectively as open black stars. We use $q^2$ bins of width 2 $\text{GeV}^2$, as for $R$ in the previous section. Since meson charge does not affect the results, except in the smallest $q^2$ bin, we show our results for the charged $B$ cases. Our results are given in detail in Tables~\ref{tab:FHl} and~\ref{tab:Ftau} in Appendix~\ref{sec:numres}. No additional uncertainties from QED effects that would affect experiment are included in the plots. 
 
 For the light, $e$ and $\mu$, lepton cases $F_H^\ell$ is small across most of the $q^2$ range but rises rapidly towards the low $q^2$ end. For small values of $m_\ell$ and $q^2$ the integrand in the numerator of $F_H^\ell$ (Eq.~\eqref{eq:F_H}) is proportional to $\beta(1-\beta^2)$ (see Eq.~\eqref{eq:ac}). For $q^2 = \mathcal{O}(4m_\ell^2)$, this has no suppression by $m_\ell^2$ and gives a large contribution to the integral. This means that great care must be taken in integrating the numerator in the region $q^2 \to 4 m_\ell^2$ (see Appendix~\ref{sec:integration}).

In  Figures~\ref{fig:FHbybin_the_e},~\ref{fig:FHbybin_the_mu} and~\ref{fig:FHbybin_the_tau}  we also compare to earlier (mainly lattice QCD) results, plotting the values for the charged meson case where both are given. There is good agreement between the different calculations in general; our results have an improvement in uncertainty. We see some tension (2.8$\sigma$) for $F_H^e$ and $F_H^\mu$ with results from FNAL/MILC '15 in the smallest $q^2$ bin for the charged $B$ case, but agree well for the neutral $B$ case. This can be seen by comparing our Table~\ref{tab:FHl} with Tables IX and X of~\cite{Du:2015tda}. For $F_H^\tau$ we see some tension with HPQCD '13~\cite{Bouchard:2013mia}. 

\begin{figure}
\includegraphics[width=0.48\textwidth]{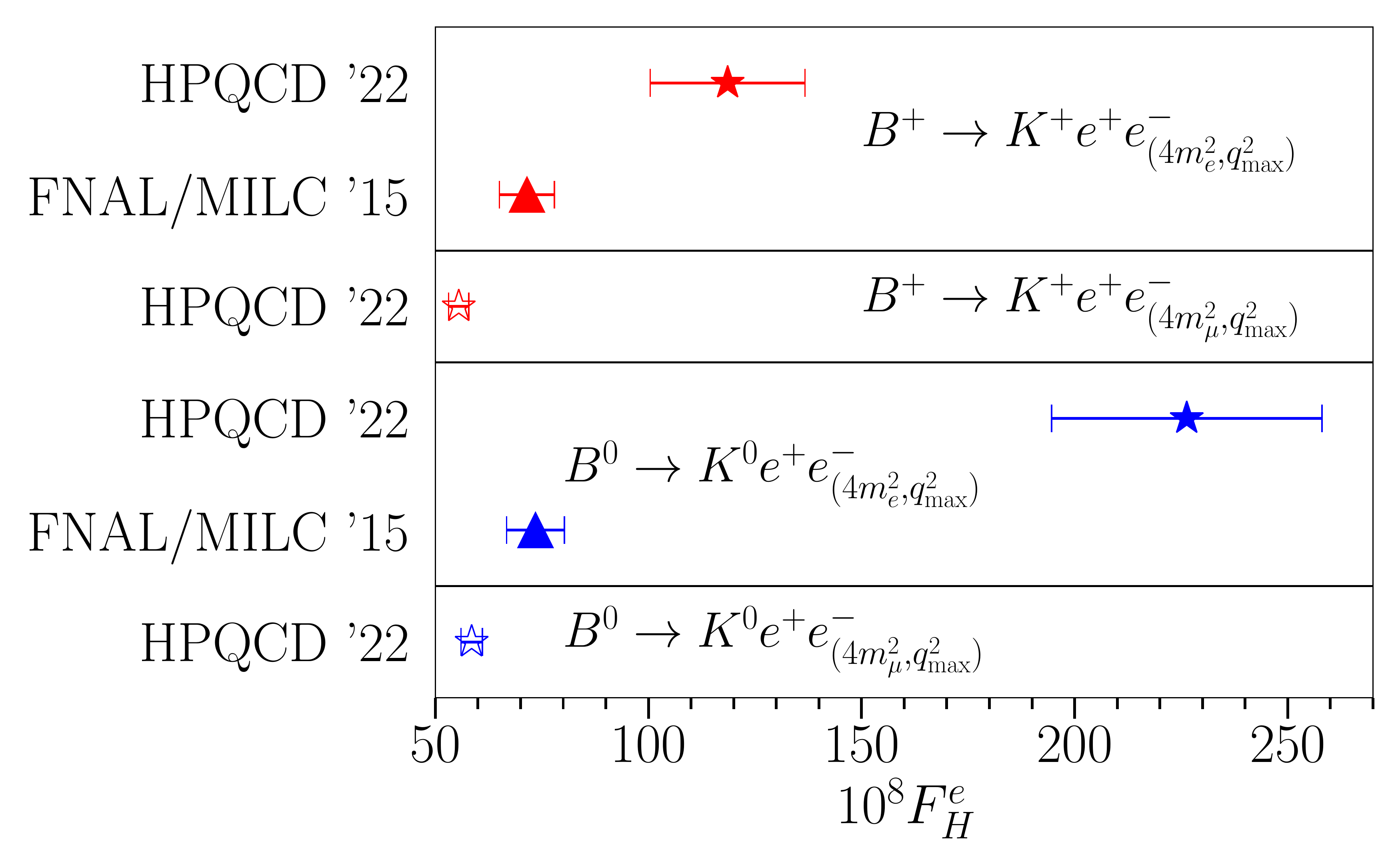}
\caption{Filled stars in the top and third-from-top panels show our results (`HPQCD '22') for $F_H^e$, integrated over the full $q^2$ range (from $4m_e^2$ to $q^2_{\mathrm{max}}$) for the charged and neutral meson cases respectively. 
Our results for this case are compared with~\cite{Du:2015tda} (filled triangles). 
We also give values for our results integrated from $4m_{\mu}^2$ to $q^2_{\mathrm{max}}$ (open stars, second-from-top and lowest panels) to illustrate the large contribution of the integral from very small $q^2$ values.}
\label{fig:FH_the_e}
\end{figure}

\begin{figure}
\includegraphics[width=0.48\textwidth]{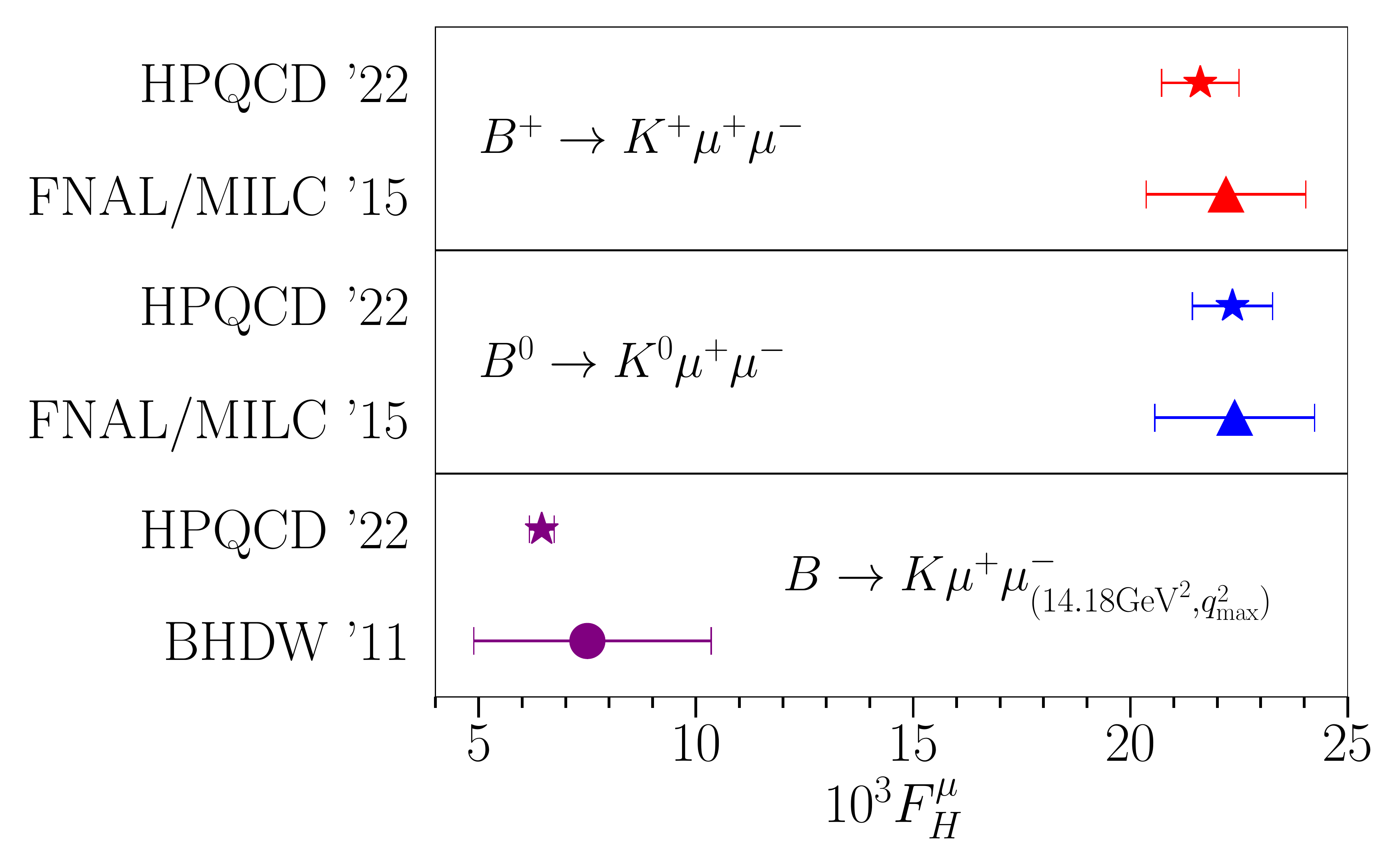}
\caption{$F_H^{\mu}$, integrated over the full $q^2$ range, for the charged and neutral meson cases, and over a reduced range from $14.18\,\text{GeV}^2$ in the case of the average of the charged and neutral cases (lowest panel). Our results are shown as the filled stars and denoted `HPQCD '22'. 
This work is compared with~\cite{Bobeth:2011nj,Du:2015tda}, shown as filled triangles and circles respectively. }
\label{fig:FH_the_mu}
\end{figure}
\begin{figure}

\includegraphics[width=0.48\textwidth]{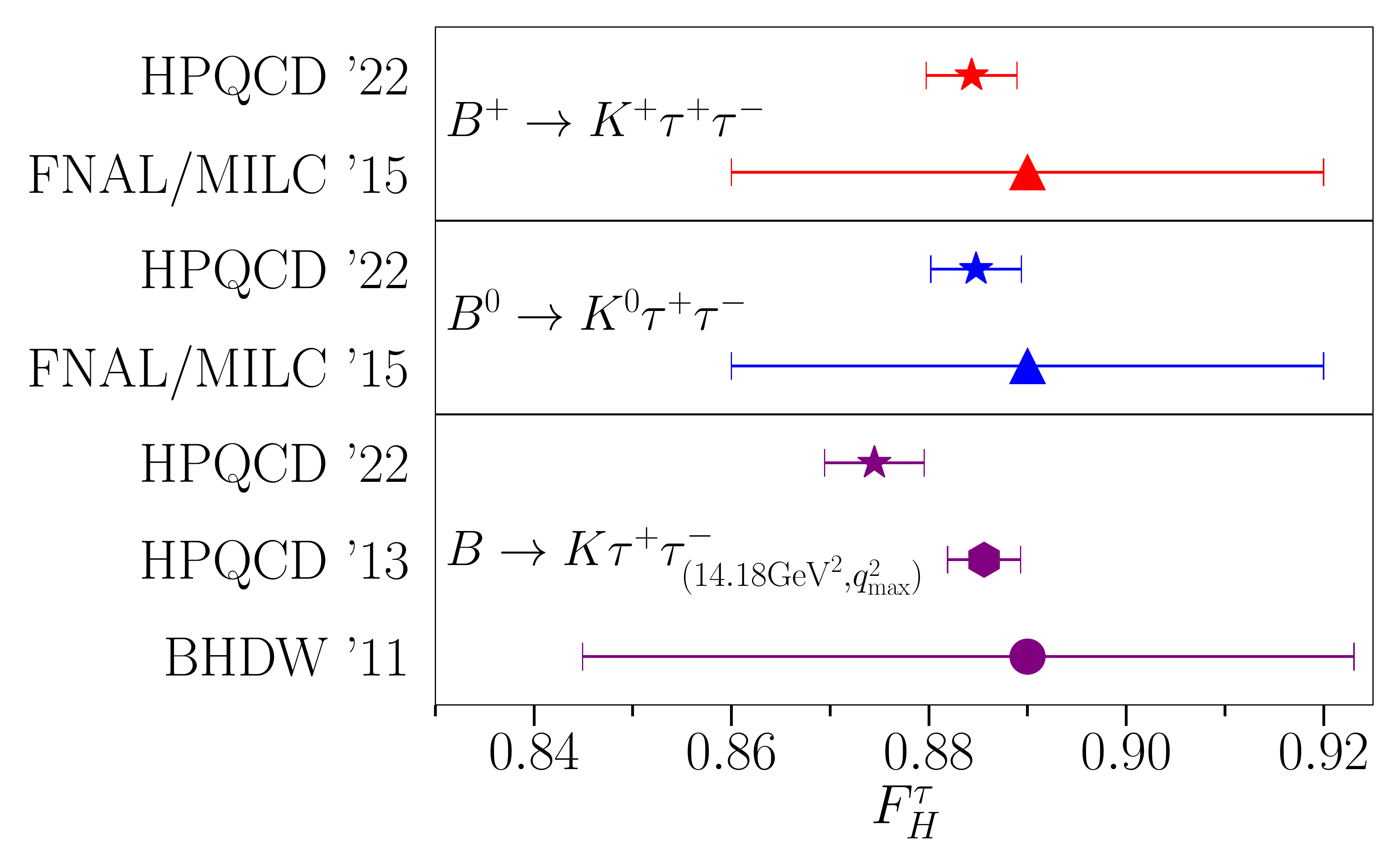}
\caption{$F_H^{\tau}$, integrated over the full $q^2$ range, for the charged and neutral meson cases, and over a reduced range from $14.18\,\text{GeV}^2$ in the case of the average of the charged and neutral cases (lowest panel). Our results are shown as the filled stars and denoted `HPQCD '22'. 
This work is compared with~\cite{Bobeth:2011nj,Bouchard:2013mia,Du:2015tda}, shown as filled hexagons, triangles and circles.}
\label{fig:FH_the_tau}
\end{figure}

Figures~\ref{fig:FH_the_e},~\ref{fig:FH_the_mu} and~\ref{fig:FH_the_tau} show $F^{\ell}_H$ as defined above, integrated over the full $q^2$ range $q^2_{\text{min}}\leq q^2\leq q^2_{\text{max}}$, except in cases where the range is stated explicitly. 
These plots are split up according to meson charge, although in most cases the difference is insignificant. A significant difference is seen for $F^e_H$ between the charged and neutral cases when integrating from $4m_e^2$ up to $q^2_{\text{max}}$. This difference comes from the nonfactorisable corrections to $C_9^{\text{eff}}$ discussed in Appendix~\ref{app:corrs}. These depend on meson charge and have a sizeable impact at very small value of $q^2$, relevant to $F^e_H$ (see below) but not for other lepton flavours or to other quantities. 

We see good agreement with previous work for $F_H^\mu$ and $F_H^\tau$, with our results showing a considerable improvement in uncertainty in most cases. 
For $F^e_H$ our results are in tension with those of FNAL~\cite{Du:2015tda} when we integrate from $4m_e^2$ up to $q^2_{\text{max}}$. Given the good agreement seen for most of the $q^2$ range in Fig.~\ref{fig:FHbybin_the_e} it seems likely that the tension is a result of behaviour in the integrand of the numerator towards the lower limit of the $q^2$ integral, discussed above.
 We illustrate this in Fig.~\ref{fig:FH_the_e} by showing results also for the case where the lower cut-off on the $q^2$ integral is $4m_\mu^2$ rather than $4m_e^2$. Doing this gives considerably lower values for $F_H^e$ because the integral in the numerator of $F_H^e$ changes; the denominator is half the total branching fraction and very insensitive to small changes in $q^2_{\text{min}}$. We conclude that determining $F^e_H$ at small $q^2$ values or integrated over the full kinematic range requires care over integration and is sensitive to small-$q^2$ corrections. However, away from small $q^2$ values there are no issues with determining $F^e_H$ (see Figure~\ref{fig:FH_the_e}) and this is the experimentally more accessible region.

\section{$B\to K \ell_1^- \ell_2^+$}\label{sec:l1l2}

\begin{figure}
\includegraphics[width=0.48\textwidth]{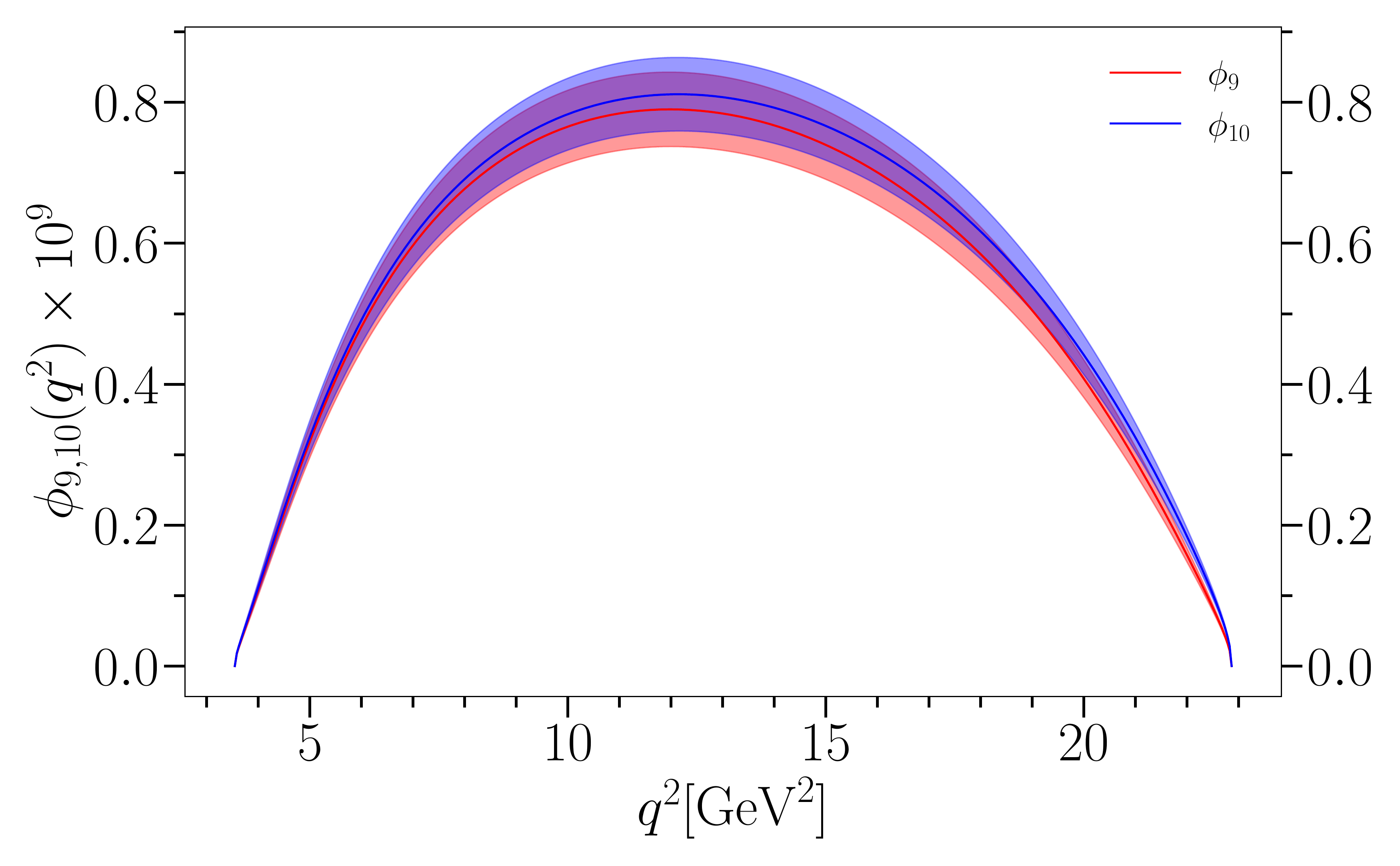}
\caption{$\phi_{9,10}(q^2)\times 10^9$ for the $B^0\to K^0\mu\tau$ decay.}
\label{fig:phi910mutau}
\end{figure}

\begin{figure}
\includegraphics[width=0.48\textwidth]{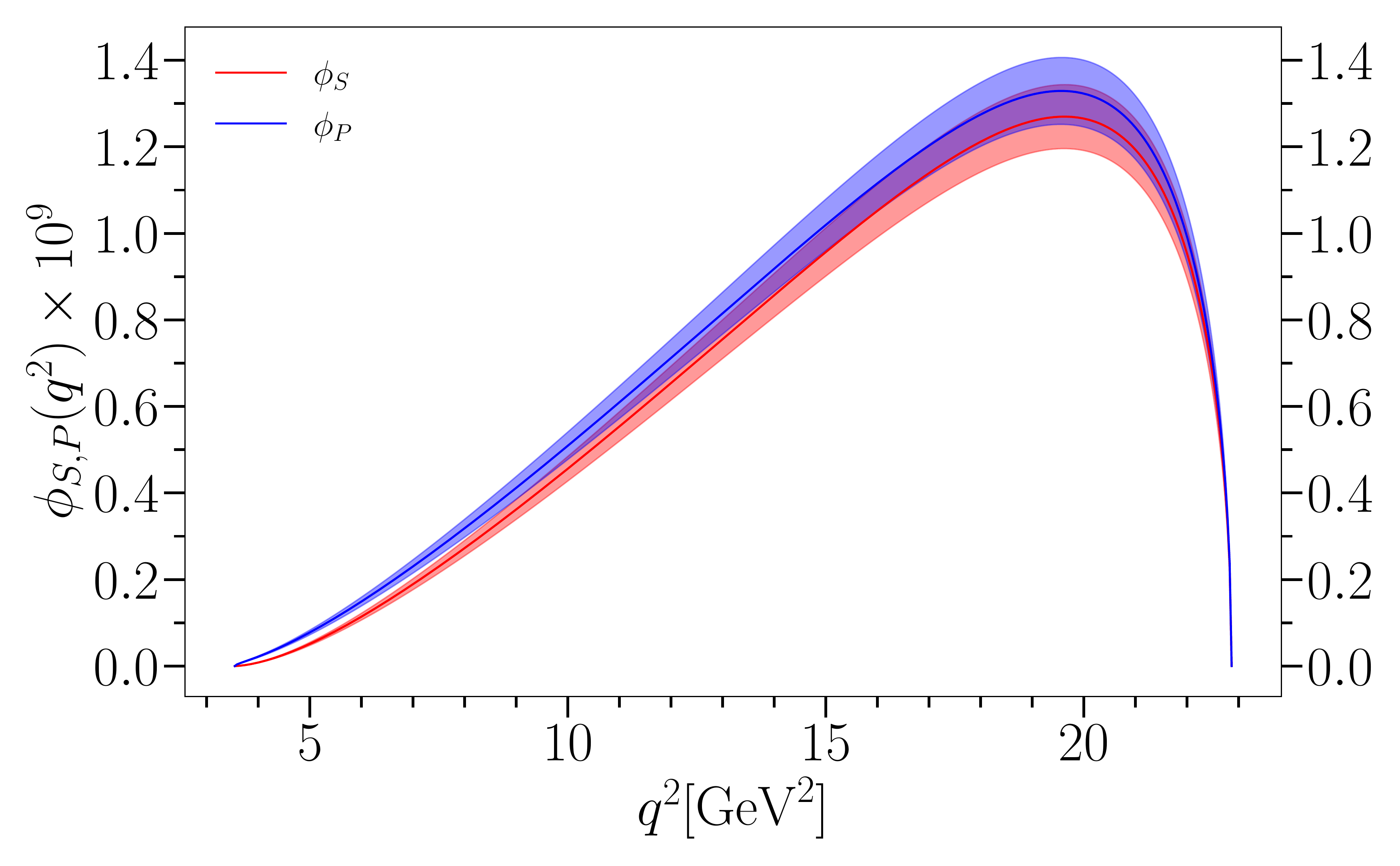}
\caption{$\phi_{S,P}(q^2)\times 10^9$ for the $B^0\to K^0\mu\tau$ decay.}
\label{fig:phiSPmutau}
\end{figure}

We also investigate the rare lepton flavour number violating (LFV) $B\to K\ell^-_1\ell^+_2$ decay, with different final state leptons $\ell_1\neq\ell_2$ (with masses $m_1$ and $m_2$). 
The differential branching fraction for this decay is given in Eq.~(9) of~\cite{Becirevic:2016zri} in terms of a number of Wilson coefficients and functions $\varphi_i(q^2)$ (Eq.~(10) of~\cite{Becirevic:2016zri}). 
The knowledge of these functions in the SM can be combined with experimental measurements to set limits on the Wilson Coefficients~\cite{Angelescu:2020uug}.

In the case that the LFV in $B\to K\ell^-_1\ell^+_2$ arises from purely vector or purely scalar operators, we can write simple forms for the branching fraction, in terms of Wilson coefficients $C^{(')}_9$, $C^{(')}_{10}$, $C^{(')}_S$ and $C^{(')}_{P}$ and lepton flavour dependent coefficients $a_K^{12}$, $b_K^{12}$, $e_K^{12}$ and $f_K^{12}$. 
\begin{equation}\label{eq:vecLFV}
  10^{9} \mathcal{B}(B\to K\ell_1\ell_2) = a_K^{12}|C_9+C_9'|^2 +b_K^{12}|C_{10}+C_{10}'|^2 
\end{equation}
for the vector case and
\begin{equation}\label{eq:scaLFV}
  10^{9}\mathcal{B}(B\to K\ell_1\ell_2) = e_K^{12}|C_S+C_S'|^2 +f_K^{12}|C_{P}+C_{P}'|^2 
\end{equation}
for the scalar. The factor of $10^9$ makes the coefficients $\mathcal{O}(10)$.

Here, we will construct the functions $\varphi_9$, $\varphi_{10}$, $\varphi_S$, $\varphi_P$, which, once multiplied by $|N_K(q^2)|^2$, can be integrated to give the coefficients above~\cite{Becirevic:2016zri} . 
\begin{equation}
  \begin{split}
    &\varphi_{9,10}(q^2)=\\
    &\frac{1}{2}|f_0|^2(m_1\mp m_2)^2\frac{(M_B^2-M_K^2)^2}{q^2}\Big(1-\frac{(m_1\pm m_2)^2}{q^2}\Big) +\\
    &\frac{1}{2}|f_+|^2\lambda(q,M_B,M_K)\Big(1-\frac{(m_1\mp m_2)^2}{q^2}-\frac{\lambda(q,m_1,m_2)}{3q^4}\Big),
  \end{split}
\end{equation}
\begin{equation}\label{eq:phiSP}
  \begin{split}
    \varphi_{S,P}(q^2) &= \frac{q^2|f_0|^2}{2(m^{\overline{\mathrm{MS}}}_b(\mu_b)-m^{\overline{\mathrm{MS}}}_s(\mu_b))^2}\times\\
    &(M_B^2-M_K^2)^2\Big(1-\frac{(m_1\pm m_2)^2}{q^2}\Big),
  \end{split}
\end{equation}
\begin{equation}\label{eq:Nforl1l2}
  \begin{split}
    &\hspace{-4.0em}|N_K(q^2)|^2=\\
    &\hspace{-2.0em}\tau_B\frac{\alpha^2_{\mathrm{EW}}G_F^2|V_{tb}V^*_{ts}|^2}{512\pi^5M_B^3} \times \\
    &\frac{\sqrt{\lambda(q,m_1,m_2)}}{q^2}\sqrt{\lambda(q,M_B,M_K)},
  \end{split}
\end{equation}
where the $\pm$s are matched with $9,10$ and $S,P$. We use our improved $f_+$ and $f_0$ form factors~\cite{BtoK} to determine these functions, including uncertainties for strong-isospin breaking discussed in Section~\ref{sec:isospin}. For the $\overline{\mathrm{MS}}$ $s$ mass in Eq.~(\ref{eq:phiSP}) we take $m^{\overline{\mathrm{MS}}}_s(\mu_b\equiv4.2\,\mathrm{GeV}) = 0.07966(80)\,\mathrm{GeV}$, run from $m^{\overline{\mathrm{MS}}}_s(\mu=3\,\mathrm{GeV})=0.08536(85)\,\mathrm{GeV}$~\cite{Lytle:2018evc}. $m^{\overline{\mathrm{MS}}}_b(\mu_b)$ is given in Table~\ref{tab-SMparams}, along with the other parameters used. 
None of the $\varphi$ functions are sensitive to interchanging $1\leftrightarrow2$, so neither are the resulting coefficients ($a_K^{12}=a_K^{21}$, etc.). 

We plot the functions $\phi_i(q^2)\equiv|N_K(q^2)|^2\varphi_i(q^2)$ in Figures~\ref{fig:phi910mutau} and~\ref{fig:phiSPmutau}, for the choice $\ell_1=\mu$, $\ell_2=\tau$, and find good agreement with similar plots in~\cite{Becirevic:2016zri}. Note that there is no issue with $c\overline{c}$ resonances in this case. The functions are insensitive to $B$ meson charge, so the neutral case is shown here - the charged case differs mainly by the overall multiplicative factor of $\tau_{B^{0/+}}$ used in Eq.~\eqref{eq:Nforl1l2}. Plots for lepton combination $e\tau$ are nearly identical to those for $\mu\tau$, and are not shown.

\begin{figure}
\includegraphics[width=0.48\textwidth]{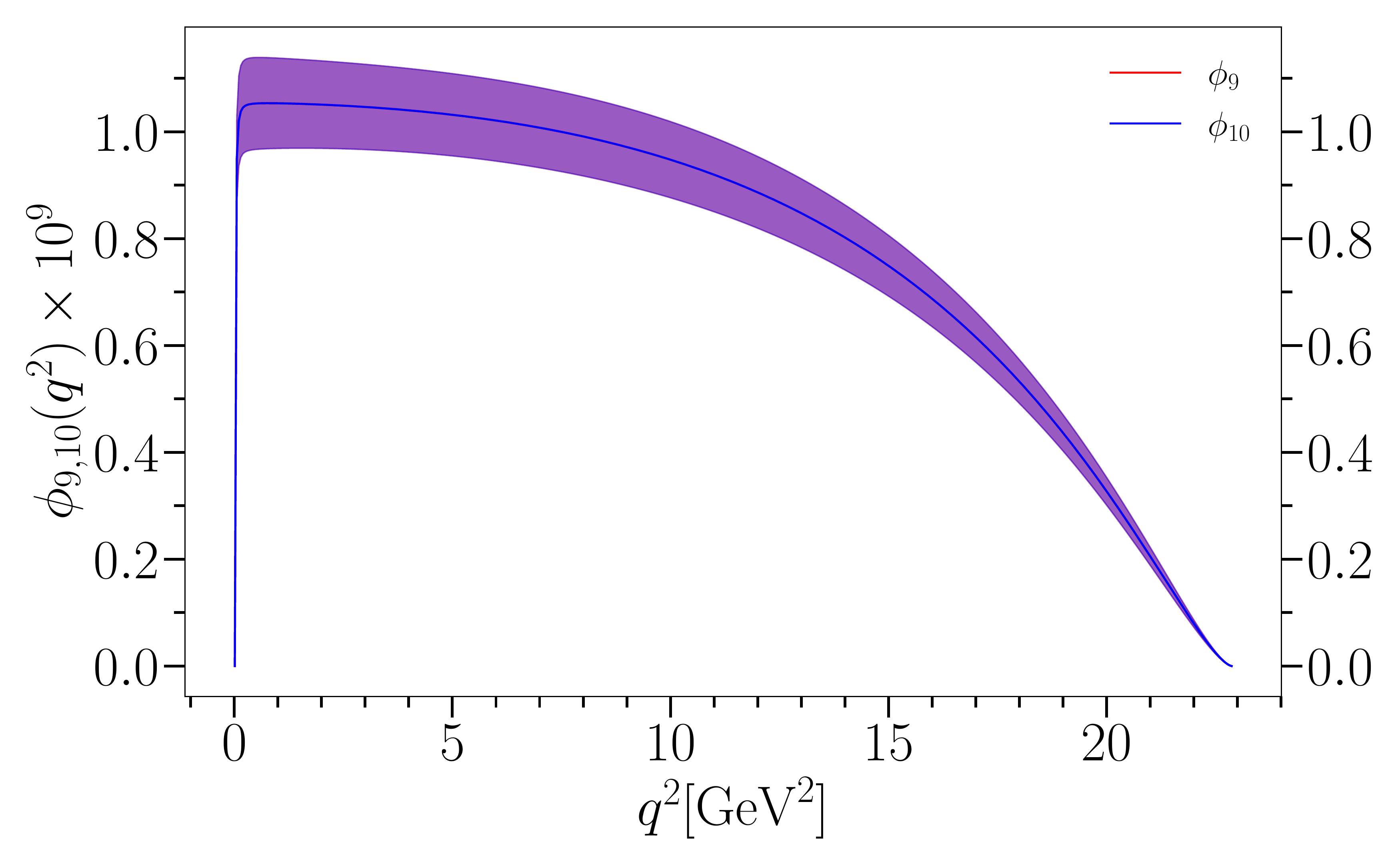}
\caption{$\phi_{9,10}(q^2)\times 10^9$ for the $B^0\to K^0e\mu$ decay.}
\label{fig:phi910emu}
\end{figure}

The shape of $\phi_{S,P}(q^2)$ varies little with lepton flavour choices, so we show only the one example in Fig.~\ref{fig:phiSPmutau}. The shape of $\phi_{9,10}$ is more sensitive, however and we plot the case of $e\mu$  in Fig.~\ref{fig:phi910emu}. It shows the functions falling much more abruptly towards the minimal $q^2=(m_1+m_2)^2$ value than when one lepton is a $\tau$. 

Integrating $\phi_{9,10,S,P}$ across the full physical $q^2$ range for each $\ell_1, \ell_2$ combination yields  $a_K^{12}$, $b_K^{12}$, $e_K^{12}$ and $f_K^{12}$ respectively, and we report their values in Table~\ref{tab:abef}. 
These are updates to the values in~\cite{Becirevic:2016zri} (which took LCSR form factors from~\cite{Ball:2004ye}). 
We agree well with the results found in~\cite{Becirevic:2016zri}, with a modest improvement in $a_K^{12}$ and $b_K^{12}$ uncertainties. Table~\ref{tab:abef} gives results for both charged and neutral mesons, which are slightly different because of the change in lifetime $\tau_{B^{0/+}}$.

The results of Table~\ref{tab:abef} can be inserted into Eqs.~\eqref{eq:vecLFV} or~\eqref{eq:scaLFV} to obtain a branching fraction for given Wilson coefficient values. 
Current experimental limits for $\mathcal{B}(B\to Ke\mu)$ are $7.0\times 10^{-9}$ for $B^+$~\cite{LHCb:2019bix} and $3.8 \times 10^{-8}$ for $B^0$~\cite{BELLE:2019xld}. These limits start to provide constraints on beyond the Standard Model (BSM) physics, for example models with leptoquarks~\cite{Hiller:2016kry}. 
Current experimental limits for $B^+\to h^+ \tau \ell$ are a few times $10^{-5}$~\cite{BaBar:2012azg} and so need to be reduced by several orders of magnitude to provide a test of this decay mode. 

\begin{table}
  \begin{center}
    \begin{tabular}{ccccc}
      \hline
      $\ell_1\ell_2$ & $a^{12}_K$ & $b^{12}_K$ & $e^{12}_K$ & $f^{12}_K$\\
      \hline
      $B^0\to K^0 e\mu$    & 17.7(1.2) & 17.7(1.2) & 24.5(1.4) & 24.5(1.4)\\
      $B^0\to K^0 e\tau$   & 11.08(69) & 11.08(69) & 13.97(79) & 13.97(79)\\
      $B^0\to K^0 \mu\tau$ & 10.85(69) & 11.26(69) & 13.44(76) & 14.41(82)\\
       \hline
      $B^+\to K^+ e\mu$    & 19.2(1.3) & 19.2(1.3) & 26.6(1.5) & 26.6(1.5)\\
      $B^+\to K^+e\tau$   & 11.99(75) & 11.99(75) & 15.14(86) & 15.15(86)\\
      $B^+\to K^+\mu\tau$ & 11.73(75) & 12.18(75) & 14.57(83) & 15.62(89)\\
      \hline     
    \end{tabular}
    \caption{Results for the coefficients of Eqs.~(\ref{eq:vecLFV}) and~(\ref{eq:scaLFV}) for $B^0\to K^0\ell_1\ell_2$ and $B^+\to K^+\ell_1 \ell_2$ decay. No additional uncertainty is included for QED effects.}
    \label{tab:abef}
  \end{center}
\end{table}

\section{$B\to K\nu\bar{\nu}$}\label{sec:nunubar}
Lastly, we study the rare decays $B\to K\nu\bar{\nu}$, which are of phenomenological interest~\cite{Descotes-Genon:2020buf}. Precise measurements in this channel could shed light on new physics models to explain lepton flavour universality violation~\cite{Altmannshofer:2020axr,BhupalDev:2021ipu}.

There is limited experimental data at the present time, and currently no experimental results have conclusively demonstrated a non-zero branching fraction. However, more experimental results are anticipated in the near future, and the upper limit on the branching fraction may soon be driven down close to theoretical determinations~\cite{Halder:2021sgd}.

We can calculate the short distance (SD) contribution to the differential branching fraction (summed over neutrino flavours)~\cite{Du:2015tda,Altmannshofer:2009ma,Buras:2014fpa}, which depends only on the vector form factor. 
\begin{equation}\label{eq:nunubarBF}
  \begin{split}
    \frac{d\mathcal{B}(B\to K\nu\bar{\nu})_{\text{SD}}}{dq^2} =& \frac{G_F^2\alpha_{\mathrm{EW}}^2(M_Z)X_t^2}{32\pi^5\sin^4\theta_W}\\
    &\times\tau_B|V_{tb}V^*_{ts}|^2|\vec{p}_K|^3f^2_+(q^2),
  \end{split}
\end{equation}
with $X_t=1.469(17)$~\cite{Brod:2010hi}, $\sin^2\theta_W=0.23121(4)$~\cite{Zyla:2020zbs} and $1/\alpha_{\rm EW}(M_Z)=127.952(9)$~\cite{Zyla:2020zbs}. Other parameters are given in Table~\ref{tab-SMparams}. We evaluate this expression using our improved vector form factor determined in~\cite{BtoK}, including uncertainties for strong-isospin breaking discussed in Section~\ref{sec:isospin}. If the reader wishes to reproduce this, $f_+(q^2)$ can be easily obtained using the code attached to~\cite{BtoK}. Note that no QED uncertainty is required in this case. 

This short distance expression is sufficient to describe the case of $B^0\to K^0\nu\bar{\nu}$ decays and dominates the branching fraction in $B^+\to K^+\nu_{\ell}\bar{\nu_{\ell}}$ decays. In the case of $B^+\to K^+\nu_{\tau}\bar{\nu_{\tau}}$ however, long distance (LD) effects from a double charged-current interaction have a noticeable contribution. These are given by~\cite{Kamenik:2009kc}
\begin{equation}
  \begin{split}
    \mathcal{B}(B^+\to K^+\nu_{\tau}\bar{\nu}_{\tau})_{\mathrm{LD}}=&\frac{|G_F^2V_{ub}V^*_{us}f_{K^+}f_{B^+}|^2}{128\pi^2M^3_{B^+}}\\
    &\times\frac{ m_{\tau}(M^2_{B^+}-m_{\tau}^2)^2(M^2_{K^+}-m_{\tau}^2)^2}{\Gamma_{\tau}\Gamma_{B^+}}.
  \end{split}
\end{equation}

We take the following parameter values to determine this branching fraction. 
$|V_{ub}|=0.00370(10)(12)$ is given from exclusive $B\to\pi\ell\nu$ decays in~\cite{Zyla:2020zbs,HFLAV:2019otj}, based on experimental data and lattice QCD form factors from~\cite{Lattice:2015tia,Flynn:2015mha}. 
Other values, $|V_{us}|f_{K^+}=0.03509(4)(4)\,\mathrm{GeV}$~\cite{Zyla:2020zbs,Rosner:2015wva,Moulson:2014cra}, $f_{B^+}=0.1894(8)(11)(3)(1)\,\mathrm{GeV}$~\cite{Bazavov:2017lyh} and $\tau_\tau=0.2903(5)\,\mathrm{ps}$~\cite{Zyla:2020zbs} are used to obtain 
\begin{equation}
\label{eq:nuLD}
\mathcal{B}(B^+\to K^+\nu_{\tau}\bar{\nu_{\tau}})_{\text{LD}} = 6.09(53)\times10^{-7} \,. 
\end{equation}
This is a minor update on the $6.22(60)\times10^{-7}$ calculated in~\cite{Du:2015tda}, and represents an adjustment of about +10\% (or roughly $1.5\sigma$) to our $\mathcal{B}(B^+\to K^+\nu_{\tau}\bar{\nu_{\tau}})_{\text{SD}}$ result.  

\begin{figure}

\includegraphics[width=0.48\textwidth]{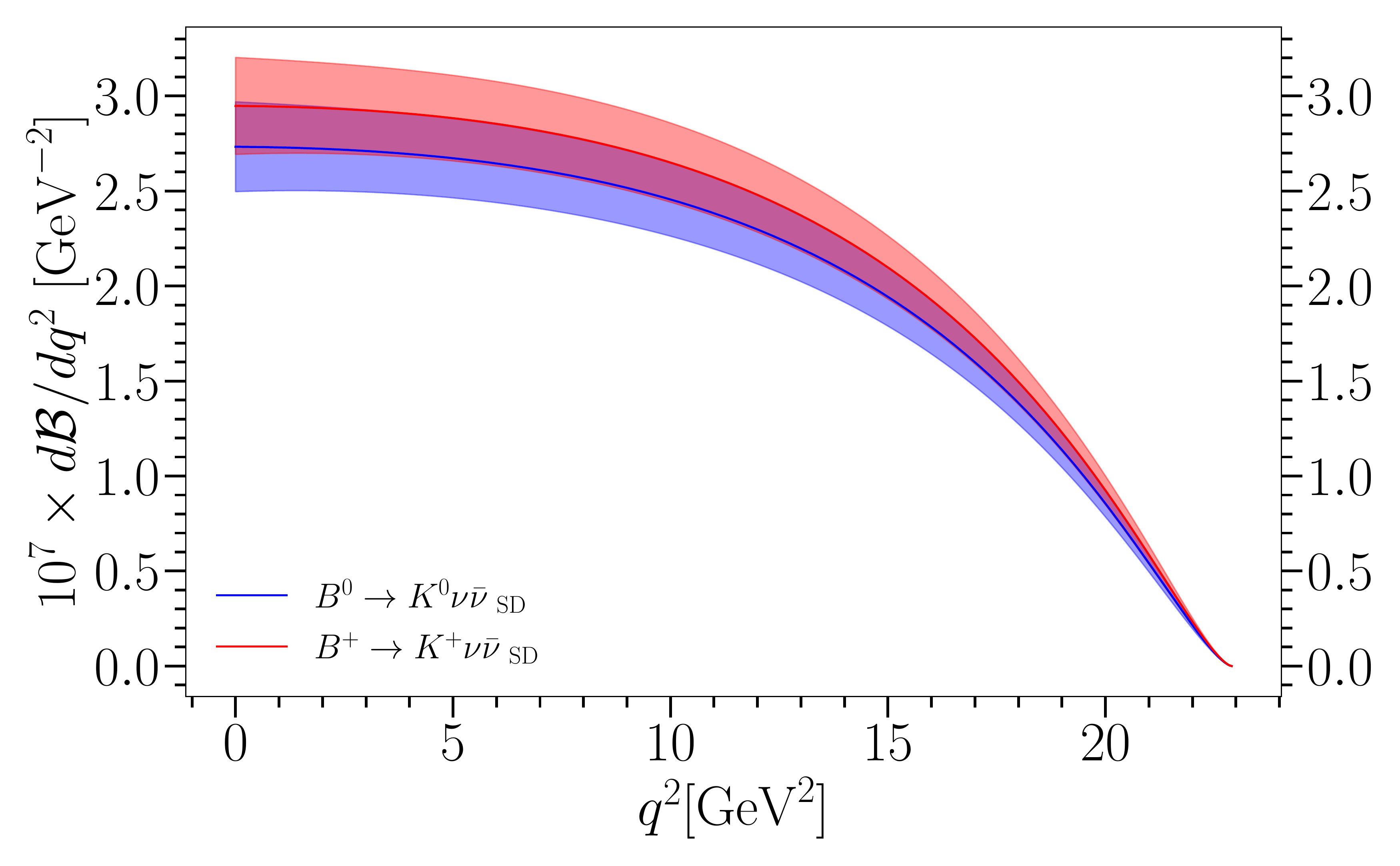}
\caption{The short distance contribution to $d\mathcal{B}(B\to K\nu\bar{\nu})/dq^2$, for both the charged and neutral $B$ meson cases.}
\label{fig:nuinqsq}
\end{figure}

Our results for the differential branching fractions for the short-distance contribution are shown in Figure~\ref{fig:nuinqsq} for both the charged and neutral $B$ meson cases. Note that there is no issue with $c\overline{c}$ resonances or nonfactorisable terms in this case so the SM calculation is very straightforward and simply requires an accurate vector form factor. From the differential branching fraction we calculate the total branching fractions, adding in the long-distance contribution to the charged case. These values are given in Table~\ref{tab-neutrino}, and show a total uncertainty of around 7\%. Of this uncertainty 5\% comes from the vector form factor, 4.4\% from the CKM elements $|V_{tb}V_{ts}^*|$ and 2.3\% from $X_t^2$. 

\begin{figure}
  \includegraphics[width=0.48\textwidth]{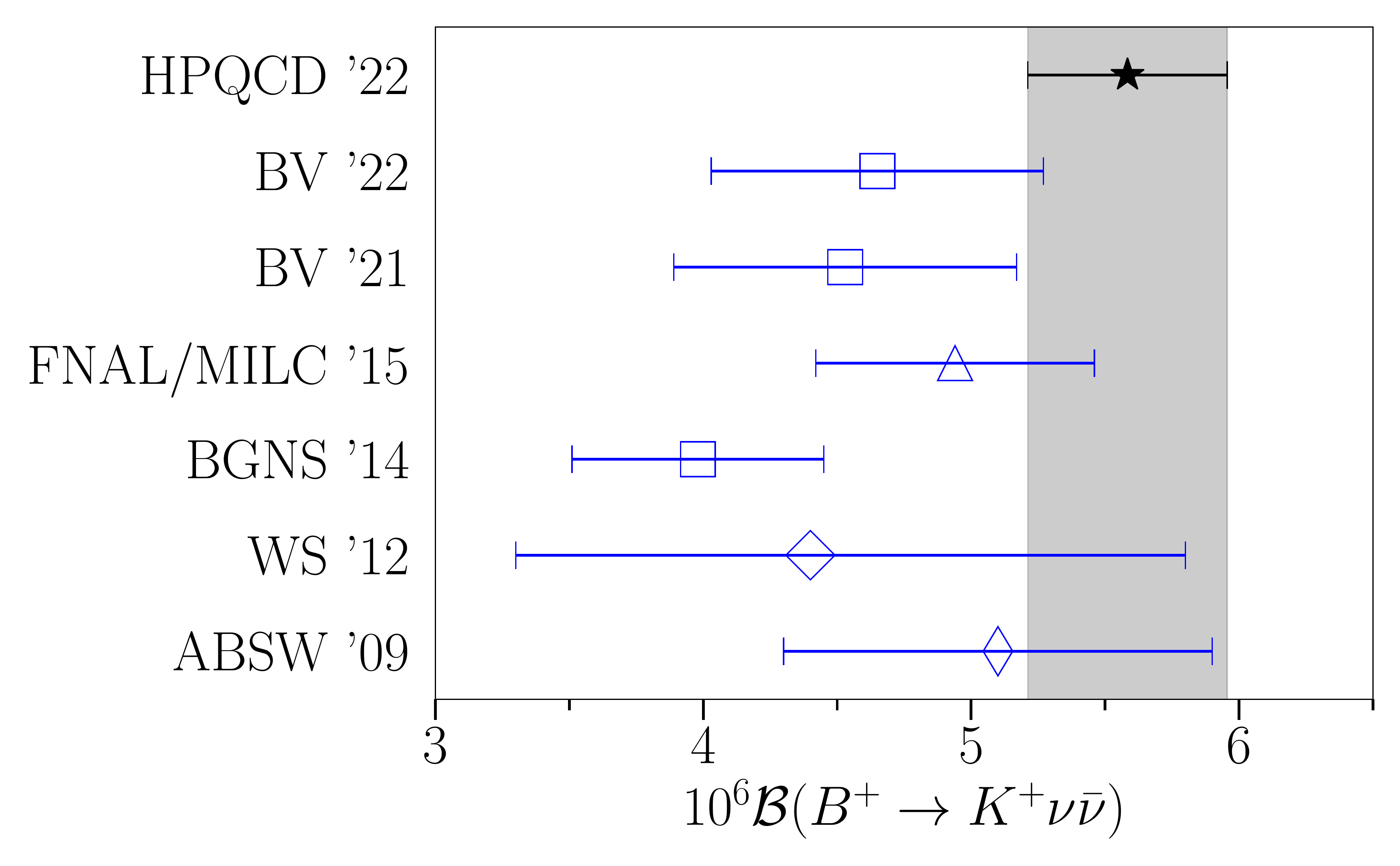}
  \caption{Our result for $\mathcal{B}(B^+\to K^+\nu\bar{\nu})$, including long distance effects, compared with other theory calculations from Table~\ref{tab-neutrino}~\cite{Altmannshofer:2009ma,Kamenik:2009kc,Wang:2012ab,Buras:2014fpa,Du:2015tda,Buras:2021nns,Buras:2022wpw}.}
  \label{fig:the_neutrino_branching}
\end{figure}

Table~\ref{tab-neutrino} also compares our values to experimental limits and earlier theoretical results. Our results sit slightly higher than previous theoretical determinations and have smaller uncertainty. The comparison of our $B^+\to K^+\nu\bar{\nu}$ branching fraction result with other theoretical values is shown in Figure~\ref{fig:the_neutrino_branching}. Results in various $q^2$ bins are presented in Table~\ref{tab:BKnunubins}.

\begin{table}
  \begin{center}
    \begin{tabular}{ccc}
      \hline
      \hline	
      Decay  		& $\mathcal{B}\times 10^{6}$		     & Reference    \\ 
      \hline
      $B^0\to K_S^0\nu\bar{\nu}$ & $<13$ (90\% CL) Exp.&  \cite{Grygier:2017tzo}\\
      \hline
      \multirow{4}{*}{$B^0\to K^0\nu\bar{\nu}$} & $<49$ (90\% CL) Exp.& \cite{Lees:2013kla}\\
                        & 4.01(49) & \cite{Du:2015tda}\\
                        & $4.1^{+1.3}_{-1.0}$ & \cite{Wang:2012ab}\\
                        & 4.60(34) & HPQCD '22\\
      \hline
      \multirow{7}{*}{$B^+\to K^+\nu\bar{\nu}$} & $<16$ (90\% CL) Exp.& \cite{Lees:2013kla}\\
                        & $<19$ (90\% CL) Exp.& \cite{Grygier:2017tzo}\\
                        & $<41$ (90\% CL) Exp. & \cite{Belle-II:2021rof} \\
                        & 5.10(80) & \cite{Altmannshofer:2009ma,Kamenik:2009kc}\\
                        & $4.4^{+1.4}_{-1.1}$ & \cite{Wang:2012ab}\\
                        & 3.98(47) & \cite{Buras:2014fpa} \\
                        & 4.94(52) & \cite{Du:2015tda}\\
                        & 4.53(64) &\cite{Buras:2021nns}\\
                        & 4.65(62) &\cite{Buras:2022wpw}\\
                        & 5.58(37) & HPQCD '22\\
      \hline
      \hline
    \end{tabular}\caption{Summary of experimental and theoretical results for the branching fraction $\mathcal{B}(B\to K\nu\bar{\nu})$, summed over neutrino flavours. Our results are denoted by `HPQCD '22' and include the long-distance contribution of Eq.~\eqref{eq:nuLD} in the charged $B$ case. Experimental results are given as 90\% confidence limit and marked `Exp.'. In the case of~\cite{Buras:2021nns}, we use $|V_{tb}V_{ts}^*|$ (Table~\ref{tab-SMparams}) to obtain the result.}
    \label{tab-neutrino}
  \end{center}
\end{table}
\begin{figure}
  \includegraphics[width=0.48\textwidth]{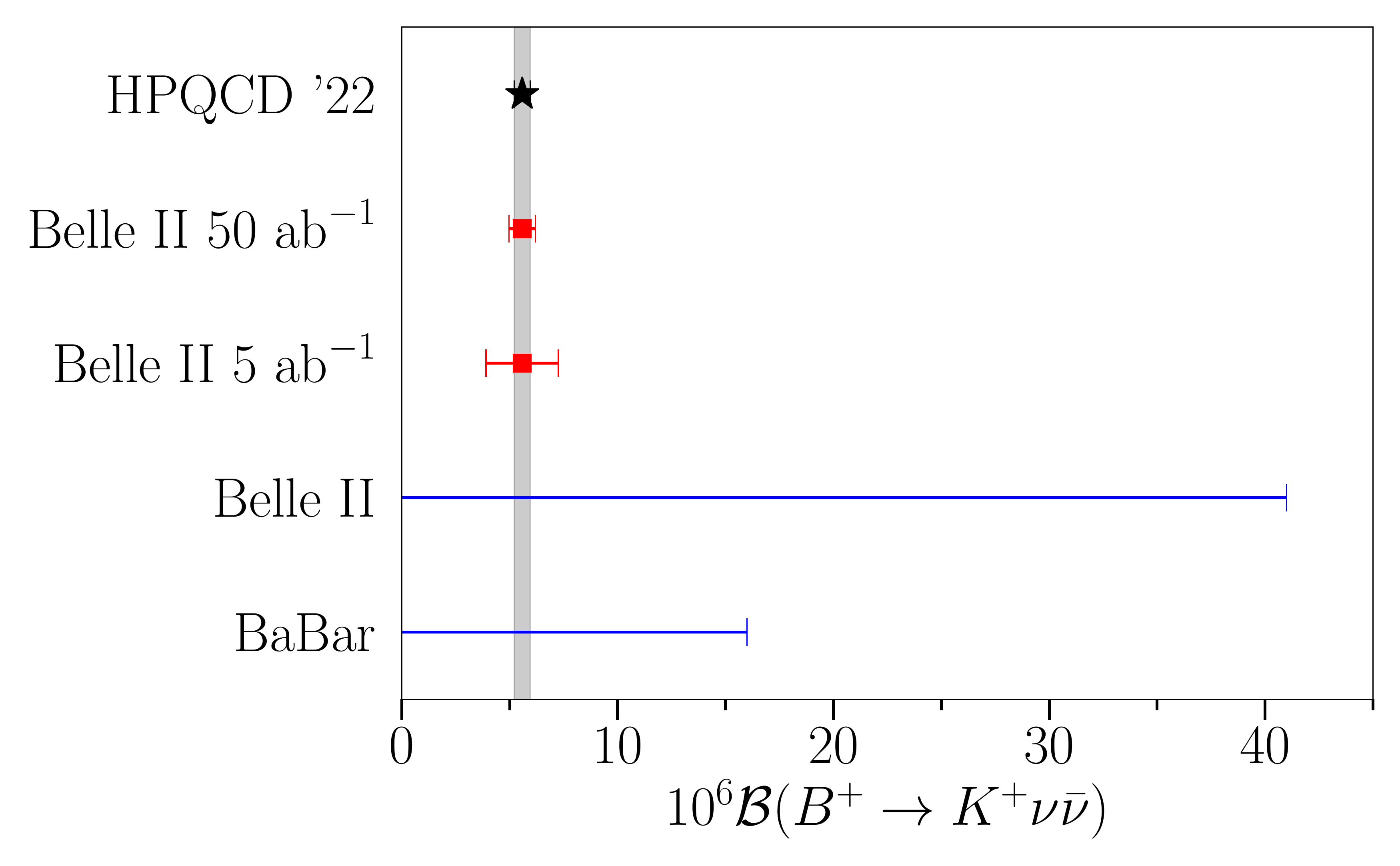}
  \caption{Our result (black filled star) for $\mathcal{B}(B^+\to K^+\nu\bar{\nu})$, summing short- and long-distance effects, compared with current 90\% confidence limits from BaBar~\cite{Lees:2013kla} and Belle II~\cite{Belle-II:2021rof} (blue lines). Red filled squares with error bars show the expected precision to be achieved by Belle II with 5~$\text{ab}^{-1}$ and 50~$\text{ab}^{-1}$~\cite{Halder:2021sgd}, assuming our result is obtained for the central value.}
  \label{fig:neutrino_branching}
\end{figure}

Current experimental bounds on the $B\to K\nu\overline{\nu}$ branching fraction are roughly an order of magnitude larger than the SM expectation given by the theory results of Table~\ref{tab-neutrino}. The uncertainty that we have achieved here, however, complements the $\approx10\%$ error expected from Belle II with 50~$\text{ab}^{-1}$~\cite{Halder:2021sgd}, allowing for a more stringent test of this quantity in future. This is demonstrated in Figure~\ref{fig:neutrino_branching}, which compares our result with the 90\% confidence limits set by BaBar~\cite{Lees:2013kla} and Belle II~\cite{Belle-II:2021rof} in Table~\ref{tab-neutrino}, as well as the precision forecast by Belle II with 5~$\text{ab}^{-1}$ and 50~$\text{ab}^{-1}$~\cite{Halder:2021sgd}, assuming a result centred on our value. 

\section{Conclusions}\label{sec:conc}
We have used our improved scalar, vector and tensor form factors calculated in $N_f=2+1+1$ lattice QCD~\cite{BtoK} to determine SM observables for  $B\to K\ell^+\ell^-$, $B\to K\ell^-_1\ell^+_2$ and $B\to K\nu\bar{\nu}$ decays. The form factors were calculated in a fully relativistic approach which enables the full $q^2$ range of the decay to be covered. This approach also gives better control of the current normalisation than in previous calculations. Our form factors then have smaller uncertainties, particularly at low $q^2$, than earlier work so that we can improve QCD uncertainties on the SM observables. We have tabulated all of our results across a variety of $q^2$ bins in Appendix~\ref{sec:numres}, for future use.

For $B\to K \ell^+\ell^-$ we determine the differential rate for both charged and neutral meson cases, including an analysis of nonfactorisable corrections at low $q^2$ (see Appendix~\ref{app:corrs}). We then compare to the wealth of experimental information (see Figures~\ref{fig:dBdqemup},~\ref{fig:dBdqemu0} and~\ref{fig:dBdqemu}) focussing on the low and high $q^2$ ranges below and above the region where $c\overline{c}$ resonances make a large contribution to experimental results, but are not included in our calculation. We allow an uncertainty for QED effects missing from our calculation in this comparison. 

Previous lattice QCD calculations by the Fermilab/MILC collaboration~\cite{Du:2015tda} saw a ~2$\sigma$ tension between the SM branching fraction and LHCb results~\cite{Aaij:2014pli} in the low and high $q^2$ regions, with the SM results being higher than experiment in both regions. Our improved form factors significantly sharpen this tension, which becomes particularly strong in the region below the charmonium resonances. For $1.1 < q^2 < 6 \text{GeV}^2$ we find a tension of $4.2\sigma$ for $B^+\to K^+ \mu^+\mu^-$ and $3.1\sigma$ for $B^0\to K^0\mu^+\mu^-$ with LHCb '14A~\cite{Aaij:2014pli} and $2.7\sigma$ for $B^+\to K^+ e^+e^-$ with LHCb '21~\cite{Aaij:2021vac} (see Table~\ref{tab:comparisons} and Fig.~\ref{fig:headline}). We see no tension with results from Belle '19~\cite{BELLE:2019xld} in this low $q^2$ region, but they have much larger uncertainties. The tension between our results and LHCb in the high $q^2$ region ($15 < q^2 < 22 {\text{GeV}}^2$) amounts to $2.7\sigma$, again with our SM results being higher than experiment. 

The tension between the SM and LHCb results in the `safe' $q^2$ regions away from narrow $c\overline{c}$ resonances points to the possibility of BSM physics that would have the effect of changing the Wilson coefficients of the effective weak Hamiltonian. Recent analyses (see, for example,~\cite{Altmannshofer:2021qrr,Gubernari:2022hxn}) focus on BSM changes to $C_9$ and $C_{10}$ and include a number of different $B$ decay processes in providing a best fit result for these BSM effects. Gubernari et al~\cite{Gubernari:2022hxn} give values $\Delta C_9^{\text{BSM}}=-1.0$ and $\Delta C_{10}^{\text{BSM}}=+0.4$ at $\mu = 4.2\,\text{GeV}$ from comparison of their SM results to experiment in three separate analyses, one of which is a combined one using $B\to K\mu^+\mu^-$ and $B\to \mu^+\mu^-$ and the other two are for $B \to K^* \mu^+\mu^-$ and $B_s\to \phi \mu^+\mu^-$. If we repeat our analysis changing $C_9$ and $C_{10}$ by these BSM shifts we see significantly lowered values for the differential rate for $B\to K \ell^+\ell^-$ because $C_9$ and $C_{10}$ now have smaller magnitudes. We find the tension with LHCb '14A in the low $q^2$ region (1.1--6 $\text{GeV}^2$) to be reduced from 4.2$\sigma$ to 1.1$\sigma$. All of the other tensions seen in Table~\ref{tab:comparisons} where the SM result exceeds the experimental value are now reduced below 1$\sigma$. At the same time a 2.2$\sigma$ tension with the Belle '19 result for $B^+\to K^+ \mu^+\mu^-$ develops in the opposite direction. We conclude that BSM shifts to $C_9$ and $C_{10}$ of -1.0 and +0.4 respectively give improved agreement between our results and the experimental values for the $B \to K\ell^+ \ell^-$ branching fractions. 

We update results for ratios of $B\to K \ell^+\ell^-$ branching fractions to leptons of different flavour, $\ell$, giving results for $q^2$ bins across the kinematic range in Section~\ref{sec:Rtheory}. The `flat term' in the lepton angular distribution is also examined in detail in Section~\ref{sec:FHtheory} for different $\ell$.  

We have also provided improved results for the form-factor-dependent pieces that, multiplied by appropriate Wilson coefficients, give the branching fraction for the lepton flavour number violating decay $B\to K \ell_1^+\ell_2^-$. Table~\ref{tab:abef} gives our results for all 3 lepton flavour pairs and for charged and neutral $B$ mesons, with uncertainties of 6--7\%. These can be used to constrain BSM models in future searches for this decay. 

Finally,  Table~\ref{tab-neutrino} gives our improved SM values for $\mathcal{B}(B^+\to K^+\nu\bar{\nu})$ and $\mathcal{B}(B^0\to K^0\nu\bar{\nu})$. We have uncertainties below 10\%, commensurate with that expected from Belle II with $50~\mathrm{ab}^{-1}$ worth of data. This decay mode, expected to reach 3$\sigma$ significance at Belle II with $5~\mathrm{ab}^{-1}$ of data, is an exciting possibility for future tests for BSM physics, since it is theoretically `cleaner' than $B \to K\ell^+\ell^-$ across the full $q^2$ range with no need for vetoed regions. 

We have shown here the improvements to $B\to K$ semileptonic decay phenomenology that result from our improved lattice QCD form factors covering the full kinematic $q^2$ range. For $B\to K \ell^+\ell^-$ our relative uncertainties are comparable with those from experiment. Our uncertainties are dominated by those from the form factors so further improvements in the lattice QCD calculation of the form factors will reduce them further. This is possible in future with increased computational power enabling smaller statistical errors and a push towards smaller values of the lattice spacing reducing residual discretisation effects. A more detailed comparison will then be possible with future experimental results with smaller uncertainties for $B\to K \ell^+ \ell^-$ and future differential rates for $B \to K\nu \overline{\nu}$. In combination with analysis of other decay processes for hadrons containing $b$ quarks, this is a promising way forward to uncover physics beyond the Standard Model.

\section{Acknowledgements}
We thank D. van Dyk, P. Owen and E. Lunghi for clarification regarding data in their papers and Rafael Coutinho for help understanding the use of PHOTOS in LHCb results. We also thank L. Grillo and R. Zwicky for useful discussions, as well as Quim Matias and S\'ebastien Descotes-Genon. Computing was done on the Cambridge Service for Data Driven Discovery (CSD3) supercomputer, part of which is operated by the University of Cambridge Research Computing Service on behalf of the UK Science and Technology Facilities Council (STFC) DiRAC HPC Facility. The DiRAC component of CSD3 was funded by BEIS via STFC capital grants and is operated by STFC operations grants. We are grateful to the CSD3 support staff for assistance. Funding for this work came from STFC.

\textit{Note added in proof} - Whilst this paper was being prepared for publication, a set of new results for $R_K$ and $R^*_K$ appeared from LHCb~\cite{LHCb:2022qnv}. These results agree with Standard Model expectations (see Sec.~\ref{sec:Rmue}) and supersede their earlier values. 

\begin{appendix}
\section{Calculating pole masses}\label{app:polemass}
The three loop relation between quark masses in the $\overline{\mathrm{MS}}$ scheme and the pole mass scheme is~\cite{Melnikov:2000qh}
\begin{equation}\label{eq:msbar_over_pole}
  \begin{split}
    \frac{\overline{m}(m)}{m} &= 1+A\Big(\frac{\alpha_s}{\pi}\Big) + B\Big(\frac{\alpha_s}{\pi}\Big)^2+C\Big(\frac{\alpha_s}{\pi}\Big)^3,\\
    A&=-\frac{4}{3},\\
    B&= 1.0414N_L-14.3323,\\
    C& = -0.65269N_L^2+26.9239N_L-198.7068,
  \end{split}
\end{equation}
where $\overline{m}$ is the $\overline{\mathrm{MS}}$ mass, $m$ is the pole mass and $N_L$ is the number of active light quarks. 
We evaluate $\alpha_s$ at scale $m$ and use $N_L = 3$ for $c$ and $N_L = 4$ for $b$.

Inverting \eqref{eq:msbar_over_pole} gives
\begin{equation}\label{eq:polemassiter}
  \begin{split}
    m =&\overline{m}(m)\Big[1-A\Big(\frac{\alpha_s}{\pi}\Big) + (A^2-B)\Big(\frac{\alpha_s}{\pi}\Big)^2\\
    &+(-A^3+2AB-C)\Big(\frac{\alpha_s}{\pi}\Big)^3\Big].
  \end{split}
\end{equation}
%
%
We solve Equation~\eqref{eq:polemassiter} iteratively, making an initial guess for $m$ then evaluating $\alpha_s(m)$ by running from $\alpha_s(5.0\,\mathrm{GeV}) = 0.2128(25)$~\cite{Chakraborty:2014aca} and $\overline{m}(m)$ by running from the $m_c^{\overline{\mathrm{MS}}}(m_c^{\overline{\mathrm{MS}}})$ or $m_b^{\overline{\mathrm{MS}}}(m_b^{\overline{\mathrm{MS}}})$ value in Table~\ref{tab-SMparams}.
Plugging these results into Equation~\eqref{eq:polemassiter} results in an updated value for $m$. 
The initial guess for $m$ is adjusted to reduce the difference between it and the value obtained from Equation~\eqref{eq:polemassiter}.
This process is repeated until the values of $m$ converge.

Using this method, we obtain the pole masses $m_c = 1.684(22)\,\mathrm{GeV}$ and $m_b=4.874(32)\,\mathrm{GeV}$. 
The perturbation series in this expression suffers from the presence of a renormalon in the pole mass~\cite{Beneke:1994sw}, so we take a $200\,\mathrm{MeV}$ uncertainty on both numbers. The effect of this uncertainty is described in Section~\ref{sec:inputs}. We note that $4m_c^2=11.34\,\mathrm{GeV}^2$ falls within the vetoed charmonium  resonance region.

\section{Corrections to $C_7^{\mathrm{eff},0}$ and $C_9^{\mathrm{eff},0}$}\label{app:corrs}
As mentioned in Section~\ref{sec:obs}, corrections need to be applied to $C_7^{\mathrm{eff},0}$ and $C_9^{\mathrm{eff},0}$ to obtain the values of $C_7^{\mathrm{eff},1}$ and $C_9^{\mathrm{eff},1}$ that enter Equation~(\ref{eq:Fs}).
The corrections are defined by
\begin{align}
  C_7^{\mathrm{eff},1} &= C_7^{\mathrm{eff},0} +\delta C_7^{\mathrm{eff}} ,\label{eq:C7eff} \\
  C_9^{\mathrm{eff},1} &= C_9^{\mathrm{eff},0} + \Delta C_9^{\mathrm{eff}}+\delta C_9^{\mathrm{eff}},\label{eq:C9eff}
\end{align}
where $C_7^{\mathrm{eff},0}$ and $C_9^{\mathrm{eff},0}$ are given in Table~\ref{tab-SMparams}. 
The corrections are discussed in Appendix B of~\cite{Du:2015tda}, which compiles results from~\cite{Bobeth:2011nj,Bobeth:2012vn,Beneke:2001at,Greub:2008cy,Grinstein:2004vb,Bobeth:2010wg,Beylich:2011aq,Bobeth:2011gi,Beneke:2000wa,Asatrian:2001de,Asatryan:2001zw,Asatrian:2003vq,Seidel:2004jh,Beneke:2004dp}.
We do not repeat formulae from that reference here, but direct the reader there for more detail; our aim here is to assess the significance of these corrections to the decay rates.
Below we outline the form of the corrections, plot them versus $q^2$ and discuss their sizes relative to  $C_7^{\mathrm{eff},0}$ and $C_9^{\mathrm{eff},0}$. 
All numerical inputs not explicitly stated below can be found in Table~\ref{tab-SMparams}. 

The leading contribution to the correction $\delta C_7^{\mathrm{eff}}$ is from $\mathcal O(\alpha_s)$ effects,
\begin{equation}\label{eq:C7OalOla}
    \delta C_7^{\mathrm{eff}} = -\frac{\alpha_s}{4\pi} \Big((C_1-6C_2)F_{1,c}^{(7)}+C_8F_8^{(7)}\Big),
\end{equation}
where the expression for $F_{1,c}^{(7)}$ is lengthy and provided in the C\verb!++! header files of~\cite{Greub:2008cy}. $F_8^{(7)}$ is given in Appendix B of~\cite{Du:2015tda}.
We use $\alpha_s(4.2\,\mathrm{GeV}) = 0.2253(28)$, which is run from $\alpha_s(5.0\,\mathrm{GeV}) = 0.2128(25)$~\cite{Chakraborty:2014aca}.
The next higher order contribution, which we neglect, is suppressed by a factor of $\lambda_u^{(s)} = \frac{V_{us}^*V_{ub}}{V^*_{ts}V_{tb}} = 0.01980(62)$~\cite{Du:2015tda} and is $\mathcal O(\alpha_s \lambda_u^{(s)})$.
The leading order contributions to the correction $\delta C_9^{\mathrm{eff}}$ are given by
\begin{equation}\label{eq:OalOla}
  \begin{split}
    \delta C_9^{\mathrm{eff}}=-&\frac{\alpha_s}{4\pi} \left(C_1F_{1,c}^{(9)}+C_2F_{2,c}^{(9)}+C_8F_8^{(9)}\right)\\
    +&\lambda_u^{(s)} \left(h(q^2,m_c)-h(q^2,0)\right)\Big(\frac{4}{3}C_1+C_2\Big).
  \end{split}
\end{equation}
We neglect the $\mathcal{O}(\alpha_s\lambda_u^{(s)})$ term, which is even smaller (see Equation~(B11) of~\cite{Du:2015tda} for more details). 
The function $h(q^2,m)$ is defined in Equation~(\ref{eq:h}) and $F_8^{(9)}$ is given in Appendix B of~\cite{Du:2015tda}. 
Expressions for $F_{1,c}^{(9)}$ and $F_{2,c}^{(9)}$ are also provided in the C\verb!++! header files of~\cite{Greub:2008cy}. 
The corrections $\delta C_7^{\mathrm{eff}}$ and $\delta C_9^{\mathrm{eff}}$ are applicable across the full $q^2$ range.
 \begin{figure}
\includegraphics[width=0.48\textwidth]{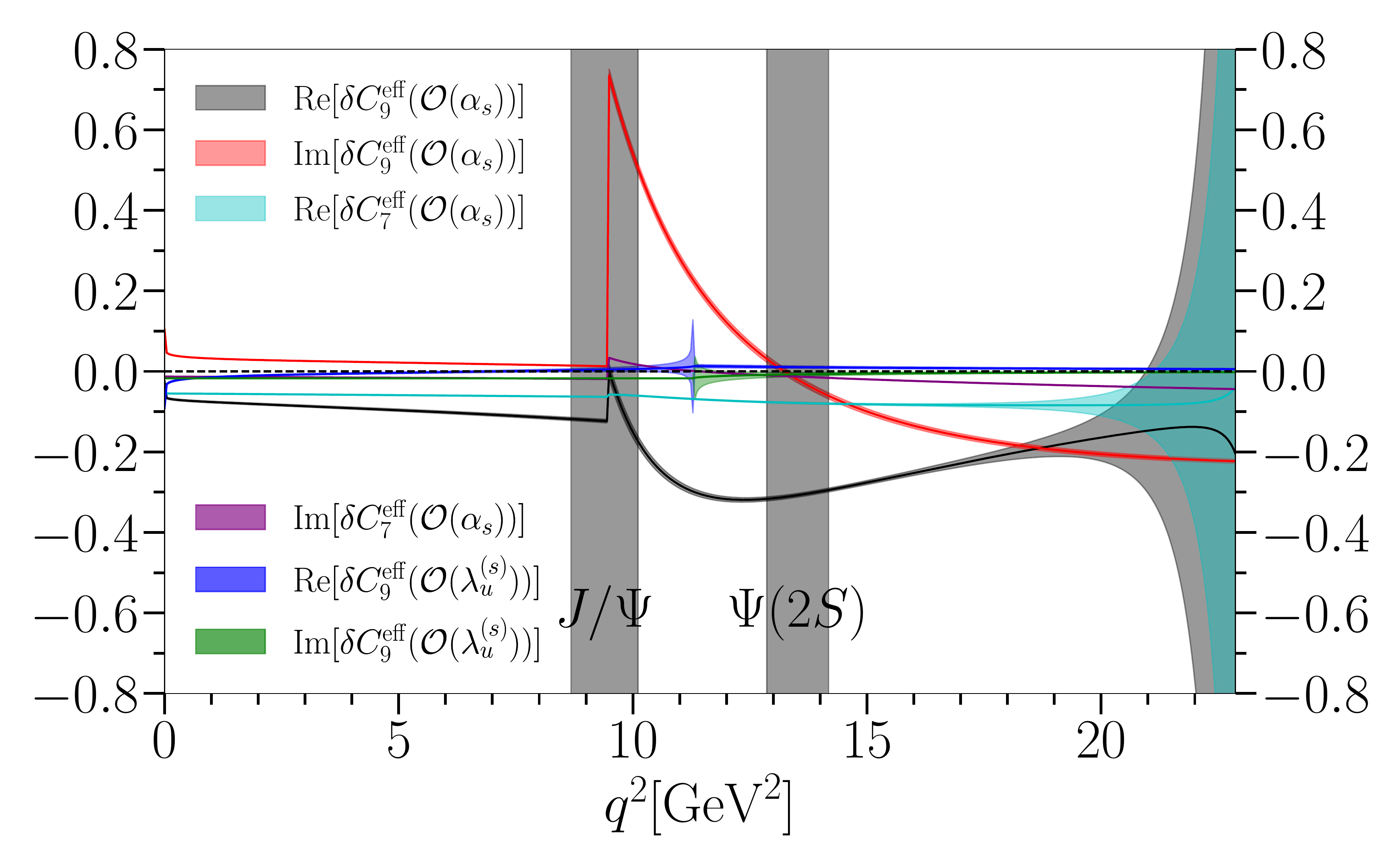}
\caption{The real and imaginary parts of $\delta C_7^{\mathrm{eff}}$ and of the $\mathcal O(\alpha_s)$ and $\mathcal O(\lambda_u^{(s)})$ contributions to $\delta C_9^{\mathrm{eff}}$, as defined in Equations~(\ref{eq:OalOla}) and~(\ref{eq:C7OalOla}).}
\label{fig:OalOla}
\end{figure}

We plot the real and imaginary parts of $\delta C_7^{\mathrm{eff}}$ and $\delta C_9^{\mathrm {eff}}$
in Figure~\ref{fig:OalOla}, showing separately the $\mathcal{O}(\alpha_s)$ and $\mathcal{O}(\lambda_u^{(s)})$ contributions to $\delta C_9^{\mathrm{eff}}$. 
The correction Re$[\delta C_7^{\mathrm{eff}}]$ is approximately 20\% of Re$[C_7^{\mathrm{eff},0}]\approx-0.3$ (Table~\ref{tab-SMparams}).
The magnitude of Im$[\delta C_7^{\mathrm{eff}}]$ is small in comparison to $|C_7^{\mathrm{eff},1}|$.
The $\mathcal{O}(\lambda_u^{(s)})$ contributions to $\delta C_9^{\mathrm{eff}}$ are negligible in comparison to $C_9^{\mathrm{eff},0}$, which is approximately 4 across the full $q^2$ range (Table~\ref{tab-SMparams}).
The $\mathcal{O}(\alpha_s)$ contributions to $\delta C_9^{\mathrm{eff}}$ peak at $q^2\approx 10\,\mathrm{GeV}^2$, due to the behaviour of the functions $F_{1,c}^{(7)}$, $F_{2,c}^{(7)}$, $F_{1,c}^{(9)}$ and $F_{2,c}^{(9)}$.
This peak occurs within the experimentally vetoed $J/\Psi$ resonance region and is largely contained within the region of $q^2$ between the $J/\Psi$ and $\Psi(2S)$ resonances, outside of which the contributions are modest.
As a result, it has minimal impact on results in the well behaved regions of $q^2$ below the $J/\Psi$ and above the $\Psi(2S)$.
The uncertainty in Re$[\delta C_7^{\mathrm{eff}}]$ and Re$[\delta C_9^{\mathrm{eff}}]$ grows rapidly towards $q^2_{\mathrm{max}}$ because of the behaviour of $F_8^{(7/9)}$~\cite{Du:2015tda}. This effect is suppressed in observables by the fact that the differential decay rate vanishes at $q^2_{\rm max}$
and so it is not noticeable in plots of observables versus $q^2$ or in uncertainties of observables in the largest $q^2$ bins.

Non-factorisable corrections are accounted for via $\Delta C_9^{\mathrm{eff}}$, which is dependent on the meson charge (see below).
Following the notation of~\cite{Du:2015tda},
\begin{equation}\label{eq:nonfac}
  \begin{split}
    &\Delta C_9^{\mathrm{eff}} = \frac{2\pi^2m_bf_Bf_K}{3M^2_Bf_+}\times\\
    &\sum_{\pm}\int_0^{\infty}\frac{d\omega}{\omega}\Phi_{B,\pm}(\omega)\int_0^1du\Phi_K(u)\big[T_{K,\pm}^{(0)}+\frac{\alpha_s}{4\pi}C_FT_{K,\pm}^{(\mathrm{nf})}\big],
  \end{split}
\end{equation}
where $C_F=4/3$, and $T_{K,+}^{(0)}=0$. 
The functions $\Phi_{B,\pm}(\omega)$, $\Phi_K$, and $T_{K,\pm}^{(0/\mathrm{nf})}$ are given in~\cite{Beneke:2001at} and Equation~(\ref{eq:nonfac}) is discussed in detail in Appendix B of~\cite{Du:2015tda}. 
We evaluate the expressions for $\Phi_{B,\pm}(\omega)$, $\Phi_K$, and $T_{K,\pm}^{(0/\mathrm{nf})}$  using the following inputs: $\omega_0^{-1} = 3(1)\, {\rm GeV}^{-1}$~\cite{Beneke:2001at}, $a_1^{K} = 0.0453(30)$, and $a_2^{K} = 0.175(50)$~\cite{Braun:2006dg} (we take $\Phi_K$ to second order). 
The non-factorisable corrections are valid for small $q^2$, and we turn them off at $q^2 = 8.68\,\mathrm{GeV}^2$, the start of the vetoed $J/\Psi$ resonance region. We do not calculate through this region. Instead, we linearly interpolate, beginning from the point where the corrections are turned off through the $\Psi(2S)$ resonance, so the differential branching fraction is a smooth function of $q^2$ (see, e.g., Figure~\ref{fig:dB_the_0}). 
\begin{figure}
\includegraphics[width=0.48\textwidth]{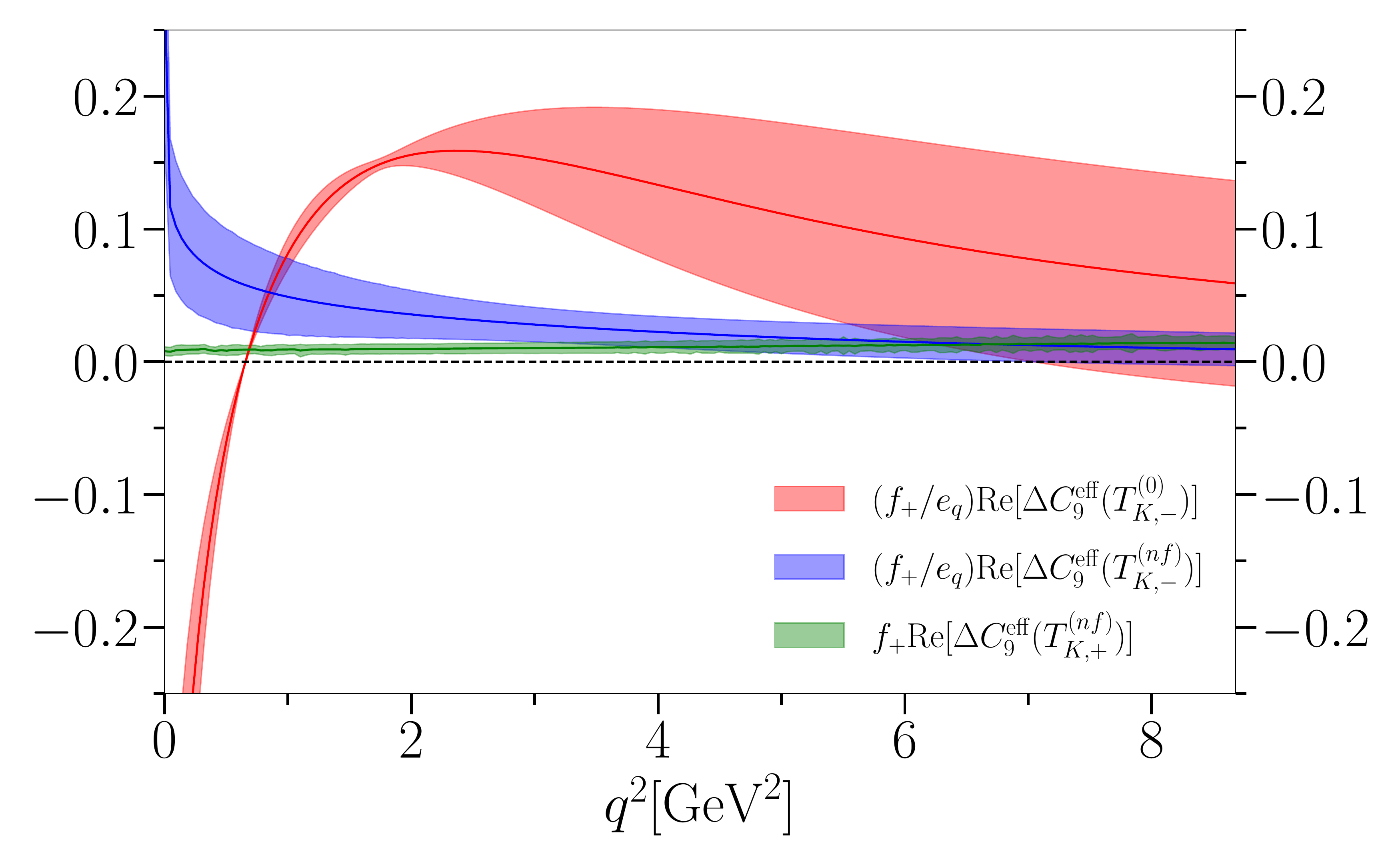}
\caption{The real parts of the contributions to $f_+ \Delta C_9^{\mathrm{eff}}$ from each of the three non-zero terms, $T_{K,-}^{(0)}$ and $T_{K,\pm}^{(\mathrm{nf})}$ (Equation~\eqref{eq:nonfac}). 
A decay channel-specific factor of $e_q \in \{2/3,-1/3\}$ is removed from $T_{K,-}^{(0)}$ and $T_{K,-}^{(\mathrm{nf})}$.}
\label{fig:ReC9}
\end{figure}
\begin{figure}
\includegraphics[width=0.48\textwidth]{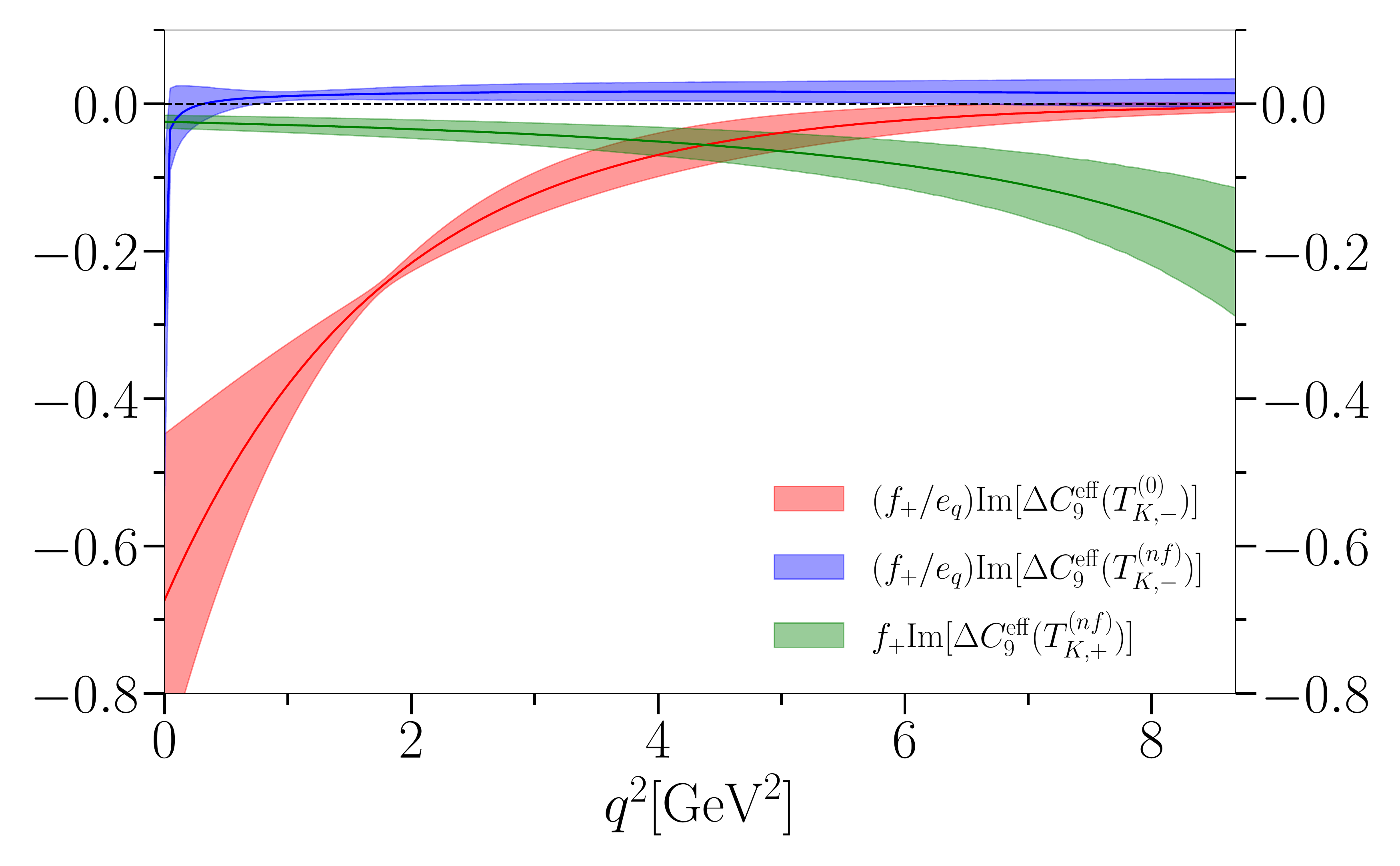}
\caption{The imaginary parts of the contributions to $f_+ \Delta C_9^{\mathrm{eff}}$ from each of the three non-zero terms, $T_{K,-}^{(0)}$ and $T_{K,\pm}^{(\mathrm{nf})}$ (Equation~\eqref{eq:nonfac}). 
A decay channel-specific factor of $e_q \in \{2/3,-1/3\}$ is removed from $T_{K,-}^{(0)}$ and $T_{K,-}^{(\mathrm{nf})}$.}
\label{fig:ImC9}
\end{figure}
 \begin{figure}
\includegraphics[width=0.48\textwidth]{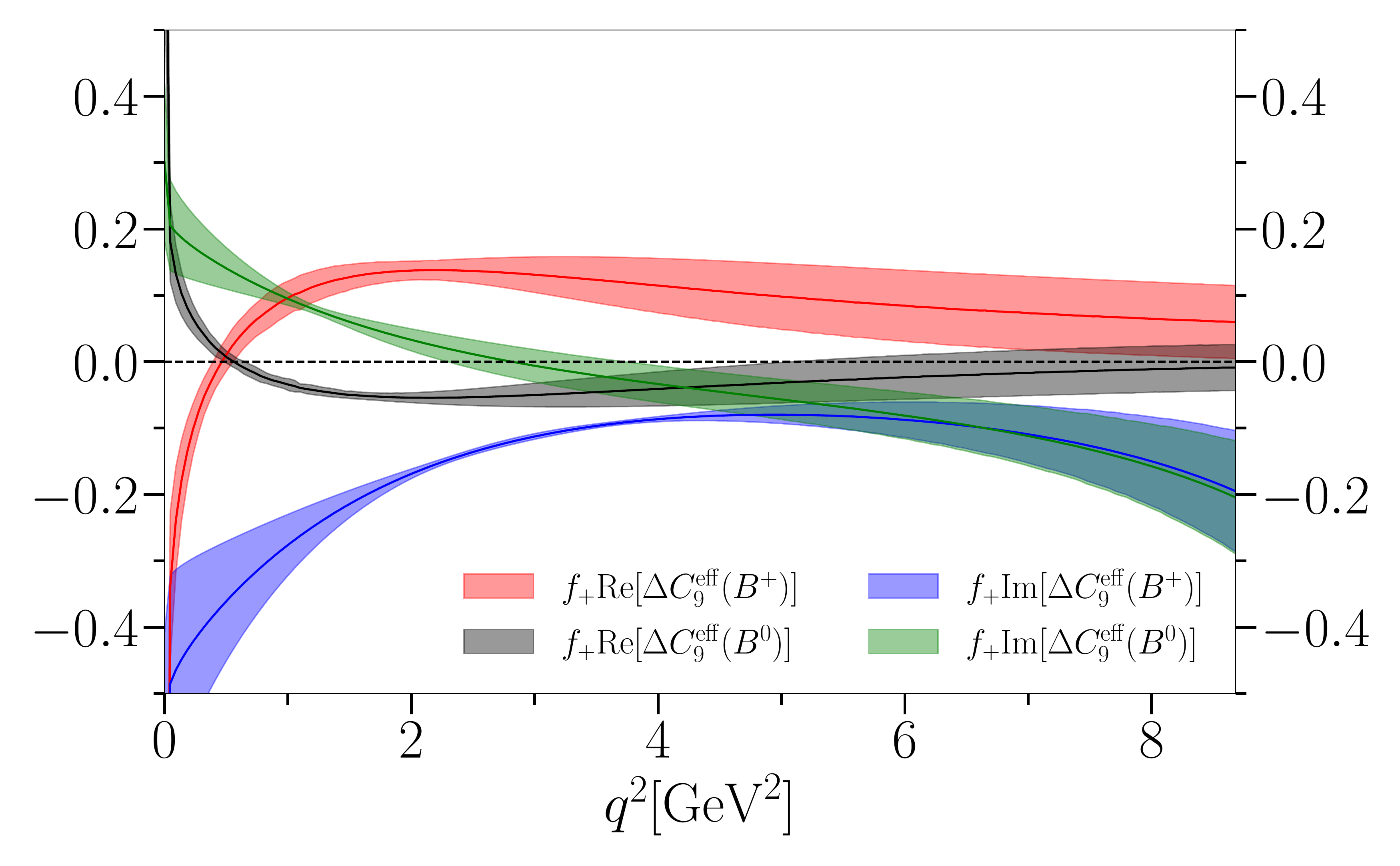}
\caption{The real and imaginary parts of $f_+\Delta C_9^{\mathrm{eff}}$ for both the $B^0$ and $B^+$ cases.}
\label{fig:DelC9}
\end{figure}

The contribution from $\Delta C_9^{\mathrm{eff}}$ to observables is via the term $f_+\Delta C_9^{\mathrm{eff},1}$ in Equation~\eqref{eq:Fs}.
This has the effect of cancelling the dependence of $f_+ \Delta C_9^{\mathrm{eff}}$ on the form factor $f_+$.
The real and imaginary parts of the three nonzero contributions to $f_+\Delta C_9^{\mathrm{eff}}$ in Equation~(\ref{eq:nonfac}), corresponding to $T_{K,-}^{(0)}$ and $T_{K,\pm}^{(0/\mathrm{nf})}$, are plotted in Figures~\ref{fig:ReC9} and~\ref{fig:ImC9}.
In these plots we remove a decay channel-specific factor of the light quark charge, $e_q$, which is $2/3$ for $B^+\to K^+$ and $-1/3$ for $B^0\to K^0$. 
Among these terms, the $T_{K,-}^{(0)}$ contribution is dominant, especially for $q^2 \lesssim 1$\,GeV$^2$.

The combined effect of these terms to $f_+\Delta C_9^{\mathrm{eff}}$ is shown in Figure~\ref{fig:DelC9}, where the real and imaginary parts are plotted separately for both the $B^0$ and $B^+$ cases.
Both the real and imaginary parts are smooth functions of $q^2$ in the region below the $J/\Psi$ resonance where they are considered ($4m_\ell^2 \leq q^2 \leq 8.68$\, GeV$^2$), and are small for $q^2 > 1\,\mathrm{GeV}^2$. 
For $q^2 <  1\,\mathrm{GeV}^2$, where the nonfactorisable corrections are largest, they also have little impact. This is because the differential decay rate falls rapidly as $q^2$ approaches $m_l^2$ for kinematic reasons (see the factor of $\beta = \sqrt{1 - 4m_\ell^2/q^2}$ in Eq.~\eqref{eq:Cbeta}). 
As a result, the corrections do not make a significant contribution to the differential branching fraction. 
\begin{figure}
\includegraphics[width=0.48\textwidth]{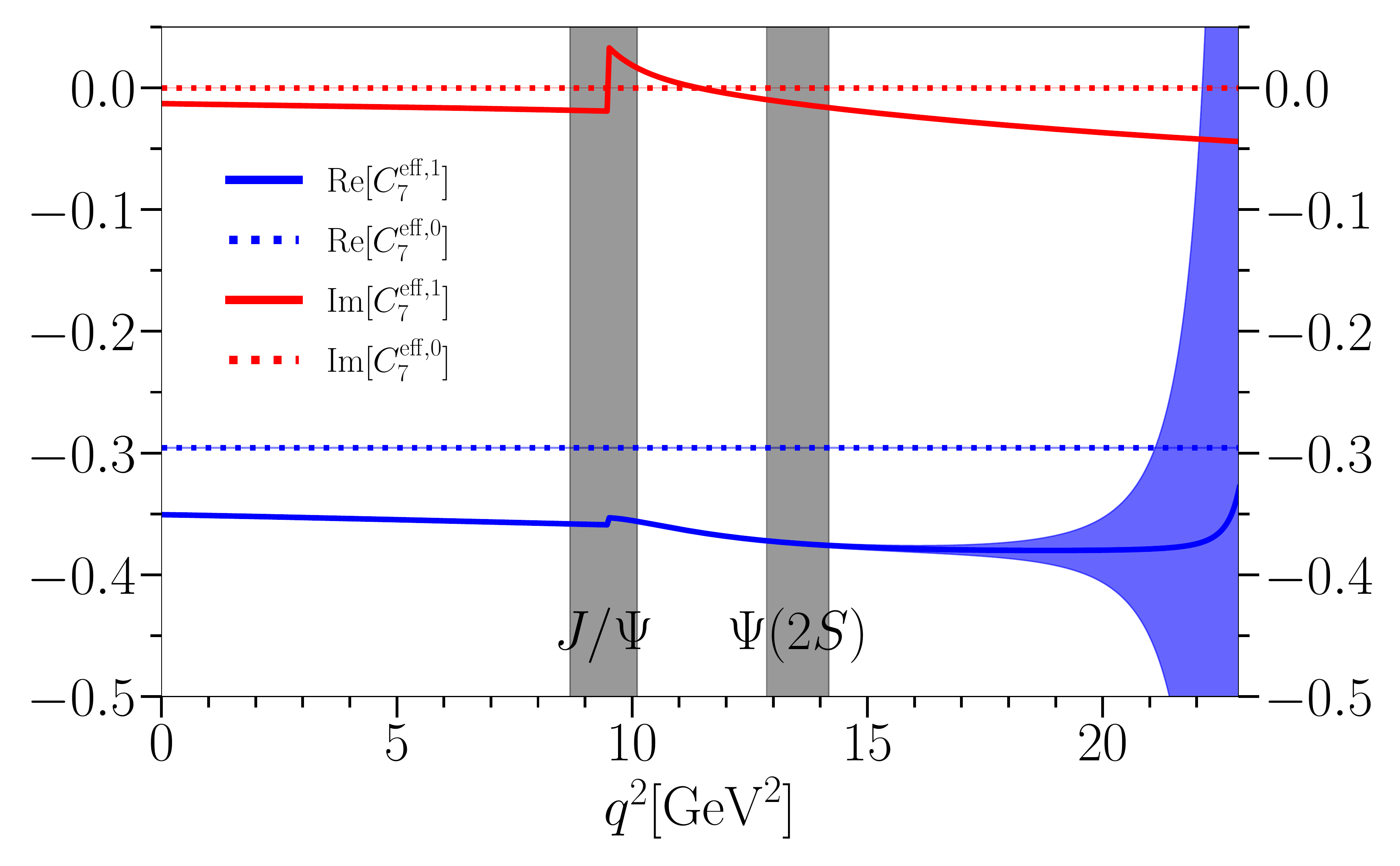}
\caption{The real and imaginary parts of $C_7^{\mathrm{eff},0}$ and $C_7^{\mathrm{eff},1}$ (see Equation~\eqref{eq:C7eff}), showing the combined effect of the nonfactorisable and $\mathcal O(\alpha_s)$ corrections. Note that this plot is independent of the meson charge.
Corrected values are shown with solid lines and dark fill colour, while uncorrected values are shown with dotted lines and light fill colour.}
\label{fig:C7eff}
\end{figure}
 \begin{figure}
   \includegraphics[width=0.48\textwidth]{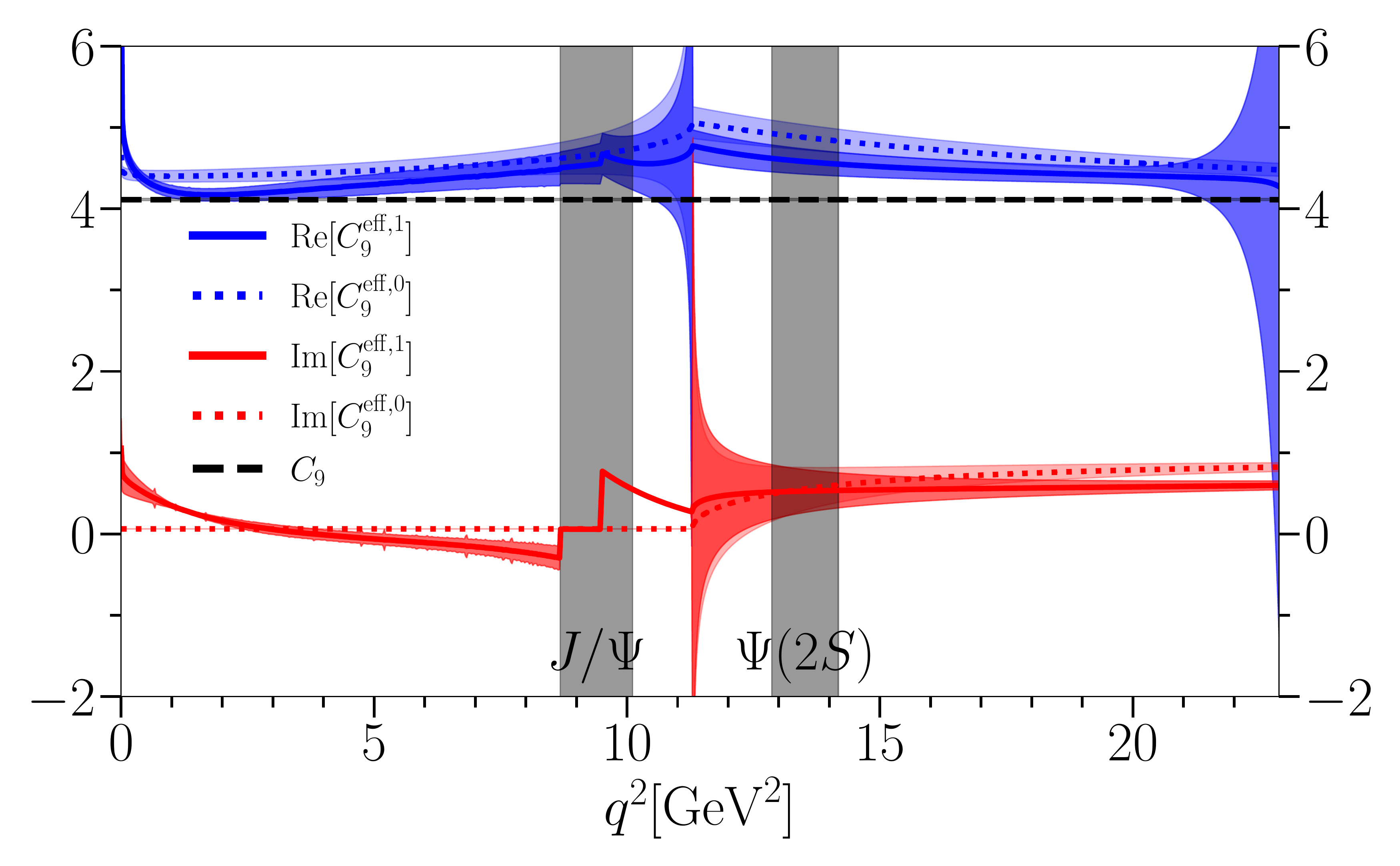}
   \includegraphics[width=0.48\textwidth]{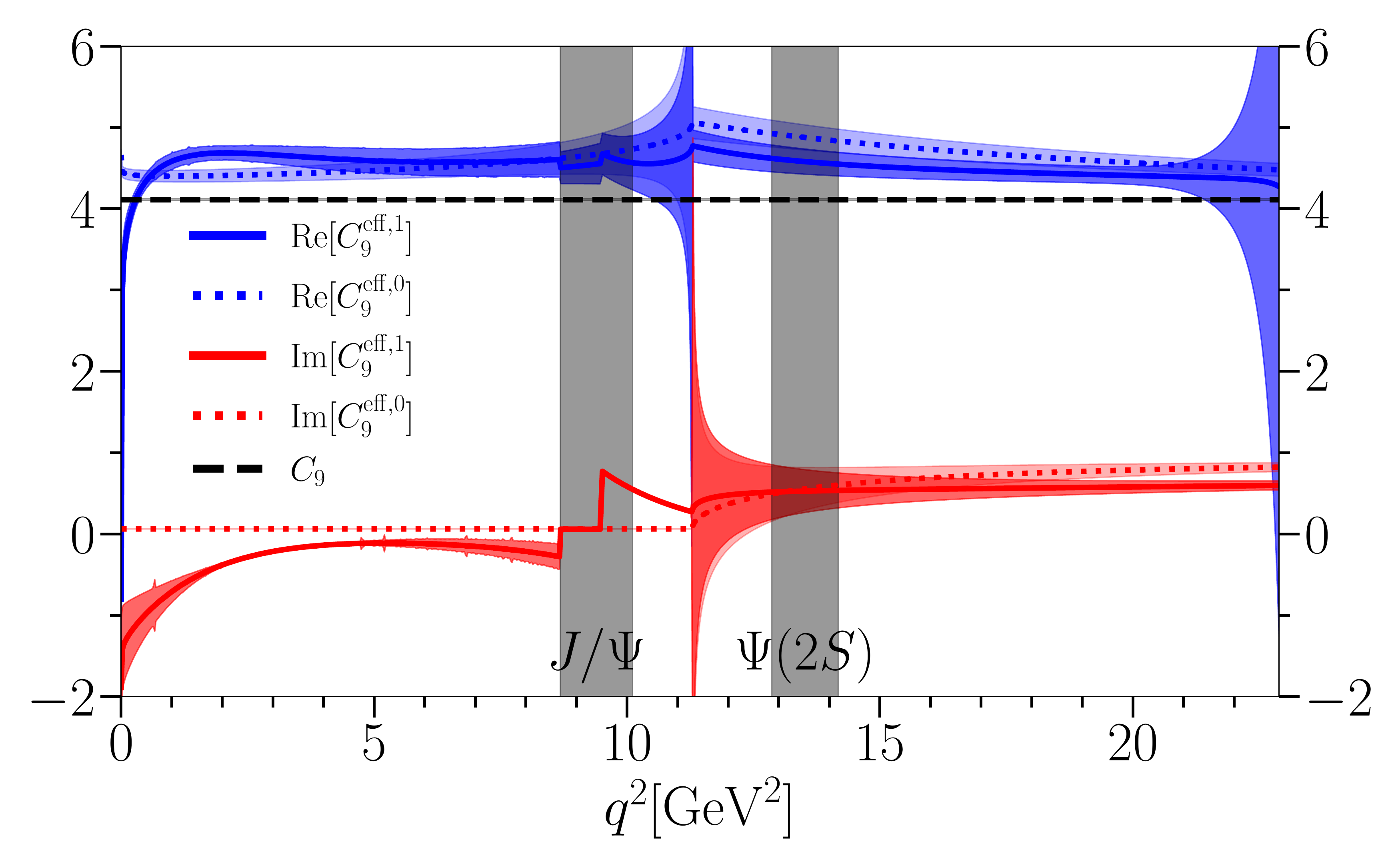}
\caption{The real and imaginary parts of $C_9^{\mathrm{eff},0}$ and $C_9^{\mathrm{eff},1}$ (see Equation~\eqref{eq:C9eff}) for $B^0\to K^0$ (top) and $B^+\to K^+$ (bottom), showing the combined effect of the nonfactorisable and $\mathcal O(\alpha_s)$ corrections.
Corrected values are shown with solid lines and dark fill colour, while uncorrected values are shown with dotted lines and light fill colour. $C_9=C_9^{\mathrm{eff},0}-Y(q^2)$ (Table~\ref{tab-SMparams}) is included for comparison.}
\label{fig:C9eff}
\end{figure}

The overall modest contributions of the above corrections are shown for $C_7^{\mathrm{eff}}$ in Figure~\ref{fig:C7eff} and for $C_9^{\mathrm{eff}}$ in Figure~\ref{fig:C9eff}.
In each plot, both corrected and uncorrected values are plotted. In the case of $C_9^{\mathrm{eff}}$, the difference between the $B^0\to K^0$ and $B^+\to K^+$ corrections, arising from the charge dependence of $\Delta C_9^{\mathrm{eff}}$, is responsible for the slightly different shapes of the differential branching fractions (see e.g. Figure~\ref{fig:dBdqemup} and~\ref{fig:dBdqemu0}).  
Cusps in $C_7^{\mathrm{eff}}$ and $C_9^{\mathrm{eff}}$ either occur within the vetoed $J/\Psi$ resonance region near $q^2\approx 10$\,GeV$^2$ or at $q^2=4m_c^2\approx 11.3$\,GeV$^2$, between the $J/\Psi$ and $\Psi(2S)$ resonances.
In this region between the resonances, we linearly interpolate the differential decay rate and are therefore unaffected by the cusp at $q^2=4m_c^2$.
The most significant effect of the corrections is an approximately 20\% shift to Re$[C_7^{\mathrm{eff},1}]$ arising from Re$[\delta C_7^{\mathrm{eff}}]$. In fact the total effect of all corrections on our results in the well-behaved regions is small (Section~\ref{sec:BtoKll_expt}), so we do not include additional uncertainty on the corrections.  
The growth of corrections and in their uncertainties at high and low $q^2$ is suppressed by kinematic factors in the decay rate, resulting in modest impact in the well-behaved regions of $q^2$ where we give our main results (Table~\ref{tab:comparisons}), as discussed.

\section{Numerical Results}\label{sec:numres}
The Tables in this appendix list our numerical results in a number of $q^2$ bins, aimed at matching those appearing in previous work and providing good coverage for future work to compare with. 
The charged and neutral meson case, as well as the average (no charge label) is given. 
Similarly the $e$, $\mu$, average of $e$ and $\mu$ (labelled as $\ell$) and $\tau$ cases are all given. 
In all cases, the average is taken (including correlations) after all other calculation and integration has taken place. 
In most cases the meson charge and/or light lepton flavour makes no significant difference. We include all cases for completeness, and ease of comparison with future results. 

\subsection{Integration}\label{sec:integration}
When performing integrals to generate the data in the following tables, as well as to generate the figures above, we use the trapezoid rule. 
That is to say,
\begin{equation}
  \begin{split}
    \int_{q^2_{\mathrm{low}}}^{q^2_{\mathrm{upp}}}&f(q^2)dq^2 \approx\\
    &\frac{\Delta q^2}{2}\Big(f(q^2_{\mathrm{low}})+2f(q^2_1)+...+2f(q^2_{N-1})+f(q^2_{\mathrm{upp}})\Big)
  \end{split}
\end{equation}
where $f$ is evaluated at $N+1$ evenly spaced $q^2$ values in total. 
In order to ensure accurate results, we start with $N=16$, and double $N$ repeatedly until the result for $2N$ differs from that for $N$ by less than $0.02\,\sigma$ (as uncertainties are given to two significant figures, this is the point at which on average the last digit of the mean has only changed by one). 
Using this method allows us to confirm that our integrals have converged, and most do so for $N\leq1024$. 
An exception is $F^e_H$. 
In this case, the integrand in the numerator (Eq.~\eqref{eq:F_H}) changes very rapidly as we approach the lower limit $4m_e^2 \approx 10^{-6}\,\mathrm{GeV}^2$ and a naive approach will not allow the integrals to converge without very large values of $N$. 
This is not the case for $F^{\mu}_H$, as the integrals do not span so many orders of magnitude for $4m_{\mu}^2\approx 0.05\,\mathrm{GeV}^2$. 
To solve this problem, we split the integral limits up into different orders of magnitude, integrating from $q^2_{\mathrm{low}}=4m_e^2$ to $10 q^2_{\mathrm{low}}$, then from $10q^2_{\mathrm{low}}$ to $100 q^2_{\mathrm{low}}$ and so on, up to $q^2_{\mathrm{upp}}$. 
In each case we increase $N$ until the result is stable, as above, and then we sum the (correlated) sub-integrals to get the final value for both the numerator and denominator. 
This method concentrates the number of evaluations of $f$ in regions where they are most needed, and so reduces the total number of evaluations required for the whole integral, making it tractable.

To assess the suitability of this method, and check that we are not introducing bias with our algorithm, we checked the integration against the vegas python package~\cite{peter_lepage_2021_4994303}, and confirmed that the results agree.
\begin{table*}
  \begin{center}
    \begin{tabular}{ c | c c c c c c }
      \hline
     $q^2$ bin & $(0.001,4)$ & $(4,8)$ & $(8,12)$ & $(12,16)$ & $(16,20)$ & $(20,q^2_{\text{max}})$ \\ [0.5ex]
      \hline
      \hline
     $10^6\mathcal{B}(B^+\to K^+\nu\bar{\nu}_{~\text{SD}})$ & 1.173(95) & 1.139(89) & 1.056(83) & 0.892(71) & 0.589(47) & 0.125(11) \\ [1.0ex]
     $10^6\mathcal{B}(B^0\to K^0\nu\bar{\nu}_{~\text{SD}})$ & 1.087(88) & 1.056(82) & 0.979(77) & 0.826(65) & 0.543(43) & 0.1133(96) \\ [1.0ex]
      \hline
    \end{tabular}
    \caption{Short distance contributions to branching fractions $\mathcal{B}^{(0/+)}=\mathcal{B}(B^{0/+}\to K^{0/+}\nu\bar{\nu})_{\text{SD}}$ (Equation~\eqref{eq:nunubarBF}) for the rare decay $B\to K\nu\bar{\nu}$ integrated over various $q^2$ bins. Numerical values for $q^2$ bins are in units of GeV$^2$.}
    \label{tab:BKnunubins}
  \end{center}
\end{table*}
\begin{table*} 
  \begin{center}
    \begin{tabular}{ c | c c c c c c c c }
      \hline
     $q^2$ bin & $(4m_{\ell}^2,q^2_{\text{max}})$ & $(0.05,2)$ & $(1,6)$ & $(2,4.3)$ & $(4.3,8.68)$ & $(14.18,16)$ & $(16,18)$ & $(18,22)$ \\ [0.5ex]
      \hline
      \hline
     $10^7\mathcal{B}(B^+\to K^+e^+e^-)$ & 6.50(50) [5.59(43)] & 0.752(65) & 1.94(16) & 0.902(76) & 1.63(15) & 0.491(44) & 0.441(39) & 0.458(41) \\ [1.0ex]
     $10^7\mathcal{B}(B^0\to K^0e^+e^-)$ & 5.79(45) [4.97(39)] & 0.655(57) & 1.64(14) & 0.751(64) & 1.44(13) & 0.454(41) & 0.407(36) & 0.419(38) \\ [1.0ex]
     $10^7\mathcal{B}(B\to Ke^+e^-)$ & 6.14(48) [5.28(41)] & 0.703(61) & 1.79(15) & 0.826(69) & 1.54(13) & 0.473(42) & 0.424(38) & 0.438(40) \\ [1.0ex]
      \hline
     $10^7\mathcal{B}(B^+\to K^+\mu^+\mu^-)$ & 6.48(50) [5.59(43)] & 0.746(64) & 1.95(16) & 0.902(75) & 1.63(15) & 0.492(44) & 0.442(39) & 0.459(41) \\ [1.0ex]
     $10^7\mathcal{B}(B^0\to K^0\mu^+\mu^-)$ & 5.77(45) [4.95(38)] & 0.648(57) & 1.64(14) & 0.751(64) & 1.44(13) & 0.455(41) & 0.408(36) & 0.421(38) \\ [1.0ex]
     $10^7\mathcal{B}(B\to K\mu^+\mu^-)$ & 6.13(47) [5.27(40)] & 0.697(60) & 1.79(15) & 0.827(69) & 1.54(13) & 0.473(42) & 0.425(38) & 0.440(40) \\ [1.0ex]
      \hline
     $10^7\mathcal{B}(B^+\to K^+\ell^+\ell^-)$ & 6.49(50) [5.59(43)] & 0.749(65) & 1.94(16) & 0.902(76) & 1.63(15) & 0.492(44) & 0.442(39) & 0.458(41) \\ [1.0ex]
     $10^7\mathcal{B}(B^0\to K^0\ell^+\ell^-)$ & 5.78(45) [4.96(39)] & 0.652(57) & 1.64(14) & 0.751(64) & 1.44(13) & 0.455(41) & 0.408(36) & 0.420(38) \\ [1.0ex]
     $10^7\mathcal{B}(B\to K\ell^+\ell^-)$ & 6.14(47) [5.27(40)] & 0.700(60) & 1.79(15) & 0.827(69) & 1.54(13) & 0.473(42) & 0.425(38) & 0.439(40) \\ [1.0ex]
      \hline
    \end{tabular}
  \end{center}
  \begin{center}
    \begin{tabular}{ c | c c c c c c c c c }
      \hline
     $q^2$ bin & $(0.1,2)$ & $(2,4)$ & $(4,6)$ & $(6,8)$ & $(15,17)$ & $(17,19)$ & $(19,22)$ & $(1.1,6)$ & $(15,22)$ \\ [0.5ex]
      \hline
      \hline
     $10^7\mathcal{B}(B^+\to K^+e^+e^-)$ & 0.736(64) & 0.786(66) & 0.761(66) & 0.740(66) & 0.495(44) & 0.380(34) & 0.284(26) & 1.91(16) & 1.16(10) \\ [1.0ex]
     $10^7\mathcal{B}(B^0\to K^0e^+e^-)$ & 0.636(55) & 0.653(56) & 0.658(57) & 0.658(58) & 0.458(41) & 0.350(31) & 0.259(24) & 1.60(14) & 1.067(95) \\ [1.0ex]
     $10^7\mathcal{B}(B\to Ke^+e^-)$ & 0.686(59) & 0.719(60) & 0.709(59) & 0.699(61) & 0.477(42) & 0.365(33) & 0.271(25) & 1.75(15) & 1.113(99) \\ [1.0ex]
      \hline
     $10^7\mathcal{B}(B^+\to K^+\mu^+\mu^-)$ & 0.733(63) & 0.786(66) & 0.762(66) & 0.741(66) & 0.496(44) & 0.381(34) & 0.285(26) & 1.91(16) & 1.16(10) \\ [1.0ex]
     $10^7\mathcal{B}(B^0\to K^0\mu^+\mu^-)$ & 0.633(55) & 0.653(56) & 0.658(57) & 0.658(58) & 0.459(41) & 0.351(31) & 0.260(24) & 1.60(14) & 1.070(95) \\ [1.0ex]
     $10^7\mathcal{B}(B\to K\mu^+\mu^-)$ & 0.683(59) & 0.720(60) & 0.710(59) & 0.700(61) & 0.477(42) & 0.366(33) & 0.273(25) & 1.76(15) & 1.116(99) \\ [1.0ex]
      \hline
     $10^7\mathcal{B}(B^+\to K^+\ell^+\ell^-)$ & 0.734(63) & 0.786(66) & 0.761(66) & 0.741(66) & 0.496(44) & 0.380(34) & 0.285(26) & 1.91(16) & 1.16(10) \\ [1.0ex]
     $10^7\mathcal{B}(B^0\to K^0\ell^+\ell^-)$ & 0.635(55) & 0.653(56) & 0.658(57) & 0.658(58) & 0.458(41) & 0.351(31) & 0.260(24) & 1.60(14) & 1.068(95) \\ [1.0ex]
     $10^7\mathcal{B}(B\to K\ell^+\ell^-)$ & 0.685(59) & 0.719(60) & 0.710(59) & 0.699(61) & 0.477(42) & 0.365(33) & 0.272(25) & 1.76(15) & 1.114(99) \\ [1.0ex]
      \hline
    \end{tabular}
  \end{center}
  \begin{center}
    \begin{tabular}{ c | c c c c c c }
      \hline
     $q^2$ bin & $(0.1,4)$ & $(4,8.12)$ & $(10.2,12.8)$ & $(14.18,q^2_{\text{max}})$ & $(10.09,12.86)$ & $(16,q^2_{\text{max}})$ \\ [0.5ex]
      \hline
      \hline
     $10^7\mathcal{B}(B^+\to K^+e^+e^-)$ & 1.52(13) & 1.55(14) & 0.847(70) & 1.40(12) & 0.903(75) & 0.910(81) \\ [1.0ex]
     $10^7\mathcal{B}(B^0\to K^0e^+e^-)$ & 1.29(11) & 1.35(12) & 0.772(64) & 1.29(11) & 0.823(68) & 0.836(75) \\ [1.0ex]
     $10^7\mathcal{B}(B\to Ke^+e^-)$ & 1.41(12) & 1.45(12) & 0.809(67) & 1.35(12) & 0.863(71) & 0.873(78) \\ [1.0ex]
      \hline
     $10^7\mathcal{B}(B^+\to K^+\mu^+\mu^-)$ & 1.52(13) & 1.55(14) & 0.847(70) & 1.40(12) & 0.904(75) & 0.913(81) \\ [1.0ex]
     $10^7\mathcal{B}(B^0\to K^0\mu^+\mu^-)$ & 1.29(11) & 1.36(12) & 0.773(64) & 1.29(11) & 0.824(68) & 0.838(75) \\ [1.0ex]
     $10^7\mathcal{B}(B\to K\mu^+\mu^-)$ & 1.40(12) & 1.45(12) & 0.810(67) & 1.35(12) & 0.864(71) & 0.876(78) \\ [1.0ex]
      \hline
     $10^7\mathcal{B}(B^+\to K^+\ell^+\ell^-)$ & 1.52(13) & 1.55(14) & 0.847(70) & 1.40(12) & 0.903(75) & 0.911(81) \\ [1.0ex]
     $10^7\mathcal{B}(B^0\to K^0\ell^+\ell^-)$ & 1.29(11) & 1.36(12) & 0.773(64) & 1.29(11) & 0.824(68) & 0.837(75) \\ [1.0ex]
     $10^7\mathcal{B}(B\to K\ell^+\ell^-)$ & 1.40(12) & 1.45(12) & 0.810(67) & 1.35(12) & 0.864(71) & 0.874(78) \\ [1.0ex]
      \hline
    \end{tabular}
    \caption{Branching fractions integrated over some commonly used $q^2$ binning schemes for the electron, muon and $\ell$, which is the average of the two. 
    In the first bin of the top panel, giving the branching fraction integrated over the full $q^2$ range, the first number is the whole integral, whilst the one which follows in square brackets is the result when the ranges 8.68-10.11\,$\text{GeV}^2$ and 12.86-14.18\,$\text{GeV}^2$ are excluded.
    Numerical values for $q^2$ bins are in units of GeV$^2$. 
    QED corrections discussed in Section~\ref{sec:QED} are not included here.}
     \label{tab:diffrateBKll}
  \end{center}
\end{table*}
\begin{table*}
  \begin{center}
    \begin{tabular}{ c | c c c c c c c c }
      \hline
     $q^2$ bin & $(4m_{\mu}^2,q^2_{\text{max}})$ & $(0.05,2)$ & $(1,6)$ & $(2,4.3)$ & $(4.3,8.68)$ & $(14.18,16)$ & $(16,18)$ & $(18,22)$ \\ [0.5ex]
      \hline
      \hline
     $10^3(R^{\mu(+)}_e-1)$ & -0.08(28) & -7.84(95) & 0.39(30) & 0.41(30) & 0.62(25) & 1.43(23) & 1.91(25) & 3.81(37) \\ [1.0ex]
     $10^3(R^{\mu(0)}_e-1)$ & -0.81(29) & -10.30(96) & 0.44(34) & 0.47(33) & 0.65(27) & 1.43(23) & 1.92(25) & 3.85(37) \\ [1.0ex]
     $10^3(R^{\mu}_e-1)$ & -0.44(26) & -9.03(90) & 0.41(32) & 0.44(31) & 0.63(26) & 1.43(23) & 1.92(25) & 3.83(37) \\ [1.0ex]
      \hline
    \end{tabular}
  \end{center}
  \begin{center}
    \begin{tabular}{ c | c c c c c c c c c }
      \hline
     $q^2$ bin & $(0.1,2)$ & $(2,4)$ & $(4,6)$ & $(6,8)$ & $(15,17)$ & $(17,19)$ & $(19,22)$ & $(1.1,6)$ & $(15,22)$ \\ [0.5ex]
      \hline
      \hline
     $10^3(R^{\mu(+)}_e-1)$ & -3.29(76) & 0.40(30) & 0.54(27) & 0.65(25) & 1.63(24) & 2.31(27) & 4.57(42) & 0.40(30) & 2.57(29) \\ [1.0ex]
     $10^3(R^{\mu(0)}_e-1)$ & -4.25(82) & 0.46(34) & 0.58(29) & 0.68(26) & 1.64(24) & 2.32(27) & 4.63(43) & 0.45(33) & 2.59(29) \\ [1.0ex]
     $10^3(R^{\mu}_e-1)$ & -3.75(79) & 0.43(32) & 0.56(28) & 0.66(25) & 1.63(24) & 2.32(27) & 4.60(42) & 0.43(32) & 2.58(29) \\ [1.0ex]
      \hline
    \end{tabular}
  \end{center}
  \begin{center}
    \begin{tabular}{ c | c c c c c c }
      \hline
     $q^2$ bin & $(0.1,4)$ & $(4,8.12)$ & $(10.2,12.8)$ & $(14.18,q^2_{\text{max}})$ & $(10.09,12.86)$ & $(16,q^2_{\text{max}})$ \\ [0.5ex]
      \hline
      \hline
     $10^3(R^{\mu(+)}_e-1)$ & -1.35(50) & 0.60(26) & 0.98(23) & 2.50(29) & 0.97(23) & 3.08(32) \\ [1.0ex]
     $10^3(R^{\mu(0)}_e-1)$ & -2.30(54) & 0.63(27) & 0.99(23) & 2.51(29) & 0.99(23) & 3.10(32) \\ [1.0ex]
     $10^3(R^{\mu}_e-1)$ & -1.80(52) & 0.61(26) & 0.98(23) & 2.50(29) & 0.98(23) & 3.09(32) \\ [1.0ex]
      \hline
    \end{tabular}
    \caption{The ratio $R^{\mu}_e$ (Equation~\ref{eq:R}) integrated over some commonly used $q^2$ binning schemes. We give results for the charged meson case, the neutral meson case and the charge-averaged case (defined as the ratio of the two charge-averaged integrals). 
    Numerical values for $q^2$ bins are in units of GeV$^2$. 
    The 1\% uncertainty from QED effects~\cite{Bordone:2016gaq} (see Sec.~\ref{sec:Rmue})  is not included in these numbers.}
    \label{tab:R-mue-vals}
  \end{center}
\end{table*}
\begin{table*}
  \begin{center}
    \begin{tabular}{ c | c c c c c c c c }
      \hline
     $q^2$ bin & $(4m_{\ell}^2,q^2_{\text{max}})$ & $(0.05,2)$ & $(1,6)$ & $(2,4.3)$ & $(4.3,8.68)$ & $(14.18,16)$ & $(16,18)$ & $(18,22)$ \\ [0.5ex]
      \hline
      \hline
     $10^8F_H^{e(+)}$ & 119(18) & 279.9(3.3) & 57.8(1.0) & 53.44(75) & 26.71(63) & 13.76(55) & 13.74(59) & 16.97(86) \\ [1.0ex]
     $10^8F_H^{e(0)}$ & 226(32) & 312.4(5.0) & 57.4(1.1) & 53.42(83) & 26.71(66) & 13.78(55) & 13.77(59) & 17.06(87) \\ [1.0ex]
      \hline
     $10^3F_H^{\mu(+)}$ & 21.61(89) & 105.0(1.1) & 24.46(43) & 22.66(31) & 11.37(26) & 5.87(23) & 5.85(25) & 7.22(36) \\ [1.0ex]
     $10^3F_H^{\mu(0)}$ & 22.35(92) & 115.1(1.5) & 24.29(44) & 22.65(34) & 11.37(28) & 5.87(23) & 5.87(25) & 7.26(37) \\ [1.0ex]
      \hline
    \end{tabular}
  \end{center}
  \begin{center}
    \begin{tabular}{ c | c c c c c c c c c }
      \hline
     $q^2$ bin & $(0.1,2)$ & $(2,4)$ & $(4,6)$ & $(6,8)$ & $(15,17)$ & $(17,19)$ & $(19,22)$ & $(1.1,6)$ & $(15,22)$ \\ [0.5ex]
      \hline
      \hline
     $10^8F_H^{e(+)}$ & 238.4(2.5) & 55.56(75) & 33.14(64) & 24.10(59) & 13.65(56) & 14.15(64) & 18.48(99) & 55.91(99) & 15.00(68) \\ [1.0ex]
     $10^8F_H^{e(0)}$ & 256.3(3.1) & 55.57(83) & 33.20(68) & 24.16(62) & 13.67(56) & 14.19(64) & 18.6(1.0) & 55.5(1.0) & 15.04(68) \\ [1.0ex]
      \hline
     $10^3F_H^{\mu(+)}$ & 94.88(90) & 23.55(31) & 14.10(27) & 10.26(25) & 5.82(24) & 6.03(27) & 7.86(42) & 23.66(41) & 6.39(29) \\ [1.0ex]
     $10^3F_H^{\mu(0)}$ & 101.4(1.1) & 23.56(34) & 14.12(29) & 10.29(26) & 5.82(24) & 6.04(27) & 7.92(42) & 23.49(42) & 6.41(29) \\ [1.0ex]
      \hline
    \end{tabular}
  \end{center}
  \begin{center}
    \begin{tabular}{ c | c c c c c c }
      \hline
     $q^2$ bin & $(0.1,4)$ & $(4,8.12)$ & $(10.2,12.8)$ & $(14.18,q^2_{\text{max}})$ & $(10.09,12.86)$ & $(16,q^2_{\text{max}})$ \\ [0.5ex]
      \hline
      \hline
     $10^8F_H^{e(+)}$ & 143.9(2.4) & 28.47(64) & 17.15(56) & 15.13(67) & 17.18(56) & 15.87(75) \\ [1.0ex]
     $10^8F_H^{e(0)}$ & 154.7(2.3) & 28.46(67) & 17.15(57) & 15.16(67) & 17.17(57) & 15.91(76) \\ [1.0ex]
      \hline
     $10^3F_H^{\mu(+)}$ & 57.97(94) & 12.12(27) & 7.31(24) & 6.44(28) & 7.32(24) & 6.75(32) \\ [1.0ex]
     $10^3F_H^{\mu(0)}$ & 61.89(89) & 12.11(28) & 7.31(24) & 6.45(29) & 7.32(24) & 6.77(32) \\ [1.0ex]
      \hline
    \end{tabular}
    \caption{ $F_H$ (Eq.~\eqref{eq:F_H}), which is double the `flat term', integrated over some commonly used $q^2$ binning schemes for the $e$ and $\mu$ final-state cases for charged and neutral meson decays. Numerical values for $q^2$ bins are in units of GeV$^2$. No uncertainty to allow for QED is included in these numbers.}
    \label{tab:FHl}
  \end{center}
\end{table*}
\begin{table*}
  \begin{center}
    \begin{tabular}{ c | c c c c c c c c c }
      \hline
     $q^2$ bin & $(4m_{\tau}^2,q^2_{\text{max}})$ & $(14.18,q^2_{\text{max}})$ & $(14.18,16)$ & $(16,18)$ & $(18,22)$ & $(15,17)$ & $(17,19)$ & $(19,22)$ & $(15,22)$ \\ [0.5ex]
      \hline
      \hline
     $10^7\mathcal{B}(B^+\to K^+\tau^+\tau^-)$ & 1.68(12) & 1.49(10) & 0.358(26) & 0.426(30) & 0.656(46) & 0.419(30) & 0.414(29) & 0.454(32) & 1.288(90) \\ [1.0ex]
     $10^7\mathcal{B}(B^0\to K^0\tau^+\tau^-)$ & 1.55(11) & 1.375(95) & 0.332(24) & 0.394(28) & 0.604(42) & 0.388(28) & 0.383(27) & 0.417(29) & 1.188(83) \\ [1.0ex]
     $10^7\mathcal{B}(B\to K\tau^+\tau^-)$ & 1.62(11) & 1.43(10) & 0.345(25) & 0.410(29) & 0.630(44) & 0.403(29) & 0.399(28) & 0.436(30) & 1.238(86) \\ [1.0ex]
      \hline
    \end{tabular}
    \caption{Branching fractions for decays to $\tau$ leptons, integrated over some commonly used $q^2$ bins. Numerical values for $q^2$ bins are in units of GeV$^2$. No uncertainty to allow for QED is included in these numbers.}
      \label{tab:BKtau}
  \end{center}
\end{table*}
\begin{table*}
  \begin{center}
    \begin{tabular}{ c | c c c c c c c c c }
      \hline
     $q^2$ bin & $(4m_{\tau}^2,q^2_{\text{max}})$ & $(14.18,q^2_{\text{max}})$ & $(14.18,16)$ & $(16,18)$ & $(18,22)$ & $(15,17)$ & $(17,19)$ & $(19,22)$ & $(15,22)$ \\ [0.5ex]
      \hline
      \hline
     $R^{\tau(+)}_e$ & 0.902(34) & 1.065(42) & 0.729(26) & 0.966(36) & 1.434(62) & 0.846(31) & 1.091(42) & 1.600(74) & 1.111(44) \\ [1.0ex]
     $R^{\tau(0)}_e$ & 0.902(35) & 1.066(42) & 0.730(26) & 0.967(36) & 1.441(63) & 0.847(31) & 1.094(42) & 1.610(74) & 1.113(44) \\ [1.0ex]
     $R^{\tau}_e$ & 0.902(35) & 1.065(42) & 0.729(26) & 0.967(36) & 1.438(63) & 0.847(31) & 1.092(42) & 1.605(74) & 1.112(44) \\ [1.0ex]
      \hline
     $R^{\tau(+)}_{\mu}$ & 0.900(34) & 1.063(42) & 0.728(26) & 0.964(36) & 1.429(62) & 0.845(31) & 1.089(42) & 1.592(73) & 1.108(44) \\ [1.0ex]
     $R^{\tau(0)}_{\mu}$ & 0.900(34) & 1.063(42) & 0.729(26) & 0.966(36) & 1.436(62) & 0.846(31) & 1.091(42) & 1.603(73) & 1.110(44) \\ [1.0ex]
     $R^{\tau}_{\mu}$ & 0.900(34) & 1.063(42) & 0.728(26) & 0.965(36) & 1.432(62) & 0.845(31) & 1.090(42) & 1.597(73) & 1.109(44) \\ [1.0ex]
      \hline
     $R^{\tau(+)}_{\ell}$ & 0.901(34) & 1.064(42) & 0.729(26) & 0.965(36) & 1.432(62) & 0.845(31) & 1.090(42) & 1.596(73) & 1.110(44) \\ [1.0ex]
     $R^{\tau(0)}_{\ell}$ & 0.901(34) & 1.064(42) & 0.729(26) & 0.967(36) & 1.438(63) & 0.846(31) & 1.092(42) & 1.606(74) & 1.112(44) \\ [1.0ex]
     $R^{\tau}_{\ell}$ & 0.901(34) & 1.064(42) & 0.729(26) & 0.966(36) & 1.435(62) & 0.846(31) & 1.091(42) & 1.601(74) & 1.111(44) \\ [1.0ex]
      \hline
    \end{tabular}    \caption{The ratio $R$ (Eq.~\eqref{eq:R}) for the $\tau$ case to the electron, muon and $\ell$, (which is the average of the two) cases, integrated over some commonly used $q^2$ bins. Numerical values for $q^2$ bins are in units of GeV$^2$. No uncertainty to allow for QED effects is included in these numbers.}
    \label{tab:Rtau}
  \end{center}
\end{table*}
\begin{table*}
  \begin{center}
    \begin{tabular}{ c | c c c c c c c c c }
      \hline
     $q^2$ bin & $(4m_{\tau}^2,q^2_{\text{max}})$ & $(14.18,q^2_{\text{max}})$ & $(14.18,16)$ & $(16,18)$ & $(18,22)$ & $(15,17)$ & $(17,19)$ & $(19,22)$ & $(15,22)$ \\ [0.5ex]
      \hline
      \hline
     $F_H^{\tau(+)}$ & 0.8843(46) & 0.8743(50) & 0.9105(32) & 0.8667(50) & 0.8543(64) & 0.8871(41) & 0.8525(57) & 0.8576(66) & 0.8656(54) \\ [1.0ex]
     $F_H^{\tau(0)}$ & 0.8848(46) & 0.8747(50) & 0.9106(32) & 0.8669(50) & 0.8551(64) & 0.8873(41) & 0.8529(57) & 0.8586(66) & 0.8661(54) \\ [1.0ex]
      \hline
    \end{tabular}    \caption{$F^{\tau}_H$ (Eq.~\eqref{eq:F_H}) integrated over some commonly used $q^2$ bins for charged and neutral $B$ meson decays. Numerical values for $q^2$ bins are in units of GeV$^2$. No uncertainty to allow for QED effects is included in these numbers.}
    \label{tab:Ftau}
  \end{center}
\end{table*}
\end{appendix}

\bibliography{BKphenopaper}
\end{document}